\documentclass[11pt, fleqn]{article}

\usepackage{amsmath}
\usepackage{amssymb}
\usepackage{amsfonts}
\usepackage{amsthm}
\usepackage{dsfont}
\usepackage{mathrsfs}

\usepackage{enumerate}

\usepackage[arrowdel]{physics}
\usepackage{tensor}

\usepackage{simplewick}
\usepackage{feynmf}

\usepackage{tabu}
\usepackage{physics}
\usepackage{mathdots}

\newcommand{\operator}[1]{\hat{#1}}
\newcommand{\identity}{\mathds{1}}
\newcommand{\R}{\mathbb{R}}
\newcommand{\C}{\mathbb{C}}
\newcommand{\Z}{\mathbb{Z}}

\newcommand{\T}{\mathbb{T}}

\newcommand{\SU}[1]{SU(#1)}

\newcommand{\Sp}[1]{Sp(#1)}
\newcommand{\Spin}[1]{Spin(#1)}
\newcommand{\E}[1]{E_{#1}}

\newcommand{\su}[1]{\mathfrak{su}(#1)}
\newcommand{\sym}[1]{\mathfrak{sp}(#1)}

\newcommand{\lie}[1]{\mathfrak{#1}}

\newcommand{\vecb}{\boldsymbol}

\newcommand{\pauli}[1]{
    \ifnum#1=1
        \operator{\sigma}_{x}
    \else
        \ifnum#1=2
           \operator{\sigma}_{y}
        \else
            \ifnum#1=3
                \operator{\sigma}_{z}
            \else
                \errmessage{Incorrect number given to pauli}
            \fi
        \fi
    \fi
}

\usepackage{geometry}
\usepackage{comment}

\usepackage{soul}
\usepackage{subcaption}

\usepackage{jheppub}
\usepackage{amsmath,amssymb,amsthm}
\usepackage{bm}
\usepackage{slashed}
\usepackage{graphicx}
\usepackage[T1]{fontenc}
\usepackage{fix-cm}

\usepackage{tabu}
\usepackage{physics}

\makeatletter
\newcommand*\bigcdot{\mathpalette\bigcdot@{.5}}
\newcommand*\bigcdot@[2]{\mathbin{\vcenter{\hbox{\scalebox{#2}{$\m@th#1\bullet$}}}}}
\makeatother

\newlength{\dummysp}
\usepackage[font=scriptsize]{caption}
\usepackage{mwe}

\def\R{{\mathbb R}}
\def\S{{\mathbb S}}
\def\Z{{\mathbb Z}}
\def\T{{\mathbb T}}

\def\tr{\,{\rm tr}\,}

 \def\eps{\varepsilon}
 
\title{The mixed  $\mathbf 0$-form/$\mathbf 1$-form anomaly in Hilbert space: pouring the new wine into old bottles}
\author{Andrew A. Cox, Erich Poppitz, F. David Wandler}
\affiliation{Department of Physics, University of Toronto, Toronto, ON M5S 1A7, Canada}
\emailAdd{aacox@physics.utoronto.ca} \emailAdd{poppitz@physics.utoronto.ca}  \emailAdd{f.wandler@mail.utoronto.ca}    

\abstract{

{\flushleft{W}}e study four-dimensional gauge theories with arbitrary simple gauge group with  $1$-form   global center symmetry and $0$-form   parity   or discrete chiral symmetry. We canonically quantize on $\T^3$, in   a fixed background field gauging the  $1$-form symmetry. We show that the mixed $0$-form/$1$-form 't Hooft anomaly  results in a central extension of the global-symmetry operator algebra. We determine this algebra in each case and show that the anomaly implies degeneracies in the spectrum of the Hamiltonian at any finite-size torus. We discuss the consistency of these constraints with  both older and  recent semiclassical calculations in $SU(N)$ theories, with or without adjoint fermions, as well as with their conjectured infrared phases.

 }

\begin{document}

\maketitle

\section{Introduction: motivation and a brief description}
\label{sec:introduction}

The anomaly matching conditions of  't Hooft  offer an important consistency check on possible scenarios for the nonperturbative behaviour of gauge theories \cite{tHooft:1979rat}. ``Traditional'' anomaly matching of continuous global symmetries acting on local operators ($0$-form symmetries) has played an important role in the construction of models of quark and lepton compositeness (see \cite{Rosner:1998wh} for  a review) and have provided checks on various nonperturbative dualities \cite{Seiberg:1994pq}. 

In recent years, new ``generalized'' anomaly matching conditions  involving discrete $0$-form and $1$-form symmetries have attracted interest. It turns out that they impose further nontrivial constraints on the infrared (IR) phases of gauge theories, including symmetry realization, the ordering of thermal phase transitions, and on the worldvolume physics of domain walls and interfaces. These studies, initiated in \cite{Gaiotto:2014kfa,Gaiotto:2017yup,Gaiotto:2017tne}, were followed by many important works (the literature is by now too large to account for here).

In this paper, we study a specific set of 't Hooft anomalies: the mixed anomalies between discrete $0$-form and $1$-form symmetries. The prime example in four dimensions is the mixed anomaly between parity (or time reversal) and  the $1$-form center symmetry of pure Yang-Mills theory at $\theta = \pi$ \cite{Gaiotto:2017yup,Gaiotto:2017tne}. A related example \cite{Gaiotto:2014kfa} is the mixed anomaly between discrete chiral symmetry and center symmetry in super Yang-Mills theory,  also present in Yang-Mills with adjoint fermions (QCD(adj))\cite{Shimizu:2017asf,Komargodski:2017smk} and in generalizations like   \cite{Anber:2019nze,Anber:2020gig,Anber:2021lzb}.

We wish to understand these anomalies from a  traditional point of view: the canonical quantization of four-dimensional (4d) gauge theories. Part of our motivation is that such an  understanding exists in two-dimensional (2d) models. Its utility, notably its immediate implications for the spectrum of the Hamiltonian, stimulates our curiosity and desire to extend a similar understanding to 4d.\footnote{We note that ref.~\cite{Delmastro:2021xox} also  found torus Hilbert spaces useful to study various  global anomalies in $d=1,2,3$.} The simplest 2d example is the Schwinger model with massless fermions of quantized charge $q \ge 2$ \cite{Anber:2018jdf}.  There is a mixed anomaly  between the $\Z_{2q}^{(0)}$ chiral symmetry and the $\Z_q^{(1)}$ center symmetry.\footnote{We use a superscript $G^{(0)}$ to denote $0$-form symmetry groups and $G^{(1)}$ for $1$-form symmetries.} The anomaly can be seen after a careful study of the symmetry algebra in the quantum theory. In 2d, this analysis can be performed in either the original fermion description \cite{Anber:2018jdf} or in the bosonized formulation \cite{Armoni:2018bga, Misumi:2019dwq}. The result is that, as in earlier quantum-mechanical examples \cite{Gaiotto:2017yup}, the anomaly is reflected in a central extension of the algebra of symmetry operators. The extended symmetry algebra can be seen to imply a $q$-fold degeneracy of all energy eigenstates.\footnote{A subtlety specific to the $d$$=$$2$ case  with a ($d$$-$$1$$=$$1$) $1$-form symmetry, as in the 2d Schwinger model, is that the degeneracy gives rise to different  so-called ``universes'' \cite{Komargodski:2020mxz,Cherman:2020cvw}, where domain walls have infinite tension \cite{Anber:2018xek}, rather than to different vacua.} There is by now substantial literature investigating the structure of this and other 2d models from varying perspectives, see e.g. \cite{Pantev:2005zs,Komargodski:2017dmc,Sulejmanpasic:2018upi,Tanizaki:2018xto,Sharpe:2019ddn,Cherman:2019hbq,Nguyen:2021naa,Smilga:2021zrw}.

We shall show that 4d gauge theories with a mixed discrete $0$-form/$1$-form anomaly also give rise to a centrally-extended symmetry algebra. An indication for this has been seen before: on $\R^3 \times \S^1$, using the semiclassical solution of deformed $SU(N)$ Yang-Mills (dYM) theory \cite{Unsal:2008ch}, a central extension of the IR-theory symmetry algebra at $\theta = \pi$ was found in \cite{Aitken:2018kky} (also, for a  particular ``mixed''  $SU(2)$ gauge theory at $\theta=\pi$, see \cite{Gaiotto:2017yup}).

We demonstrate here that such central extensions due to anomalies are quite general. They   lead to a symmetry algebra similar to \cite{Aitken:2018kky} and to  the  2d Schwinger model \cite{Anber:2018jdf},  depending on the anomaly in question. In the bulk of the paper, we show in  detail how the central extension of the algebra emerges from the canonical quantization of 4d theories on $\T^3$, for all simple gauge groups with nontrivial centers,  namely  $SU(N), Sp(N), Spin(N), E_6$, and $E_7$. 

For the experts,  in the remainder of the Introduction we give a brief account of our work. Here, we  use the mixed anomaly between parity and the $\Z_N^{(1)}$ center symmetry at $\theta = \pi$ in pure $SU(N)$ Yang-Mills theory as an illustration. 

Let us first recall the by now usual description of the mixed anomaly  \cite{Gaiotto:2017yup,Gaiotto:2017tne}. One begins by turning on a $2$-form background gauge field for the $1$-form $\Z_N^{(1)}$ symmetry.  
This can be done in the continuum formalism of \cite{Kapustin:2014gua} or by turning on a flat background 2-form (i.e. plaquette) $\Z_N$ gauge field on the lattice---a set of intersecting center vortices (see \cite{Greensite:2011zz} for an introduction). In the Euclidean set-up, it is well known from either formalism that a generic $\Z_N$ $2$-form background, say on $\T^4$, has non-integer topological charge (see \cite{Gaiotto:2017yup,Gaiotto:2017tne} and the older work \cite{tHooft:1981sps,vanBaal:1982ag}, reviewed in \cite{GonzalezArroyo:1997uj}). Because of this non-integer charge, the center-symmetry background violates the invariance under $2\pi$ shifts of the $\theta$-angle. As these shifts are  part of the parity transformation at $\theta = \pi$, parity is explicitly broken.  This breaking of parity is a classic example of a mixed 't Hooft anomaly: the non-dynamical background for the $1$-form symmetry breaks the $0$-form symmetry. 

We now describe how we see the mixed anomaly in the canonical formalism on a spatial $\T^3$. We begin by turning on a fixed $2$-form center-symmetry $\Z_N$ background in the spatial directions. It is well known that such a background amounts to changing the co-cycle conditions for the $SU(N)$ transition functions on $\T^3$ to ones appropriate to an $SU(N)/\Z_N$ bundle \cite{vanBaal:1982ag}. Equivalently, one is led to consider a fixed 't Hooft magnetic flux (or twist) sector, labeled by a three-vector\footnote{We use $\vec{a}$ to denote vectors in $\R^3$, and reserve the bold-face symbol $\vecb{a}$ for weight-lattice vectors.} $\vec m$ with  integer (mod $N$) components \cite{tHooft:1979rtg}.

Next, we quantize the $\T^3$ gauge theory with boundary conditions twisted by $\vec{m}$. Following  the earlier work of refs.~\cite{tHooft:1981sps,vanBaal:1984ra,GonzalezArroyo:1988dz}, we proceed to explicitly define the  operators performing $1$-form center symmetry transformations in the spatial directions,  the operators performing a spatial reflection, and  the operators performing $2\pi$ shifts of the $\theta$-angle. Most importantly, the operator generating the $1$-form center symmetry in a direction parallel to $\vec{m}$ does not commute with the operator generating $2\pi$ shifts of the $\theta$ angle.
We then show that  this non-commutativity leads, at $\theta = \pi$, to a central extension of the algebra of the parity and center symmetries.\footnote{The $\vec{m} \ne 0$ (mod $N$) twist breaks charge conjugation (for $N>2$) but preserves parity, so in the following we discuss $P$ and not $CP$. In $SU(N)$, the  parity and $\Z_N^{(1)}$ (along $\vec{m}$) $1$-form symmetry generators at $\theta=0$ obey a $D_{N}$ group algebra, which   is centrally extended at $\theta=\pi$  (as per the usual notation of e.g. \cite{Ramond:2010zz}, $D_N$ is the dihedral group of order $2N$). For the  mixed chiral-center anomaly, the  centrally-modified algebra is  as found in the 2d Schwinger model \cite{Anber:2018jdf}.} For even $N$, the centrally extended algebra at $\theta = \pi$ implies a 2-fold degeneracy of the energy eigenstates for any size $\T^3$. In the infinite volume limit, assumed to be unique, this degeneracy indicates the spontaneous breaking of parity. 
The  spontaneous breaking of parity at $\theta=\pi$ in pure $SU(N)$ gauge theory is borne out by older large-$N$ \cite{Witten:1980sp} and other arguments, reviewed in \cite{Gabadadze:2002ff}, and by explicit semiclassical calculations\footnote{The results   reported in this paper also  originate  from work on extensions of our recent study \cite{Poppitz:2020tto}  of anomalies in the calculable $\R^3 \times \S^1$ framework.} on $\R^3 \times \S^1$ from the past decade 
 \cite{Unsal:2012zj,Poppitz:2012nz,Anber:2013sga,Bhoonah:2014gpa,Anber:2017rch,Aitken:2018mbb}. Finally, there is  recent lattice evidence for $SU(2)$ \cite{Kitano:2021jho}.

Let us now make a comment pertaining to the second line of our title. The double-degeneracy at $\theta=\pi$, exactly as implied by the centrally-extended algebra, was seen in semiclassical  calculations of  the instanton-induced splitting of 't Hooft electric flux energies in the background twisted by $\vec{m}$, in the framework of the ``femto-universe,''  where the entire $\T^3$ is taken smaller than $\Lambda^{-1}$. These calculations date back to the 1980's (see \cite{vanBaal:1984ra,vanBaal:2000zc} and section \ref{sec:algebrathetadiscussion}). In fact, as we shall see below, all ingredients needed to see the mixed anomaly and the emergence of a centrally-extended algebra at $\theta = \pi$  are contained therein. We stress, however, that the extension of the algebra and its interpretation as reflecting the mixed 't Hooft anomaly involving parity and the $1$-form symmetry is new. 

To end the introduction, we express our hope that the new wine will, in time, improve in the old bottles. The relationship between newer and  older developments does not appear to be widely appreciated and we believe that working out the details pertaining to the new insights will, apart from possible pedagogical advantage, benefit the further explorations of gauge dynamics and implications of anomalies. Some possible  venues for future studies are discussed in the text.

\section{Outline and summary}

\label{sec:outline}

Here, we offer a guide through the various sections of this paper, along with  only a  brief  mention of the results. Readers interested primarily in the implications of the anomaly in Hilbert space can proceed to sections \ref{sec:algebrathetadiscussion} and \ref{sec:algebrachiraldiscussion}, where the consequences of the centrally-extended symmetry algebras for $SU(N)$ gauge groups are discussed.

The bulk of the paper is section \ref{sec:T3canonical}, devoted to a detailed description of 
the quantization of  $SU(N)$ gauge theories on $\T^3$ in the fixed background of a $2$-form $\Z_N$ gauge field of the $1$-form $\Z_N^{(1)}$ global symmetry. Section \ref{sec:T3fluxbasics} explains all the basic ingredients needed to derive the parity/center-symmetry operator algebra at $\theta = \pi$ in the pure gauge theory, as well as the parity/chiral algebra in the theory with $n_f$ massless adjoint Weyl fermions, QCD(adj). 
In particular, section \ref{mequals1} introduces the minimal 't Hooft flux background $\vec{m} = (0,0,1)$, which we find very instructive and useful to illustrate the extension of the global-symmetry algebras in all cases.

The centrally-extended algebra of the parity and center-symmetry operators at $\theta = \pi$ is derived in section \ref{sec:algebratheta}.  The main implication of the extended algebra is that, for even-$N$, there is a two-fold degeneracy of all energy eigenstates on $\T^3$ at $\theta = \pi$, while for odd-$N$, one finds a global inconsistency between $\theta = 0$ and $\theta=\pi$. A more detailed discussion is given in section \ref{sec:algebrathetadiscussion}. There, we also review the older and more recent  semiclassical calculations on $\T^3$ and $\R^3 \times \S^1$ and outline a few directions for future studies.

We next turn to QCD(adj) and study the algebra of the discrete chiral symmetry and center-symmetry operators. It is derived in section \ref{sec:algebrachiral} and a discussion of its implications is given in section \ref{sec:algebrachiraldiscussion} (again, we recommend that the interested reader jump to  section \ref{sec:algebrachiraldiscussion}). Here, the extended symmetry algebra implies  an $N$-fold degeneracy of all $\T^3$ energy eigenstates. We discuss the consistency of this degeneracy with the proposed phases of QCD(adj) with various $n_f$ on $\R^3$ and outline some directions for future study.

We then consider the same classes of theories---pure gauge theories at $\theta=0$ and $\pi$ and theories with massless Weyl adjoints---but with arbitrary gauge groups with nontrivial centers. The fractional topological charges for general gauge groups are calculated in several voluminous appendices.\footnote{The appendices are structured as follows. In appendix \ref{appx:groupconventions},  we summarize the relevant group-theory data, in particular our choice of ``convenient representations,'' where the center acts faithfully, and of the ``convenient co-weights'' used to represent the generators of the center of each group, see table \ref{tab:centers}. In appendices \ref{appx:normalizingQ} and \ref{appx:Qforallgroups}, we explicitly calculate the fractional topological charge on $\T^4$ for all gauge groups with a center. These are summarized in table \ref{tab:top_charges}.  For completeness and possible future uses, in table \ref{tab:top_charges} we also  give the fractional charges on non-spin manifolds. These are calculated in appendix \ref{appx:nonspin} using the $\mathbb{CP}^2$ background.}  The results of table \ref{tab:top_charges} are known, but are derived here in an explicit physicist-friendly manner. The same results are also shown in table \ref{tab:top_charges1} of section \ref{sec:anomalyall}, in a form adapted to the canonical quantization  on $\T^3$ in the $2$-form gauge field background.

Armed with these results, in section
\ref{sec:anomalyall} we outline the canonical quantization for general gauge groups in a 't Hooft flux background.
 In section
\ref{sec:parityall}, we find the centrally extended algebra of the parity and center-symmetry operators at $\theta = \pi$ and discuss its implications. The pattern  follows the one found for $SU(N)$:  for groups whose center is of an even order ($Sp(2k+1), Spin(2k), E_7$) the $\theta=\pi$ algebra implies  a double degeneracy of all  eigenstates of the $\T^3$ Hamiltonian, while for groups with an odd-order center ($E_6$)  it results in a global inconsistency. There is no degeneracy/global inconsistency for $Sp(2k)$ and $Spin(2k+1)$.

In section
\ref{sec:chiralall}, we perform a similar analysis for the discrete chiral symmetry and center-symmetry operators for all gauge theories with $n_f$ massless adjoint Weyl fermions. We find that the corresponding  centrally extended algebra  implies that groups with $\Z_2$ or $\Z_2 \times \Z_2$ centers have a double degeneracy of all $\T^3$ energy eigenstates, while groups with a $\Z_3$ or $\Z_4$ center have a three- or four-fold degeneracy, respectively. 

We end with  a few comments on  future studies. We 
believe that the Hilbert space interpretation of the mixed $0$-form/$1$-form anomaly and the associated degeneracies will prove useful in studies of gauge theory dynamics. 
For example, the degeneracies between different electric-flux states might be useful in lattice studies, especially for the $\theta=\pi$ theories. As also mentioned in the body of the paper, the degeneracies have interesting implications for the controlled semiclassical studies on $\R^3 \times \S^1$ and it would be of interest to confront them with explicit calculations. It would also be desirable to improve our understanding of the implications of the anomaly in the infinite-volume limit,  both in the semiclassically calculable domain and in more general cases, notably ones believed to flow to conformal field theories.

\section{Quantization on $\T^3$ in a $2$-form gauge background and the anomaly: $\mathbf{SU(N)}$}
\label{sec:T3canonical}

In this section, we describe the canonical quantization of pure Yang-Mills theory on $\T^3$ with twisted boundary conditions, corresponding to introducing a  fixed $2$-form $\Z_N$ gauge background field.  We   use  results of 
\cite{tHooft:1981sps,vanBaal:1982ag,vanBaal:1984ra,GonzalezArroyo:1988dz},\footnote{Pierre~van~Baal's~Ph.D.~thesis \cite{vanBaal:1984ra} is at \begin{center} \url{https://www.lorentz.leidenuniv.nl/research/vanbaal/DECEASED/HOME/PHD/thesis.html}.\end{center} Chapter III contains relevant unpublished results. See also \cite{GonzalezArroyo:1988dz}.
 }  but  attempt to make our presentation as self-contained as possible.
 
The addition of adjoint fermions obeying the same boundary conditions on $\T^3$ as the gauge fields is trivial. The various modifications necessary will be mentioned when we discuss the chiral symmetry. Related discussions appear in the   calculation of the Witten index of 4d ${\cal N} =1$ supersymmetric Yang-Mills theory \cite{Witten:1982df,Witten:2000nv}.

In order to simplify the presentation, in this section we describe the quantization appropriate to $SU(N)$ gauge theories with boundary conditions twisted by $\Z_N$ ``magnetic flux'' $\vec{m}$. In later sections, we shall discuss other gauge groups with nontrivial centers. In most cases, the centers are cyclic groups, $\Z_k$, with $k=2,3,4$, and the effect of the   corresponding twists amounts to simply replacing $\Z_N$ by the appropriate $\Z_k$ in all results for $SU(N)$.\footnote{The only exception, $Spin(4N)$, has a $\Z_2 \times \Z_2$ center and requires two twist vectors. This modification will also be easily accommodated.}

\subsection{Canonical quantization in the $2$-form $\Z_N$ magnetic flux background}
\label{sec:T3fluxbasics}

\subsubsection{Preliminaries}
Before we discuss the canonical quantization on $\T^3$, we discuss some general features about gauge fields on $\T^4$.
For all the following, we will be considering a $SU(N)$ gauge connection, $A$. We also introduce shorthand notation for the action of a gauge transformation, $U$, acting on $A$: 
\[
U \circ A = U AU^{-1} - iU d U^{-1}.
\]
Considering $A$ on a compact Euclidean space(time), it is impossible to consistently define $A$ globally; instead, one must define it on local coordinate patches which are connected with transition functions. These are elements of the gauge group defined on the overlaps between patches. On a 4d torus, it is possible to expand one coordinate patch to cover the entire space(time) parameterized by $[0,L_0] \times [0,L_1] \times [0,L_2]  \times [0,L_{3}]$, where $L_\mu$ is the circumference of the $x^\mu$-direction (these are Euclidean spacetime directions and we shall sometimes interchangeably use $L_0$ to denote $\beta$, the inverse temperature). The transition functions can then be understood as boundary conditions. We denote the boundary condition around the $x^\mu$-direction by an $SU(N$) group element $\Omega_\mu$ such that 
\[
A(x^\mu = L_\mu) = \Omega_\mu \circ A(x^\mu = 0)
\]
Notice that in general $\Omega_\mu$ is a function that depends on all the space(time) coordinates except for $x^\mu$ and that, as implied by consistency, $\Omega_\mu$ also transform under gauge transformations, $\Omega_\mu \rightarrow U(x^\mu=L_\mu) \Omega_\mu U(x^\mu=0)$. There is also a consistency condition on the boundary conditions that follows from the usual co-cycle condition of transition functions:
\[
\Omega_\mu(x^\nu = L_\nu) \Omega_\nu(x^\mu = 0) = \Omega_\nu(x^\mu = L_\mu) \Omega_\mu(x^\nu = 0)
\] 

Introducing the 2-form background field for the center symmetry changes this condition. In particular, it introduces the following $\Z_N$ phases:
\begin{equation}
\label{eqn:twisted_bc}
\Omega_\mu(x^\nu = L_\nu) \Omega_\nu(x^\mu = 0) = \Omega_\nu(x^\mu = L_\mu) \Omega_\mu(x^\nu = 0) e^{i {2\pi \over N} n_{\mu\nu}}
\end{equation}
The integers (mod $N$) $n_{\mu\nu}$ are completely determined by the background field. Note that $n_{\mu\nu}$ is antisymmetric.  For the purposes of canonical quantization, we find it useful to break up $n_{\mu\nu}$ into spatial and temporal parts, via the following definitions:
\begin{equation}\label{defk}
k_i \equiv n_{i0}
\end{equation}
and 
\begin{equation}\label{defm}
n_{ij} \equiv \eps_{ijk} m_k.
\end{equation}
For use below note that, with $\epsilon_{0123}=1$, we have ${\rm{Pf}}(n) = {1 \over 8} \eps_{\mu\nu\lambda\sigma} n_{\mu\nu} n_{\lambda\sigma} = - \vec{k} \cdot \vec{m}$.

The relation to the formalism of \cite{Kapustin:2014gua} can be briefly stated as follows. Given an explicit 2-form $\Z_N$ gauge field, $C^{(2)}$, defined there, one can find $n_{\mu\nu}$ by integrating $C^{(2)}$ over the $\mu\nu$-plane (which forms a closed torus). This integral results in \[\oint C^{(2)} = \frac{2\pi n_{\mu\nu}}{N} + 2\pi \Z.\] Here the antisymmetry of $n_{\mu\nu}$ is a product of the choice of orientation of the $\mu\nu$ two-torus. 

\subsubsection{Hilbert space and $\Z_N^{(1)}$ center symmetry}
{\flushleft{N}}ow, to canonically quantize the $A$ field in the presence of a center background, we follow a series of steps:
\begin{enumerate}
\item Pick boundary conditions that satisfy \eqref{eqn:twisted_bc} on the spatial torus $\T^3$. Notice that here we only need to consider the spatial part $\vec{m}$ of $n_{\mu\nu}$, the temporal part will come in later. It turns out \cite{vanBaal:1982ag} that any choice of boundary condition that give the same $\vec{m}$ are necessarily equivalent up to a gauge transformation. In particular, it is always possible to find constant matrices, $\Gamma_i \in SU(N)$, such that the co-cycle conditions \eqref{eqn:twisted_bc} are satisfied by $\Omega_i = \Gamma_i$, i.e. they read 
\begin{equation}\label{gammacocycle}
\Gamma_k \;\Gamma_l = \Gamma_l\; \Gamma_k \;e^{i {2 \pi \over N} \eps_{klj} m_j}.
\end{equation} 
As an example, consider the  ``clock and shift'' matrices obeying $W_P W_Q = \omega W_Q W_P$ with $\omega = e^{2\pi i / N}$:
\begin{equation}\label{wpq}
\begin{split}
W_P = \alpha \begin{pmatrix} 0 & 1 & 0 &  \dots  & 0 \\ & 0 & 1 & \dots & \\ \vdots & & \ddots & & \vdots \\  & & & 0 & 1 \\ 1 & & \dots & & 0 \end{pmatrix},~~
W_Q =  \beta \begin{pmatrix} 1 & & & & \\ & \omega & & & \\ & & \omega^2 & & \\ & & & \ddots & \\  & & & & \omega^{N-1} \end{pmatrix},
\end{split}
\end{equation}
where $\alpha$ and $\beta$ are constants that ensure $\det W_P = \det W_Q = 1$.  Boundary conditions with transition functions of the form $\Gamma_i = W_Q^{q_i} W_P^{p_i}$ then correspond to $\vec{m} = \vec{p} \times \vec{q}$.

 One can find (though not uniquely) suitable $\vec{p}, \vec{q} \in \Z^3$ for any $\vec{m} \in \Z^3$, so boundary conditions of this form will always suffice  \cite{tHooft:1981nnx,vanBaal:1982ag}. From now on in this paper, make the choice of constant $\Omega_i = \Gamma_i$. Notice that the choice of constant boundary conditions implies $A = 0$ is a valid background.

\item Borrowing notation from 't Hooft \cite{tHooft:1981sps}, construct a Hilbert space of $A$ fields that satisfy the chosen boundary conditions and the gauge condition $A_0 = 0$.\footnote{This gauge condition may appear to not allow for non-trivial Polyakov loops; however, the Polyakov loop will be determined by imposing temporal boundary conditions.} This results in the large Hilbert space: \begin{eqnarray}\label{bc}
&& \mathcal{H} =  \left\{ \ket{A} \right. \\&& \left. | A(L_1,y,z)  = \Gamma_1 \circ A(0,y,z), A(x,L_2,z)  = \Gamma_2 \circ A(x,0,z), A(x,y,L_3)  = \Gamma_3 \circ A(x,y,0)  \right\}, \nonumber 
\end{eqnarray} where $\ket A$ stands for an eigenvector of the ``position'' operator $\hat A(\vec{x})\ket{A} = \ket{A} A(\vec x)$.
Consider the set of gauge transformations preserving the boundary conditions (\ref{bc}) \begin{eqnarray}
\label{eqn:gt_def}
&&\left\{U:SU(N) \rightarrow  \T^3 \right. \\ &&\left. | U(L_1,y,z)   = \Gamma_1 U(0,y,z) \Gamma_1^{-1}, U(x,L_2,z)  = \Gamma_2 U(x,0,z) \Gamma_2^{-1}, U(x,y,L_3)  = \Gamma_3 U(x,y,0)  \Gamma_3^{-1}\right\}. \nonumber
\end{eqnarray}
A gauge transformation $U$ uniquely determines an operator on the large Hilbert space by the relation
\begin{equation}
\hat{U} \ket{A} = \ket{U \circ A}~.
\end{equation} 
Gauss' law requires that the physical states $\ket\psi \in \mathcal{H}$ obey $\hat U \ket\psi = \ket\psi$, i.e. are invariant under gauge transformations $U$, which obey (\ref{eqn:gt_def}) and are homotopic to the identity.

In addition to gauge transformations homotopic to the identity, maps from $\T^3$ to $G$ are also characterized by their instanton number $\nu$, associated\footnote{An explicit example for a $\nu=1$ map $\T^3 \rightarrow SU(2)$, obeying the boundary conditions (\ref{eqn:gt_def}), is $T_3^2$, with  $T_3$ of eqn.~\eqref{t3su2}. } with $\pi_3(G)$. These ``large''
gauge transformations do not leave physical states invariant but act as 
\begin{equation}\label{Htheta}
{\cal{H}}^{phys.}_\theta = \left\{\ket{\psi} \in \mathcal{H} \,:\, \hat{U} \ket{\psi} = e^{-i\theta \nu} \ket{\psi}, \forall U\right\}
\end{equation}
where $\nu$ is the instanton number associated with the transformation $U$ ($\nu$ vanishes for the ``small'' gauge transformations \eqref{eqn:gt_def}). ${\cal{H}}^{phys.}_\theta $ defines the physical Hilbert space, where all vectors have  definite theta angle. 

\item In terms of the position,\footnote{\label{note1}In this section, we use fundamental hermitean generators  with $\tr T^a T^b = \delta^{ab}/2$ and $[T^a,T^b]=i f^{abc}T^c$. In form notation, to be used later, $A = A_\mu^a T^a dx^\mu$, $F = d A + i A \wedge A$.}  $\hat A_i^a(\vec{x})$,   and momentum, $\hat \Pi_i^a(\vec{x}) = - i {\delta \over \delta A_i^a(\vec{x})}$, operators, the Hamiltonian in the physical Hilbert space is 
 \begin{equation} \label{hamiltonian}
 \hat{H} = \int\limits_{\T^3} d^3 x \left( {g^2} \;\tr \hat \Pi_i  \hat \Pi_i + {1 \over  g^2}\;\tr \hat B_{i} \hat B_{i}\right), ~~[ \hat\Pi_i^a(\vec{x}), \hat A_j^b(\vec{y})] = -i \delta^{ab} \delta_{ij} \delta^{(3)}(\vec{x}-\vec{y}).
 \end{equation}
Here $\hat{B}_i = {1 \over 2} \eps_{ijk} \hat F_{jk}$, $\hat F_{ij}$ is   given in footnote \ref{note1}, and  the operators $\hat \Pi_i (\vec{x})$ and $\hat{A}_i (\vec{x})$ obey the  boundary conditions (\ref{bc}) twisted by  $\Gamma_j$. 

The perturbative expansion of the spectrum of  $\hat{H}$ in a small $\T^3$ was studied in \cite{GonzalezArroyo:1988dz} (for $\vec{m} \sim (0,0,1)$ or $(1,1,1)$). See also \cite{vanBaal:2000zc} for nonperturbative instanton-based results that we shall return to later. We stress that our focus here is not on calculational aspects, which can become technically involved. Instead we focus on the representation of the symmetries and their anomalies in Hilbert space.
 \item On $\T^3$, in addition to transformations used to define the physical Hilbert space ${\cal{H}}^{phys.}_\theta$, one can perform transformations on the fields (here, ``$C\: $'' stands for center, for reasons explained below) that look like gauge transformations
 \begin{equation}\label{oldomega}
 A \rightarrow A' = C[\vec k, \nu] \circ A, \end{equation} with $SU(N)$ group elements $C[\vec k, \nu]$. They  preserve the boundary conditions (\ref{bc}) but themselves do not obey (\ref{eqn:gt_def}). Instead, they obey (\ref{eqn:gt_def}) only up to a center element:
\begin{equation}
\label{eqn:Ti_bcs}
\begin{split}
C[\vec{k},\nu] (L_1,y,z) = &e^{i \frac{2\pi k_1}{N}} \;\Gamma_1 C[\vec{k},\nu] (0,y,z) \Gamma_1^{-1}~, \\
C[\vec{k},\nu] (x,L_2,z) = &e^{i \frac{2\pi k_2}{N}} \;\Gamma_2 C[\vec{k},\nu] (x,0,z) \Gamma_2^{-1} ~,\\
C[\vec{k},\nu](x,y,L_3) = &e^{i \frac{2\pi k_3}{N}} \;\Gamma_3 C[\vec{k},\nu] (x,y,0) \Gamma_3^{-1} ~,
\end{split}~
\end{equation}
which guarantees that $A$ and $A'$ of (\ref{oldomega}) obey the same boundary conditions \eqref{bc}. Thus, $C[\vec k, \nu]$ maps states of $\mathcal{H}$ to states of $\mathcal{H}$. The label $\nu$ indicates that the instanton number of $C$ can be nonzero. In the  literature $C[\vec{k},\nu]$ with $\vec{k} \ne 0$ have been often called ``improper gauge transformations'' (or ``central conjugations'' in \cite{Luscher:1982ma}).  The modern terminology is that \eqref{oldomega} with \eqref{eqn:Ti_bcs} represent the action of global $1$-form symmetries. That this is so is clear from the fact that the only gauge invariant operators they act on are winding Wilson loops. For example, the gauge invariant Wilson loop winding once in $x^l$,\footnote{The insertion of the transition function $\Gamma_l$ in ${\cal{W}}_l$ is required by  invariance under (\ref{eqn:gt_def}).}
\begin{equation}\label{wilsonloop}
{\cal{W}}_l \equiv \tr[{\cal{P}} e^{- i \int\limits_0^{L_l} dx^l A_l} \; \Gamma_l], 
\end{equation}
is multiplied by $e^{i 2 \pi k_l/N}$ upon the action of $C[\vec{k}, \nu]$.

For the discussion that follows, it will be useful to define the three generators of the $1$-form center symmetry, $\hat T_{i}$, by their action on vectors in $\mathcal{H}$ as follows:
\begin{equation}
\label{eqn:Ti_bcs2}
\begin{split}
\hat T_1 \ket A = &\ket{C[(1,0,0), 0] \circ A} \\
\hat T_2 \ket A = &\ket{C[(0,1,0), 0] \circ A} \\
\hat T_3 \ket A = &\ket{C[(0,0,1), 0] \circ A} ~,
\end{split}
\end{equation}
where $(1,0,0)$, etc., denote the components of $\vec{k}$.
The above definition is somewhat open-ended as the $C[\vec{k},0]$ used to define $\hat T_i$ can be multiplied by any small gauge transformation and still satisfy (\ref{eqn:Ti_bcs}). Moreover, the operators $\hat T_i$ must map physical states to physical states. Note however, that for any gauge transformation $U$, the transformation $U' = T_i^\dagger U T_i$ satisfies the conditions of \eqref{eqn:gt_def} and hence is a gauge transformation. Thus, for any physical states $\ket\psi$ and any gauge transformation $U$ we have
\begin{equation}
\hat{U} T_i \ket{\psi} = \hat T_i \hat{U}' \ket{\psi} = e^{-i \theta \nu} \; \hat T_i \ket{\psi}. 
\end{equation}
This demonstrates that $\hat T_i$ map physical states to physical states and that they are well defined on physical states.

Before we continue, we comment on the relation to the modern understanding of $p$-form symmetries in $d$ spacetime dimensions.  These symmetries are represented by topological operators defined on codimension-($p+1$) surfaces in spacetime \cite{Gaiotto:2014kfa}. While this property is not immediately obvious from \eqref{eqn:Ti_bcs2}, we note that one can, instead, use canonical momenta and coordinates to define the unitary operator $\hat T_i$ by an exponential of an integral of an  operator over a $2$-surface in $\R^3$. We will not need such a definition here,\footnote{An analogous definition can be explicitly seen in the 2d Schwinger model, where the $1$-form symmetry is generated by a local operator, as in e.g. \cite{Armoni:2018bga}, or using the Kogut-Susskind lattice Hamiltonian \cite{Kogut:1974ag}. For a related continuum discussion, see also \cite{Reinhardt:2002mb} and the appendix of  \cite{Anber:2015wha}.}  as (\ref{eqn:Ti_bcs2}) suffices for our purposes.

\item When the spatial boundary conditions are twisted by a nonzero $\vec{m}$, the operators $\hat T_i$, and the related\footnote{The operators $\hat C$ are defined analogously to \eqref{eqn:Ti_bcs2} by their action on $\ket A$ via the functions $C[\vec{k}, \nu]$, as in \eqref{oldomega}.} $\hat C[\vec{k},\nu]$ have fractional winding number $\T^3 \rightarrow G$ \cite{tHooft:1981sps}. The winding number    is familiar from Skyrmion physics
\begin{equation} \label{winding1}
  Q[C]={1 \over 24 \pi^2} \int_{\T^3} \tr (C d C^{-1})^3~.
  \end{equation}
and its  fractional nature in the $\vec{m} \ne 0$ background can be explicitly demonstrated as follows. Consider the topological charge on the Euclidean  $\T^4$,
\begin{equation}\label{Qtop}
Q = {1 \over 8 \pi^2} \int  \tr F \wedge F = {1 \over 64 \pi^2} \int d^4 x F_{\mu\nu}^a F_{\lambda\sigma}^a \epsilon^{\mu\nu\lambda\sigma} = \int d^4 x \partial_\mu K^\mu~,
\end{equation} where we defined $
K^\mu = {1 \over 16 \pi^2} \epsilon^{\mu\nu\lambda\sigma} \left( A_\nu^a \partial_\lambda A_\sigma^a - {1 \over 3} f^{abc} A_\nu^a A_\lambda^b A_\sigma^c\right)$.\footnote{For completeness, we defined $f^{abc}$ the usual way, see footnote \ref{note1}.}
Using Stoke's theorem, and assuming that the background $A$ obeys, on the spatial $\T^3$, boundary conditions given by our choice of constant transition functions $\Gamma_i$, we can simplify the topological charge to\begin{equation}\label{K0def}
Q = \int_{\T^3} K_0\left(A\big|_{x^0 =L_0} \right) -  K_0\left(A\big|_{x^0 = 0} \right),~  ~K_0(A) \equiv {1 \over 8 \pi^2} \tr( A\wedge F -{i \over 3} A \wedge A \wedge A),
\end{equation}
Here $2 \pi K_0(A)$ is the Chern-Simons form, normalized to   shift by $2\pi$ under gauge transformations with unit $\T^3 \rightarrow G$ winding number (see \eqref{qomega} below). Now consider a gauge field $A$ on $\T^4$, obeying the spatial boundary conditions \eqref{bc}, and a time-direction twist by $C$, $A\big|_{x^0 =L_0} = C[\vec{k},\nu]\circ A\big|_{x^0 = 0}$ and observe that its topological charge \eqref{K0def} equals the winding number (\ref{winding1}) of  $C$:
\begin{equation}
\begin{split}\label{qomega}
Q[C] &= \int_{\T^3} K_0\left( C \circ A \right) -  K_0\left(A\right) \\
&= {1 \over 24 \pi^2} \int_{\T^3} \tr (C d C^{-1})^3  + { 1  \over 8 \pi^2} \int_{\T^3} \; d \tr (i A \;d C^{-1} C) = {1 \over 24 \pi^2} \int_{\T^3} \tr (C d C^{-1})^3  ~.
\end{split}
\end{equation}
The boundary term in the second line of \eqref{qomega}  vanishes owing to the  boundary conditions (\ref{bc}, \ref{eqn:Ti_bcs}) and the fact that the transition functions $\Gamma_i$ are constant.

In words, we found that the winding number (\ref{winding1}) of the map $C[\vec{k}, \nu]: \T^3 \rightarrow G$ is, by reversing the chain from \eqref{qomega} to (\ref{Qtop}),   equivalent to the topological charge of a field configuration $A$ on $\T^4$, twisted by $C$ in the time direction and by $\Gamma_i$  in space.
Thus, the $\T^4$ transition functions of this field configuration are $\Omega_\mu = (C, \Gamma_1, \Gamma_2, \Gamma_3)$. We now notice that owing to the properties of $C[\vec{k},\nu]$, the integers $\vec{k}$ play the role of twists $n_{i0}$ in the time direction.\footnote{For example, use \eqref{eqn:Ti_bcs} to find $C(x_1=L_1,y,z) \Gamma_1 = \Gamma_1 C(x_1=0,y,z) e^{- i 2 \pi k_1/N}$. Comparing with \eqref{eqn:twisted_bc}, we conclude $n_{01} = - k_1$, as per \eqref{defk}.}
As the topological charge depends only on the twists $n_{\mu\nu}$ and the usual integer instanton number, $\nu$, we can use the result  from \cite{tHooft:1981sps,vanBaal:1982ag} (or consult appx.~\ref{appx:fractionalcharge}) to find the winding number (\ref{winding1}):
\begin{equation}\label{qomega2}
Q[C[\vec{k},\nu]] = -\frac{1}{N} \text{Pf}(n) + \nu   =   \frac{\vec{m}\cdot \vec{k}}{N} + \nu
\end{equation} 
The preceding argument is especially  helpful to find the fractional part of $Q$, as it determined solely by the twists $n_{\mu\nu}$. An explicit expression for $C[\vec{k},\nu](x,y,z)$ (up to small gauge transformations) would   allow us to directly calculate (\ref{qomega}) and yield both the fractional and integer parts,  see also \cite{Witten:2000nv}.\footnote{A concrete example might be useful. Consider the  $Q=1/2$ map $\T^3 \rightarrow SU(2)$, explicitly defined by $T_3(\vec{x})$ in \eqref{t3su2} below. $T_3(\vec{x})$ obeys the boundary conditions (\ref{eqn:Ti_bcs}) with $\vec{m} = (0,0,1)$ and $\vec{k} =(0,0,1)$.  Clearly, $T_3$ and $(T_3)^3$ have the same $n_{\mu\nu}$, but the latter has $Q = 3/2$.}

\item 
In what follows, it suffices to work with the operators generating the $\Z_N^{(1)}$ center symmetry, $\hat T_l$, $l=1,2,3$, from \eqref{eqn:Ti_bcs2},   which, from the discussion above, we define to have 
\begin{equation}\label{chargeofcenter}
Q[T_l] = {m_l \over N}~.
\end{equation}
 For simplicity, we further assume that $m_l$ and $N$ are co-prime, so that $e^{i 2 \pi Q[T_l]}$ is of order $N$, i.e.~$N$ is the smallest power of $\hat T_l$ with an integer topological charge, so that $Q[T_l^N] = m_l$. 
Let $\ket \psi$ denote a state in the physical Hilbert space ${\cal{H}}^{phys.}_\theta$  which is an eigenstate of $\hat T_l$.  From the above, we have that $\hat T^N$ represents a gauge transformation of unit instanton number, thus $\hat T_l^N \ket \psi = \ket \psi e^{- i \theta m_l}$. Then, it must be that \begin{equation}\label{centereigenstate}
\hat T_l \ket \psi = \ket \psi e^{i {2 \pi \over N} e_l - i \theta {m_l \over N}} = \ket\psi e^{i {2 \pi \over N}(e_l - {\theta \over 2\pi} m_l) }~, ~l=1,2,3.
\end{equation}
Here, $e_l$ is a (mod $N$) integer called $\Z_N$ ``electric flux.''\footnote{We stress again that we are working in the theory with fixed $\vec{m}$, so the label $\vec{m}$ is implicit in $\ket\psi$.} The name is justified with the following reasoning \cite{tHooft:1979rtg}: consider a state $\hat{\cal{W}}_l \ket \psi$, obtained from $\ket\psi$ by the action of a fundamental Wilson loop  (\ref{wilsonloop}) winding once in the $x^l$ direction. Then, using $ \hat T_l \hat{\cal{W}}_l \hat T_l^{-1} = e^{i {2 \pi \over N}} \hat{\cal{W}}_l$ and \eqref{centereigenstate}, it follows that $\hat T_l \hat{\cal{W}}_l \ket \psi = \hat{\cal{W}}_l  \ket\psi e^{i {2 \pi \over N} (e_l +1) - i \theta {m_l \over N}}$, i.e. acting with $\hat{\cal{W}}_l$ on the state $\ket \psi$ increases $e_l$ by one unit. Since $\hat{\cal{W}}_l$   inserts an electric flux tube winding in the $x^l$ direction, the interpretation of $e_l$ as electric flux follows. 
Thus, $\hat T_l$  measures the amount of $\Z_N$ electric flux carried by a given state. Electric flux free energies are used as order parameters for confinement, see \cite{Greensite:2011zz}.

As $\hat T_l$ commute with the Hamiltonian \eqref{hamiltonian}, they can be simultaneously diagonalized. Thus, all energy eigenstates on $\T^3$ are labelled by three integers, $\vec{e}$, the $(\Z_N)^3$ discrete electric fluxes. As already mentioned, electric flux energies have been studied analytically, for small $\T^3$, in the ``femto-universe'' framework, or for ``large'' volumes $L_i \gg \Lambda^{-1}$ via numerical simulations. Beginning with L\" uscher's work \cite{Luscher:1982ma}, which took $\vec{m}=0$,  this has been explored for various choices of $\vec{m}$, see the  review \cite{vanBaal:2000zc}.
\end{enumerate}

\subsubsection{An important commutation relation} 

Now we have all the information to begin  discussing the mixed $0$-form/$1$-form anomaly. 
Define the operator 
\begin{equation}\label{defofV}
\hat V_{\alpha}[\hat{A}] = e^{{i \alpha} \int_{\T^3} K_0 (\hat{A})},
\end{equation}
where $K_0$ is given in \eqref{K0def}. From (\ref{qomega}), we know that $\int_{\T^3}K_0$ shifts by an integer $\nu$ under large gauge transformations with instanton number $\nu$. Also, recall that, for any $\ket\psi$ in ${\cal{H}}^{phys.}_\theta$  of (\ref{Htheta}), under a gauge transformation with instanton number $\nu$, we have
$\hat U_\nu \ket \psi = \ket\psi e^{-i \theta \nu}$, hence $\hat U_\nu (\hat V_\alpha \ket\psi) = (\hat V_\alpha \ket\psi) e^{-i (\theta -\alpha)\nu}$. Thus, the operator $\hat V_{\alpha}$ shifts the $\theta$ angle by $-\alpha$.
Further, using \eqref{qomega} and  \eqref{chargeofcenter}, we can find the commutation relation of $\hat V_{\alpha}$ with the center symmetry generators:
\begin{equation}\label{vtcommutator1}
\hat T_l \; \hat V_\alpha[\hat{A}] \; \hat T_l^{-1} = \hat V_\alpha [\hat C[k_i = \delta_{il}, 0] \circ \hat{A}]  =  e^{ i \alpha  \int_{\T^3} \left[K_0 (\hat C[k_i = \delta_{il}, 0] \circ \hat{A}) - K_0 (\hat{A})\right]} \; \hat V_{\alpha} = e^{i \alpha {m_l\over N}} \;\hat V_{\alpha}.
\end{equation}
Another commutation relation involving $\hat V_\alpha$  follows from (\ref{hamiltonian}) and (\ref{defofV}):
\begin{equation}\label{PiVcommutator}
[\hat\Pi_i^a(\vec{x}),\hat V_\alpha] ={ \alpha \over 8 \pi^2} \hat B_i^a(\vec{x})\; \hat V_\alpha~.
\end{equation}
For our purposes, the most important consequence of (\ref{vtcommutator1}) is a relation crucial for our analysis of the anomaly
\begin{equation}
\label{vtcommutator}
\hat T_l \; \hat V_{2\pi} = e^{i 2\pi {m_l\over N}} \;\hat V_{2\pi} \;\hat T_l~,
\end{equation}
showing that $2\pi$ shifts of $\theta$ do not commute with the $1$-form center symmetry in the  $2$-form $\Z_N$ gauge field background labelled by $\vec{m}$.
The relation (\ref{vtcommutator}) is behind  the mixed 't Hooft anomaly between the 1-form center symmetry and 0-form symmetries that involve $2\pi$ shifts of $\theta$, such as the parity symmetry at $\theta = \pi$ or the discrete chiral symmetry in the presence of adjoint fermions. Satisfying these non-trivial algebras requires non-trivial vacuum structure, so we gain useful insight into the IR physics by studying these algebras.

\subsubsection{The case of $\vec{m} = (0,0,1)$}
\label{mequals1}

Before elaborating on these anomalies, we shall write down more explicit details for $\Gamma_i$ and $\hat T_i$ for the choice $\vec{m} = (0,0,1)$.  This is the case considered in \cite{Witten:1982df}, and we found it to be an instructive example.

Following our steps above, we first must find the constant transition functions, or twist matrices, $\Gamma_i$, entering (\ref{bc}, \ref{eqn:gt_def}). Since $\Gamma_3$ must commute with the other two for this choice of $\vec{m}$, we can take it to be the identity. The others are  the ``clock and shift'' matrices \eqref{wpq}:
\begin{equation}
\begin{split}
\Gamma_1 = W_P, ~\Gamma_2 =  W_Q, ~\Gamma_3 = \identity_N, ~ ~\Gamma_1 \Gamma_2 = e^{i {2 \pi \over N}} \Gamma_2 \Gamma_1~, \end{split}
\end{equation}
thus, by (\ref{gammacocycle}), $n_{12} = m_3 = 1$, as desired.
Witten noticed that for this choice of boundary conditions, we can take $T_1$ and $T_2$ constant. In particular, the choices $T_1 = \Gamma_2^{-1}$ and $T_2 = \Gamma_1$ work.\footnote{E.g., by the first relation in (\ref{eqn:Ti_bcs}), $T_1$ has to obey $T_1 = e^{i 2 \pi/N} \Gamma_1 T_1 \Gamma_1^{-1}$, satisfied by $T_1 = \Gamma_{2}^{-1}$, etc.} The fact that these operators are so simple is not surprising, since $m_1 = m_2 = 0$ ensures that they enjoy a trivial algebra with $V_{2\pi}$, as per (\ref{vtcommutator}). The same algebra implies that
$T_3$ is bound to be more complicated. As an explicit example, in $SU(2)$ where $\Gamma_1 \propto \sigma^1$ and $\Gamma_2 \propto \sigma^3$, we find
\begin{equation}\label{t3su2}
T_3(\vec{x}) = e^{i\frac{\pi}{2} \frac{y}{L_2} \sigma^3} e^{i\frac{\pi}{2} \frac{x}{L_1} \sigma^1} e^{-i\pi \frac{z}{L_3} \sigma^3} e^{-i\frac{\pi}{2} \frac{x}{L_1} \sigma^1} e^{-i\frac{\pi}{2} \frac{y}{L_2} \sigma^3},
\end{equation}
which can be seen to obey (\ref{eqn:Ti_bcs}) with $\vec{k} = (0,0,1)$. As  alluded to several times above, from  \eqref{qomega} one can explicitly calculate $Q[(T_3)^n]=n/2$. We will not give an explicit form of $T_3$ for $N>2$, but they do exist.
The algebra \eqref{vtcommutator} now becomes 
\begin{equation}\label{t3vcommutator}
\hat T_3\;  \hat V_{2\pi} = e^{i \frac{2\pi}{N}}\hat V_{2\pi} \; \hat T_3,
\end{equation}
so, recalling \eqref{centereigenstate}, $2\pi$ shifts of $\theta$ change the eigenvalues of $\hat T_3$, $e_3 \rightarrow e_3 +1$.

\subsection{The algebra of parity and $\Z_N^{(1)}$ operators: $\theta = 0$ vs. $\theta = \pi$}
\label{sec:algebratheta}

The parity operation acts on $A$ in the following way:
\begin{equation}\label{parity1}
A(x,y,z)  \rightarrow A^P(x,y,z) =  -\Gamma_P A(L_1-x,L_2-y,L_3-z) \Gamma_P~,
\end{equation}
Here the matrix $\Gamma_P \in SU(N)$,  $\Gamma_P^2 = \pm 1$, is required in order that $A^P(x,y,z)$ also obey the boundary conditions \eqref{bc}. This requires
\begin{equation}
\Gamma_P \Gamma_i \Gamma_P = e^{i\phi} \Gamma_i^{-1}.
\end{equation}
where $e^{i\phi}$ can be any $\Z_N$ phase. With our boundary conditions of the form $\Gamma_i = W_Q^{q_i} W_P^{p_i}$, this is fulfilled by the anti-diagonal matrix 
\begin{equation}\label{gammap1}
\Gamma_P = \gamma \begin{pmatrix} 0 &  \cdots &  1 \\ \vdots & \iddots & \vdots \\ 1 & \cdots & 0  \end{pmatrix},
\end{equation}
with $\gamma$ a factor ensuring $\det \Gamma_P = 1$. Let $\hat P_0$ denote the operator that implements the transformation (\ref{parity1}) on our large Hilbert space. The subscript denotes that this is the correct parity symmetry operator for $\theta = 0$. Notice that also $\hat P_0^2 = 1$ as required.

By considering the above action of $\hat P_0$  and $\hat T_i$ on an arbitrary eigenstate of $A$ in the large Hilbert space, it follows that $\hat P_0 \hat T_i \hat P_0$ acts as a center symmetry transformation $\hat{T}_i' $ with
\begin{equation}\label{tprime}
{T}_i' (x,y,z) = \Gamma_P T_i(L_1 - x, L_2 - y, L_3 - z) \Gamma_P.
\end{equation} 
Now recall that $T_i(\vec{x})$ obeys the boundary conditions \eqref{eqn:Ti_bcs} with $(\vec{k})_j = \delta_{ij}$. Therefore,   \eqref{tprime} implies   that $\hat{T}_i'(\vec{x})$ corresponds to a transformation with $(\vec{k})_j = -\delta_{ij}$ and $\nu = 0$, hence $\hat{T}_i'$ is gauge equivalent to $T_i^\dag = T_i^{-1}$. Therefore, on the space of physical states, we have the  $D_N$ commutation relation\footnote{\label{dihedralreps}The dihedral  group $D_N$ is  defined by \eqref{eqn:P0_algebra} plus $\hat P_0^2 = 1, \hat T_i^N = 1$. $D_N$ has one- and two-dimensional irreducible complex representations. In our notation, the one-dimensional representations correspond to taking $\hat P_0 = \pm 1$ and $\hat T_3 = 1$ or $e^{i \pi}$ ($e_3 = 0$ or $N/2$) for even-$N$,  while $\hat T_3=1$ ($e_3=0$) for odd-$N$. The other representations are parity-partner doublets \cite{Ramond:2010zz}. All this simply  follows from the action of $\hat P_0$ on fluxes.}
\begin{equation}
\label{eqn:P0_algebra}
\hat P_0 \; \hat T_i\; \hat P_0 = \hat T_i^\dag.
\end{equation}
Hence, $\hat P_0$ changes the sign of the eigenvalues of $\hat T_i$, the electric fluxes: $\vec{e} \rightarrow -\vec{e}$. 
Note also that $\hat P_0$ does not change the sign of the magnetic field, $\hat P_0 \hat B_i(x,y,z) \hat P_0 = \Gamma_P \hat B_i(L_1-x, L_2-y, L_3-z) \Gamma_P$, but changes the sign of $\hat \Pi_i$, the electric field. 

In order to study invariance under parity in our formalism, it is convenient to
  move the $\theta$-angle dependence  from the states in ${\cal{H}}^{phys.}_\theta$ , eqn.~(\ref{Htheta}), to the Hamiltonian. This is accomplished by conjugating the latter with $\hat V_{ \theta}$ and working in the Hilbert space ${\cal{H}}^{phys.}_{\theta=0}$ (the $\theta$-dependent Hamiltonian  $\hat H_\theta$ has the same spectrum in the space ${\cal{H}}^{phys.}_{\theta=0}$ as the $\theta$-independent Hamiltonian $\hat H_{\theta=0}$ has in ${\cal{H}}^{phys.}_\theta$ ). 
  The $\theta$-dependent Hamiltonian (\ref{hamiltonian}) then becomes, making use of \eqref{PiVcommutator}:\begin{equation}\label{hamiltoniantheta}\hat H \rightarrow \hat H_\theta \equiv \hat V_{\theta} \hat H \hat V_{\theta}^\dagger =  \int\limits_{\T^3} d^3 x \left( {g^2\over 2} \;( \hat \Pi_i^a -{\theta \over 8 \pi^2} \hat B_i^a)(\hat \Pi_i^a -{\theta \over 8 \pi^2} \hat B_i^a)+ {1 \over 2 g^2}\; \hat B_{i}^a \hat B_{i}^a\right).\end{equation}
For $\theta = 0$, $\hat P_0$, defined via (\ref{parity1}), is the operator generating the parity symmetry: from the remarks after (\ref{eqn:P0_algebra}) it follows that 
 $\hat H_{\theta=0}$ commutes with $\hat P_0$. However, for $\theta \neq 0$, this transformation flips the sign of the theta term, as it reverses the sign of $\hat \Pi_i$, thus parity cannot be a symmetry for almost all non-zero values of $\theta$, with $\theta=\pi$ being the notable exception.  Thus, consider the  action of $\hat P_0$ on the Hamiltonian \eqref{hamiltoniantheta} with $\theta=\pi$ \begin{equation}\label{parityhpi}\hat P_0 \hat{H}_{\theta=\pi} \hat P_0 =  \int\limits_{\T^3} d^3 x \left( {g^2\over 2} \;( \hat \Pi_i^a +{1 \over 8 \pi } \hat B_i^a)(\hat \Pi_i^a +{1 \over 8 \pi} \hat B_i^a)+ {1 \over 2 g^2}\; \hat B_{i}^a \hat B_{i}^a\right) = \hat H_{\theta = - \pi}~. \end{equation}
Now act with $\hat V_{2 \pi}$ on (\ref{parityhpi}), using \eqref{PiVcommutator} as $\hat V_{2 \pi} \hat \Pi_i^a \hat V_{2\pi}^{-1} = \hat \Pi_i^a - {1 \over 4 \pi} \hat B_i^a$, to find
\begin{equation}\label{hamiltoniantheta2}
\hat V_{2\pi} \hat P_0  \hat{H}_{\theta=\pi} \hat P_0 \hat V_{2\pi}^{-1} = \hat{H}_{\theta=\pi}.
\end{equation} In other words,  parity  at $\theta=\pi$ is generated by the operator\begin{equation} \label{paritypi}
 \hat P_\pi = \hat V_{2\pi} \hat P_0~.
\end{equation} Notice that $\hat P_0 \hat V_{2\pi}  \hat P_0 = \hat V_{2\pi}^{-1}$, so $\hat P_\pi^2 = 1$ as required for a parity symmetry. Finally, to
 find the commutator of $\hat P_\pi$ with the center generators,  we use the algebras \eqref{vtcommutator} and \eqref{eqn:P0_algebra}:
\begin{equation}
\label{eqn:Ppi_algebra}\hat T_j \; \hat P_\pi = e^{\frac{2\pi i}{N}m_j} \hat P_\pi \; \hat T_j^\dag.
\end{equation}
Hence, $\hat P_\pi$ sends $\vec{e}$ to $\vec{m} - \vec{e}$. The algebra \eqref{eqn:Ppi_algebra} is a central extension of the $D_N$ algebra  \eqref{eqn:P0_algebra}. 

To see the implications of the algebras \eqref{eqn:P0_algebra} and \eqref{eqn:Ppi_algebra}, consider, with no loss of generality,  the  background $\vec{m} = (0,0,1)$ of section~\ref{mequals1}. 
Let us summarize our knowledge of the parity and center symmetries in this background. 
The operators $\hat T_1$ and $\hat T_2$ commute with the Hamiltonian, as well as with $\hat P_\pi$ and $\hat T_3$. The interesting part of the algebra is: 
\begin{equation}\label{algebra1}
[\hat{T}_3, \hat{H}_{\theta = \pi}] =0~, ~~ [\hat{P}_\pi,\hat{H}_{\theta = \pi}]= 0~,~~
\hat{T}_3 \hat P_\pi = e^{i {2 \pi \over N}} \hat P_\pi \hat T_3^\dagger~,
\end{equation}
where  $\hat P_\pi^2 =1$ and $\hat T_3^N =1$, where we recall that we are working in ${\cal{H}}^{phys.}_{\theta=0}$.
Clearly, every energy eigenstate can also be labeled by the value of discrete electric flux, $e_3$\footnote{As well as by $e_1$ and $e_2$, the eigenvalues of $\hat T_{1,2}$. However,  the symmetry algebra does not imply degeneracies between states labeled by different $e_1$ and $e_2$, as $\hat T_{1,2}$ commute with $\hat P_\pi$, $\hat H_{\theta = \pi}$, and $\hat T_3$. Hence to avoid cluttering, we omit denoting the energy eigenstate by  $|E, e_1, e_2, e_3\rangle$.} (of course, finding what values of $e_3$  a given energy eigenstate has requires solving for the spectrum). Let us denote the energy eigenstate by $|E, e_3\rangle$, where $\hat{H}_{\theta = \pi}| E, e_3\rangle = | E, e_3 \rangle E$ and $\hat T_3 | E, e_3\rangle = | E, e_3\rangle e^{i {2 \pi \over N} e_3}$. By (\ref{algebra1}), the state $\hat P_\pi | E, e_3\rangle $ is also a eigenstate of $\hat H_{\theta = \pi}$ of the same energy $E$. In addition, from the last commutator in (\ref{algebra1}), it obeys $\hat T_3 (\hat P_\pi | E, e_3\rangle) = (\hat P_\pi | E, e_3\rangle) e^{i {2 \pi \over N}(1-e_3)}$, i.e. is an eigenstate of $\hat T_3$ of electric flux $1-e_3$. Note that this could be the same state, should it happen that $1-e_3 = e_3 \;(\text{mod} N)$, see below.

Thus, we have shown that the algebra (\ref{eqn:Ppi_algebra}) implies that the eigenstates of $\hat H_{\theta = \pi}$ on $\T^3$ with boundary conditions twisted by $\vec{m} = (0,0,1)$ have certain degeneracies. In particular, parity relates eigenstates of $\hat H_{\theta=\pi}$ of the same energy $E$
\begin{equation}\label{paritydegeneracy}
\hat P_\pi: | E, e_3\rangle \rightarrow |E, 1-e_3 \;({\rm mod} N) \rangle~.
\end{equation}
The implications of the above equation are different for even and odd $N$ as we discuss below. 

\subsubsection{Discussion}
\label{sec:algebrathetadiscussion}
{\flushleft{L}}et us now comment on the implications of the algebra \eqref{algebra1} and eqn.~\eqref{paritydegeneracy}, as well as on their manifestation in various calculable setups. Unless stated otherwise, the comments below refer to the $\theta=\pi$ theory in the $\vec{m} = (0,0,1)$ background.
\begin{enumerate}
\item
Remembering that the electric flux $e_3$ is defined (mod $N$), it follows that 
if $N$ is even, there are no parity invariant states. This implies that all the eigenstates of $\hat H_{\theta=\pi}$ are at least doubly degenerate. In particular, the vacuum states must spontaneously break the parity symmetry.
This double degeneracy occurs at a  finite $\T^3$ of any size. Exact degeneracies of states related by a symmetry are usually not expected in finite volume, but by now there are similar examples in quantum mechanics and 2d field theories, all related to  anomalies, as in \cite{Gaiotto:2017yup,Gaiotto:2017tne,Kikuchi:2017pcp}. One expects that tunnelling amplitudes, which usually lift the degeneracies at finite volume, vanish due to delicate cancellations involving complex phase factors.\footnote{Here, these should arise due to the  twist $\vec{m}\ne 0$ and, possibly, various analytic continuations, e.g. \cite{Behtash:2015kna}.} It would be interesting to see these cancellations in an explicit controlled calculation in the 4d theory at hand. 

The double degeneracy of the spectrum of $\hat H_{\theta=\pi}$ for even $N$ is a consequence of the parity-center symmetry anomaly reflected in (\ref{eqn:Ppi_algebra}, \ref{algebra1}). In the infinite volume limit, one expects that local physics is independent of the twist $\vec{m}$ and that   the double degeneracy   persists and is manifested as spontaneous breaking of parity in the $\R^3$ theory. 

In the $\R^3$-limit, it is natural to expect that two of the pairwise degenerate  electric-flux states related by $\hat P_\pi$ (the ones of lowest energy, finite as $L_i \rightarrow \infty$, after subtracting UV divergences)  become  the two parity-breaking vacua of the theory. There is no reason for the other $N/2 -1$ parity-partner electric-flux sectors to have the same minimum energy. These are expected to become higher-energy  degenerate  pairs of vacua whose (meta-)stability is a complicated dynamical issue.\footnote{Semiclassical calculations on $\R^3 \times \S^1$ have explicitly exhibited such metastable or unstable vacua \cite{Bhoonah:2014gpa,Anber:2017rch,Aitken:2018mbb}.}

\item
If $N$ is odd, there is no anomaly\footnote{\label{counterterm}Formally, one can redefine $T_3 = e^{ -i {2 \pi k\over N}} T_3'$, with $2 k = N-1$. For odd $N$, this preserves   $\text{det} \;T_3=1$ and removes the phase from  $[\hat T_3, \hat P_\pi]$ in  \eqref{algebra1}. This reflects the freedom to add a 4d local counterterm \cite{Gaiotto:2017yup}.} and there is a parity invariant state, namely the state with electric flux $|{e_3={(N+1)/2}}\rangle$. Hence, it is possible to avoid the spontaneous symmetry breaking. However, notice that the only parity invariant state at $\theta =\pi$ is different from the parity invariant state at $\theta = 0$, the state with $|e_3 = 0 \rangle$, recalling \eqref{eqn:P0_algebra} and footnote \ref{dihedralreps}. Thus, there is a global inconsistency between these two theories meaning there must exist level crossing, becoming a phase transition in the $\R^3$ limit, as $\theta$ is changed. 

\item The double degeneracy (global inconsistency) at $\theta=\pi$ forced upon us by (\ref{algebra1}) has been explicitly seen in the limit of a small $\T^3$, the ``femto-universe'' with $L_i \ll \Lambda^{-1}$ \cite{vanBaal:1984ra,vanBaal:2000zc}, where the gauge coupling is small and a semiclassical weak-coupling expansion is under control.

Again, we consider $\vec{m} = (0,0,1)$. In the small-$L_i$ limit, focusing on the lowest-lying states, one neglects spatially dependent modes of $A_i$. One then constructs states of (classically) zero energy, as first done in calculations of the Witten index    \cite{Witten:1982df}, and then studies the perturbative and nonperturbative corrections to their energies. 
Clearly, with constant transition functions, $A_j = 0$  obeys the twisted boundary conditions \eqref{bc} and has  zero classical energy, $B_i^a =0$. This classical background corresponds to a state in the physical Hilbert space that we denote\footnote{This state is obtained after averaging $\ket{A_j=0}$ over appropriate   gauge transformations, as in \eqref{Htheta}.}   $\ket{[0]}$. There are a total of $N$ classical static backgrounds  $A_j^\alpha$ ($\alpha = 0,1,\dots,N-1$)  that also have zero energy, $B_i^a =0$, obey the boundary conditions \eqref{bc}, but are not gauge transformations of $A_j^0=0$. The corresponding classical backgrounds are $A_j^\alpha \equiv - i T_3^\alpha \partial_j T_3^{- \alpha}$ and the states are given by $\hat T_3^\alpha \ket{[0]}$.
Eigenstates of $\hat T_3$ can be obtained by projecting \begin{equation}\label{e3states}
\ket{e_3} =  {1 \over {N}} \sum_{\alpha = 0}^{N-1} e^{-i\frac{2\pi}{N} e_3 \alpha} \; \hat T_3^\alpha \ket{[0]}, ~~ \hat T_3 \ket{e_3} = e^{i\frac{2\pi}{N} e_3 } \ket{e_3}.
\end{equation}
Therefore, the classically degenerate zero-energy states $\ket{e_3}$ also satisfy, at $\theta = \pi$, $\hat P_\pi \ket{e_3} = \ket{1-e_3}$ and, at $\theta = 0$, $\hat P_0 \ket{e_3} = \ket{-e_3}$. The $N$ states \eqref{e3states} remain degenerate to any finite order of perturbation theory \cite{Luscher:1982ma,GonzalezArroyo:1988dz} but tunnelling effects lift the degeneracy. To leading order  in the semiclassical expansion,\footnote{The splittings are due to fractional instantons on $\T^3 \times \R$, of action $8\pi^2/(g^2 N)$ and topological charge $1/N$. The $g^{-4}$ prefactor in \eqref{elfluxsplitting} is due to the four translational zero modes of the instantons. There is no size modulus as the size of the instantons is fixed by $L_i$. There are no analytic solutions known, although it is argued that they exist and that their action saturates the self-dual bound $8\pi^2 \over g^2 N$ \cite{vanBaal:2000zc}. Clearly, this makes  higher-order calculations difficult, for recent progress see \cite{Gonzalez-Arroyo:2019wpu}.} the electric flux energies become $e_3$- and $\theta$-dependent \cite{vanBaal:1984ra,vanBaal:2000zc}:
\begin{equation}\label{elfluxsplitting}
E(\theta, e_3) =-  {C e^{- {8 \pi^2 \over g^2 N}} \over L g^4} \cos \left({2 \pi \over N} e_3 - {\theta \over N} m_3 \right),
\end{equation}
where we restored $m_3$ dependence and ignored perturbative corrections ($C$ is a numerical constant and $L$ denotes the length of the torus sides, which are taken equal).\footnote{Ref.~\cite{vanBaal:2000zc}  gives an expression equal to  our $E(\theta,e_3)-E(0,0)$. We prefer the  form   in (\ref{elfluxsplitting}) as it emphasizes the contribution of  the various semiclassical objects and can be compared to a similar expression in dYM \cite{Unsal:2012zj,Aitken:2018mbb}. Following remarks of \cite{Unsal:2020yeh}, a virtually identical formula can be obtained in dYM, but the details will not be given here.  }
A look at the electric flux energies \eqref{elfluxsplitting} shows complete agreement with (\ref{algebra1}): all levels are doubly degenerate at $\theta=\pi$ and even $N$, and there is a global inconsistency between $\theta=0$ and $\theta=\pi$ at odd $N$.

\item To connect to the Euclidean formalism, note that the double degeneracy of the energy eigenstates due to \eqref{algebra1}   imposes restrictions on  the partition function twisted by a center transformation in the time direction. Consider 
\begin{equation}\label{zthermal}
Z[k,1]= \tr(e^{- \beta \hat H_{\theta=\pi}}\; \hat T_3^k),
\end{equation} where the trace is over the physical Hilbert space
${\cal{H}}^{phys.}_{\theta=0}$ with our chosen twist $\vec{m}=(0,0,1)$. In the Euclidean formalism (in the continuum or on the lattice), this is the path integral of the $\theta=\pi$ theory in a particular $2$-form gauge field background  of  topological charge $k/N$. Inserting $\hat P_\pi^2=1$ in the trace and using \eqref{algebra1}, we obtain the relation\begin{equation}\label{twistedZ}
Z[k, m_3] = Z[-k, m_3] \; e^{i {2 \pi k m_3 \over N}},
\end{equation} expressing the 't Hooft anomaly in the path integral (here, we imagine restricting to even-$N$, as for odd $N$ one can add a counterterm, as per footnote \ref{counterterm}; note that we also restored explicit $m_3$-dependence).
The expression (\ref{twistedZ}) is formal, as the Hilbert space trace \eqref{zthermal} diverges and needs a proper definition. Assuming that this is provided, note that a simple solution of (\ref{twistedZ}) is $Z[k,m_3]=e^{i {\pi k m_3\over N}} \; \Xi$, with $\Xi$ an undetermined even function of $k$. 

For example, in the case of the ``femto-universe,'' the partition function (\ref{zthermal}) in the $k,m_3$ background of only the two lightest fluxes, $e_3 = 0$ and $e_3= 1$, with energies given in (\ref{elfluxsplitting}), is of this form, with $\Xi = e^{- \beta E_\text{vac}} \; 2 \cos {\pi k m_3 \over N}$. More generally, this solution of (\ref{twistedZ}) can be thought of as the partition function of the IR TQFT whose states correspond to the two vacua with spontaneously broken parity.\footnote{One can relate the anomalies represented by (\ref{twistedZ}) (and by \eqref{twistedZadj} for the discrete chiral symmetry) to the variations of appropriate 5d ``invertible TQFTs,'' or ``anomaly theories,'' see e.g. \cite{Wan:2018zql,Wan:2018djl,Wan:2019oyr,Cordova:2019jqi,Cordova:2019bsd}, but we shall not discuss this here.}

\item Another calculable regime studied more recently is that of deformed Yang-Mills (dYM) theory on $\R^{3} \times \S^1$, for $\S^1$ size $L$ obeying $\Lambda N L \ll 2 \pi$ \cite{Unsal:2008ch}. This can be viewed as a $\T^3$ gauge theory, with added appropriate massive adjoint fermions (see  also \cite{Myers:2007vc,Myers:2009df})  considered in the limit $L_{1,2} \rightarrow \infty$ with  $L_3 = L$ kept small. Here, the semiclassical expansion is  significantly friendlier than in the femto-universe, at least  to leading order.
We shall not review the work on  $\theta$-dependence in dYM, as there is  extensive recent literature \cite{Unsal:2012zj,Poppitz:2012nz,Bhoonah:2014gpa,Anber:2017rch,Aitken:2018mbb,Bonati:2018rfg,Bonati:2019kmf}. The upshot is that, to leading order in the semiclassical expansion, spontaneous breaking of parity is found at $\theta =\pi$ in all cases.
In addition, a deformed algebra similar to \eqref{algebra1} was also found in  dYM, within the abelian IR theory on $\R^3 \times \S^1$ valid at energies below $1/(NL)$ \cite{Aitken:2018kky}.

 It would be of  interest to understand its precise relation to (\ref{eqn:Ppi_algebra}), e.g. by taking $\vec{m}\ne 0$ on an asymmetric $\R \times \T^3$. 
The importance of studying $\vec{m}\ne 0$ backgrounds  was also stressed, with a different motivation, in \cite{Unsal:2020yeh,Unsal:2021cch}. In fact, it should be possible to use the discussion there to  explain why the centrally-extended algebra found in \cite{Aitken:2018kky} in the dYM framework coincides with the one in discrete flux backgrounds on $\T^3$ of this paper. We also stress the striking  similarity between the $N$ electric flux energies in the femto-universe of  eqn.~(\ref{elfluxsplitting}) and the  energies of the $N$ (meta) stable vacuum states in dYM \cite{Aitken:2018mbb}, suggestive of a close relation between the   semiclassical expansions in the two limits, a subject worthy of further investigation.
\end{enumerate}

To conclude, in this section we showed that the anomaly and global inconsistency structures between the parity and center symmetries are completely reproduced in the algebra of the symmetry operators in the canonically quantized theory with twisted boundary conditions. This has immediate consequences on the vacuum structure and, therefore, symmetry breaking pattern of the theory as we reviewed above.

In the next section, we perform a similar analysis for the chiral symmetry in theories with adjoint fermions.

\subsection{The algebra of discrete chiral and $\Z_N^{(1)}$ operators}
\label{sec:algebrachiral}

Consider now QCD(adj), the $SU(N)$ gauge theory with $n_f \le 5$ massless adjoint Weyl fermions (the six-flavour theory is not asymptotically free). The fact that the  fermions are in the adjoint representation means that all the machinery we have developed surrounding boundary conditions is unchanged. In particular, the fermions obey the same boundary conditions (\ref{bc}) as the gauge field.  In Hilbert space the fermions are represented by creation and annihilation operators $\hat\lambda_\alpha^a$, $\hat{\lambda}^{a \; \dagger}_{\dot\alpha}$ ($a=1,..., N^2-1$; $\alpha, \dot\alpha$ are $SL(2,C)$ indices in the convention of \cite{Wess:1992cp}) obeying the anticommutation relations
\begin{equation}\label{fermionccr}
\{  \hat{\lambda}^{a \; \dagger}_{\dot\alpha}(x) \bar\sigma^{0 \; \dot\alpha \alpha}, \hat\lambda_\beta^b(y)\}  =    \delta^{(3)}(x-y)\delta^{ab} \delta^\alpha_\beta~.
\end{equation} For brevity, we do not display the flavour index; in all our formulae below, flavour is assumed to be summed over. 
The Hamiltonian (\ref{hamiltonian}) acting on the physical Hilbert space\footnote{The Gauss' law constraint and the definition of the physical Hilbert space \eqref{Htheta} is modified by  adding the fermions, but we will not need an explicit expression.} is  modified to
 \begin{equation}
 \label{hamiltonian2}
 \hat{H} = \int d^3 x \left( {g^2 \over 2} \; \hat \Pi_i^a \hat \Pi_i^a + {1 \over 2 g^2}\; \hat B_{i}^a \hat B_{i}^a - i \hat{\lambda}^{a \; \dagger} \bar\sigma^j \partial_j \hat\lambda^a +  i \hat{\lambda}^{a \; \dagger} \bar\sigma^j f^{abc}\hat A_j^b \hat\lambda^c~\right).
 \end{equation}
 Since the adjoint fermions obey \eqref{bc}, the $\Z_N^{(1)}$ center-symmetry generators $\hat T_i$ commute with the Hamiltonian.

Classically, the $n_f$ Weyl fermions have a $U(n_f)$ ($0$-form) global chiral symmetry. However, in the quantum theory, this is broken by the triangle anomaly to $\frac{\Z_{2n_fN} \times SU(n_f)}{\Z_{n_f}}$. In what follows, we shall only consider the discrete chiral symmetry which is defined as the center of the full unbroken chiral symmetry, that is $\Z_{2n_fN}$.  
The classical $U(1) \in U(n_f)$ chiral current operator $\hat{j}^\mu_f = \hat{\lambda}^{a \; \dagger} \bar\sigma^\mu\hat\lambda^a$, with a sum over $a$ and flavour understood, has an anomaly given by the (Heisenberg picture) operator equation
\begin{equation}\label{anomaly1}
\partial_\mu \hat j^\mu_f = \partial_\mu(\hat{\lambda}^{a \; \dagger} \bar\sigma^\mu \hat\lambda^a) = 2 n_f N \partial_\mu \hat K^\mu~.
 \end{equation}
This allows one  to define a conserved but gauge variant current which we label $\hat J^\mu_5$ for historical reasons:\footnote{See \cite{Adler:1970qb} for the calculation of the relevant field-current and current-current equal-time commutators.}
 \begin{equation}
{\hat{J}}^\mu_5 = \hat j^\mu_f - 2 n_f N \hat K^\mu~.
 \end{equation}
The corresponding $U(1)$ charge operator,  $\hat Q_5 = \int d^3 x {\hat{J}}^0_5 = \int d^3 x \hat{j}_f^0 - 2 n_f N \int d^3 x \hat K^0$, commutes with the Hamiltonian  but is not gauge invariant. However,
the unitary operator representing a $\Z_{2 n_f N}^{(0)}$ subgroup of the chiral symmetry is gauge invariant\footnote{The discussion that follows parallels the  one in the charge $q>1$ Schwinger model \cite{Anber:2018jdf}. In particular, the algebra \eqref{eqn:chiral_algebra} with $m_j=1$, for one  chosen $j$, is identical to the one found there. }
\begin{equation} \label{chiraloperator}
\hat X_{\Z_{2 n_f N}^{(0)}} = e^{i {2 \pi \over 2 n_f N} \hat Q_5} =   e^{i {2 \pi \over 2 n_f N} \int d^3 x \hat j^0_f} \; \hat V_{2\pi}^{-1}~,\end{equation}
with $\hat V_{2\pi}$ from (\ref{defofV}). 
Since the fermions are adjoint and the operator $\int d^3 x \hat j^0_f$ contains a trace in its definition, the fermion part of the chiral symmetry operator  commutes with the $1$-form center symmetry generators $\hat T_j$. Hence, the algebra between $\hat X_{\Z_{2 n_f N}^{(0)}}$ and the $\hat T_j$ is exactly the same as between $\hat V_{2\pi}$ and $\Z_N^{(1)}$ symmetry generators $\hat T_j$ of eqn.~(\ref{vtcommutator})
\begin{equation}
\label{eqn:chiral_algebra}
\hat T_j \; \hat X_{\Z_{2 n_f N}^{(0)}} = e^{-i {2 \pi \over N} m_j} \; \hat X_{\Z_{2 n_f N}^{(0)}} \; \hat T_j .
\end{equation}
This implies that the discrete chiral symmetry transformation results in a shift $\vec{e} \rightarrow \vec{e}- \vec{m}$. 

We can now return to our example of $\vec{m} = (0,0,1)$. We have, as in the pure gauge theory, that $\hat T_{1,2}$ commute with the Hamiltonian and the chiral symmetry generator $\hat X_{\Z_{2 n_f N}^{(0)}}$. Similar to \eqref{algebra1}, the interesting part of the algebra is
\begin{equation}
\label{eqn:chiral_algebra2} [\hat T_3, \hat H] = 0,~~ [\hat X_{\Z_{2 n_f N}^{(0)}}, \hat H] =0, ~~
\hat T_3 \; \hat X_{\Z_{2 n_f N}^{(0)}} = e^{-i {2 \pi \over N}} \; \hat X_{\Z_{2 n_f N}^{(0)}} \; \hat T_3 .
\end{equation}
As $\hat H$ commutes with $\hat T_3$, as before, we can label energy eigenstates as $| E, e_3\rangle$. Clearly, the algebra \eqref{eqn:chiral_algebra2} then requires that 
\begin{equation}\label{degeneracychiral}
\hat X_{\Z_{2 n_f N}^{(0)}} \ket{E, e_3} = \ket{E, e_3 - 1}.
\end{equation}
Therefore, the discrete chiral symmetry transformation cyclically permutes all $N$ electric flux states. This suggests an $N$-fold degeneracy and the spontaneous breaking of the discrete chiral symmetry, $\Z_{2n_fN}\rightarrow \Z_{2n_f}$. This matches the effects of the mixed anomaly in the usual picture, where the introduction of a non-trivial center background introduces fractional topological charges that also break $\Z_{2n_fN}\rightarrow \Z_{2n_f}$.\footnote{We also note that there other anomalies of the  discrete chiral symmetry that we do not study here, notably its mixed anomaly with gravity, see \cite{Cordova:2019bsd,Cordova:2019jqi}.}

Assuming, as in section~\ref{sec:algebratheta}, that the infinite volume limit is unique and independent of the boundary-condition twist $\vec{m}$, the $N$-fold degeneracy found at finite $\T^3$ above  implies that the  $N$-fold degeneracy of ground states persists in the $\R^3$ theory and the discrete chiral symmetry is spontaneously broken  to at least $\Z_{2n_f}$ (as we discuss below, there can be other degeneracies emerging in the $\R^3$ limit).

\subsubsection{Discussion}
\label{sec:algebrachiraldiscussion}
{\flushleft{W}}e now make some comments regarding the main result of this section, eqn.~\eqref{degeneracychiral}.
\begin{enumerate}
\item The result \eqref{degeneracychiral}  about the degeneracy between eigenstates of $\hat{H}$ is based on the deformed algebra \eqref{eqn:chiral_algebra2} reflecting the mixed chiral-center anomaly. As such, it is  general, but  provides no insight as to the nature of, say, the vacuum states on $\T^3$ that break the discrete chiral symmetry. Barring a solution of  the theory, this is a complicated dynamical question. Here, we will offer some limited\footnote{Recall that QCD(adj) has no supersymmetry for $n_f >1$ and our classical discussion below is subject to quantum corrections, which we ignore.} insight into the nature of the $N$ classical ground states on a small $\T^3$. According to (\ref{degeneracychiral}) these $N$ states are interchanged by the discrete chiral operator $\hat X_{\Z_{2 n_f N}^{(0)}}$. We now take   $\vec{m}=(0,0,1)$, as in section~\ref{mequals1}, and perform an analysis of the states of lowest classical energy on a small  $\T^3$, with $L_i$ smaller than the inverse strong coupling scale. 

Let us begin with the fermions.  The $\lambda_\alpha^a$ obey boundary conditions twisted by $\Gamma_1 \sim W_P, \Gamma_2 \sim W_Q$. The lowest energy states must have a constant fermion background, since non-constant fermion modes have a Kaluza-Klein mass of at least $1 / L_i$, which is large on a small $\T^3$. In order to satisfy our boundary conditions, this constant must be $\su{N}$ valued and commute with $\Gamma_1 = W_P$ and $\Gamma_2 = W_Q$. Any matrix that commutes with $W_Q$ must be diagonal, and any diagonal matrix that commutes with $W_P$ must take the form $\chi I$. Such a matrix is in $\su{N}$ if and only if $\chi = 0$. Hence, we must have $\lambda_\alpha^a = 0$ for all  flavours in our deep IR states. Thus, in what follows we ignore the fermions, taking them in their Fock-vacuum state.

This means that we can focus on the gauge fields. Their zero-energy states were already analyzed in section~\ref{sec:algebratheta} and we simply borrow the results here. The $N$ degenerate states of zero energy of eqn.~\eqref{e3states}, $\ket{e_3}$ also satisfy
\begin{equation}\hat V_{2\pi}^{-1} \ket{e_3} = \ket{e_3 - 1}, ~\text{hence}~ ~\hat X_{\Z_{2 n_f N}^{(0)}} \ket{ e_3} = \ket{e_3 - 1},
\end{equation}
where we used the fact that $\hat V_{2\pi}$ is  the bosonic part of the chiral symmetry operator \eqref{chiraloperator}. 
Thus, it is the purely bosonic zero-energy states  \eqref{e3states} that are interchanged under the chiral symmetry, as required by  \eqref{degeneracychiral}. The fact that bosonic states transform under the chiral symmetry is due to the anomaly (which led to \eqref{chiraloperator}). We also note that this is similar to how the bosonic dual photons transform under the discrete chiral symmetry in QCD(adj)   in the calculable regime
on $\R^3 \times \S^1$ \cite{Unsal:2007jx}.

\item As in section \ref{sec:algebrathetadiscussion}, we can  connect with the Euclidean path integral formalism via  the partition function of the theory twisted by a center transformation in the time direction. Consider, as in (\ref{zthermal}), the partition function
\begin{equation}\label{zthermal2}
 Z[k,1] = \tr(e^{- \beta \hat H}\; \hat T_3^k),
 \end{equation} where the trace is over the physical Hilbert space  with $\vec{m}=(0,0,1)$. As before,  \eqref{zthermal2} defines the thermal partition function of the adjoint theory in a particular $2$-form gauge field background  of  topological charge $k/N$.
Inserting $\hat X_{\Z_{2 n_f N}^{(0)}}^{-1} \hat X_{\Z_{2 n_f N}^{(0)}}=1$ in the trace and using \eqref{eqn:chiral_algebra2}, we obtain (again restoring $m_3$)
\begin{equation}\label{twistedZadj}
Z[k, m_3] = Z[k, m_3] \; e^{i {2 \pi k m_3 \over N}},
\end{equation} expressing the 't Hooft anomaly in the path integral. 

As opposed to the $\theta=\pi$ partition function (\ref{twistedZ}), the only solution of (\ref{twistedZadj}) with $k \ne 0 (\text{mod}  {N\over \text{gcd}(N,m_3)})$ is $Z[k,m_3]=0$. From the gauge theory path-integral perspective, this can be understood  by recalling that QCD(adj) has $2 n_f k m_3$ zero modes in the background with topological charge $k m_3/N$. \footnote{
We could also study the partition function with a $(-1)^F$ insertion, which becomes the Witten index for $n_f=1$. Eqn.~(\ref{twistedZadj}) also holds for the   partition function twisted by $(-1)^F$. The vanishing of $Z[k, m_3]$ in the high-temperature limit is explained in \cite{Anber:2018jdf,Anber:2018xek}.
}

\item
The $\Z_{2n_fN}\rightarrow \Z_{2n_f}$ breaking pattern leading to $N$ ground states on $\R^3$ is realized by the known IR behaviour of the theory with $n_f=1$ (super-Yang-Mills). This breaking pattern is also seen in a setup where the IR dynamics for any $n_f \le 5$ can be solved using semiclassical tools, namely on $\R^3 \times \S^1$ at a small-size $\S^1$  \cite{Unsal:2007jx}. Here, the continuous $SU(n_f)$ chiral symmetry is not broken, but the discrete symmetry is broken to $\Z_{2 n_f}$. Similar scenarios have also been proposed on $\R^3$ for various $n_f$ \cite{Anber:2018iof,Poppitz:2019fnp}.

\item  The $N$-fold degeneracy implied by (\ref{eqn:chiral_algebra2}) is also consistent with the ``vanilla'' scenario for the realization of the continuous chiral symmetry on $\R^3$, where $SU(n_f) \rightarrow SO(n_f)$. This breaking is due to the formation of a bilinear fermion condensate $\langle \lambda^{a \alpha}_I \lambda^{a}_{\alpha J}\rangle \sim \delta_{IJ}$, where $I,J$ are $SU(n_f)$ flavour indices. This condensate breaks the discrete chiral symmetry to $\Z_2$, see the study  \cite{Cordova:2018acb}. This symmetry-breaking pattern   is believed to be realized at least for a range of ``small enough'' $n_f \ge 2$.

\item
An interesting question that we shall not attempt to address here is about the fate of the mixed chiral-center anomaly in theories that, on $\R^3$, are thought to flow to fixed points in the IR.  
In particular, QCD(adj), a theory with such an anomaly, has been argued to exhibit conformal IR behaviour for sufficiently ``large'' $n_f$, although this has not been shown without the trace of a doubt for any $n_f$.\footnote{The lattice literature on the subject is quite voluminous, beginning with \cite{Catterall:2008qk,Hietanen:2008mr,DelDebbio:2009fd}, while  \cite{Athenodorou:2021wom}  has the most recent update and references.} 

In particular, for $n_f = 5$, it has been argued (see e.g. \cite{Poppitz:2009uq} and references therein) that the coupling  $g_{*}$  at the IR fixed point of the two-loop beta-function is ``small,'' with $g_{*}^2/(4 \pi) \sim 0.13$, so that the theory appears ``Banks-Zaks-ish,'' a ``weakly-coupled'' conformal field theory.\footnote{The multiple use of quotation marks is to indicate the uncertain nature of this argument. As opposed to the Banks-Zaks limit \cite{Banks:1981nn}, the fixed-point coupling can not be made arbitrarily small by adjusting $N_f$ and $N_c$ and there is no controlled expansion.}
Accepting this picture, ref.~\cite{Poppitz:2009uq} suggested that this ``semiclassical calculability'' of  the $n_f=5$ theory on $\R^3 \times \S^1$  implies that the discrete chiral symmetry is broken, $\Z_{10 N}\rightarrow \Z_{10}$, at any size $\S^1$, with a mass gap that goes to zero as $L \rightarrow \infty$. It might be interesting 
 to study the theory on a large (asymmetric) $\T^3$ with nonzero $\vec{m}$ in order to understand the implications of the algebra (\ref{eqn:chiral_algebra2}) in theories which flow to fixed points in the $\R^3$ 
limit.\end{enumerate}

\section{The mixed anomaly  for all other gauge groups with a center}
\label{sec:anomalyall}
The discussion in this section will closely follow the study of the mixed $0$-form/$1$-form anomaly in $SU(N)$ and we shall therefore be brief. In section \ref{sec:parityall}, we consider the parity-center mixed anomaly/global inconsistency at $\theta=0$ or $\pi$, the deformation of the center-parity algebra and the degeneracy that occurs in each case. In the following section \ref{sec:chiralall} we  do the same for the chiral symmetry. 

We begin with
table \ref{tab:top_charges1}, where we show all the group theory data that we will need for our analysis of the mixed parity-center and parity-chiral anomalies in the theories with general gauge groups. The fractional topological charges are derived in appendix  \ref{appx:Qforallgroups}.  In the last column, we also show the order of the discrete $\Z_p$ chiral symmetry in the theory with $n_f$ massless adjoint Weyl fermions,\footnote{For super-Yang-Mills these are all given in e.g. \cite{Anber:2014lba}, and here we simply multiply them by $n_f$.} which has a mixed anomaly with the corresponding center symmetry.

For the purpose of canonical quantization, the third column in the table, giving the fractional value of topological charge in terms of $\vec{m}$ and $\vec{k}$, called $Q_\text{top}[\vec{m}, \vec{k}]$ there, is the most important one. 
For each gauge group, without loss of generality, we shall consider quantization in the $\vec{m}=(0,0,1)$ background of section \ref{mequals1}. As usual,  $\vec{m}$ is defined modulo $p$, where $p$ is the order of the cyclic center-symmetry group. For $Spin(4N)$, we take $\vec{m}^+ = \vec{m}^- = (0,0,1)$. 

Quantization proceeds in complete analogy with the $SU(N)$ case and we shall not repeat the steps here. Again, the transition functions can be taken to be constant matrices $\Gamma_{i}$ obeying the appropriate generalization of \eqref{gammacocycle}. Such constant twist matrices exist and can be explicitly constructed  by embedding the $SU(2)$ (or $SU(4)$/$SU(3)$/ for  $Spin(4N+2)$/$E_6$/) matrices in the corresponding convenient representation \cite{Witten:2000nv}. The matrix $\Gamma_P$ which determines the parity transformation, see eqn.~\eqref{parity1}, can also be constructed from \eqref{gammap1} using the same embedding.\footnote{For example, in $Sp(N)$ one can take $\Gamma_1 = i \sigma_1 \otimes \identity_N$, $\Gamma_2 = i \sigma_3 \otimes \identity_N$, $\Gamma_3 = \identity_{2N}$, $\Gamma_P = i \sigma_1 \otimes \identity_N$.}
The explicit form of $\Gamma_i$ and $\Gamma_P$ plays no role in the commutation relations that we are interested in.

\label{sec:groupall}
\begin{table}[h]
    \centering
    {\tabulinesep=1.0mm
    \begin{tabu}{lllll}
        \hline
        Group &Center  & $Q_\text{top}\pmod{1}$& $= Q_\text{top}[\vec{m}, \vec{k}]$ & $\Z_p^{\text{discrete chiral}}$\\ \hline
        $\SU{N}$  & $\Z_N$ &$-\frac{1}{N}\operatorname{Pf}(n)$ & $= {1 \over N}\; \vec{m}\cdot \vec{k} $ & $\Z_{2 n_f N} $  \\
        $\Sp{N}$ & $\Z_2$&$\frac{N}{2}\operatorname{Pf}(n)$ & $ =- {N  \over 2}\; \vec{m}\cdot \vec{k}$ & $ \Z_{2 n_f (N+1)}$ \\
        $\Spin{8N}$& $\Z_2^+ \times \Z_2^-$ &$\frac{1}{2} \left({1\over 4} {\epsilon_{\mu\nu\lambda\sigma} n^+_{\mu\nu} n^-_{\lambda\sigma}} \right)$ &$=-\frac{1}{2}(\vec{m}^+ \cdot \vec{k}^- + \vec{m}^- \cdot \vec{k}^+) $ & $ \Z_{2 n_f (8 N -2)}$\\
        $\Spin{8N+4}$&$\Z_2^+ \times \Z_2^-$  &$\frac{1}{2}\left(\operatorname{Pf}(n^+)+\operatorname{Pf}(n^-)\right)$ &
        $ = -{1 \over 2}(\vec{m}^+ \cdot \vec{k}^+ + \vec{m}^- \cdot \vec{k}^-)$ & $ \Z_{2 n_f (8 N +2)} $  \\
        $\Spin{4N+2}$ &$\Z_4$ &$\frac{1+2N}{4}\operatorname{Pf}(n) $ &$= - {1 + 2N\over 4} \;\vec{m} \cdot \vec{k}$  & $   \Z_{8 n_f N}$\\
        $\Spin{2N+1}$& $\Z_2$ &$0$ &$=0$& $  \Z_{2 n_f (2 N -1)}$ \\
        $\E{6}$& $\Z_3$ & $\frac{1}{3}\operatorname{Pf}(n) $ &$= - {1 \over 3} \; \vec{m}\cdot \vec{k}$ & $ \Z_{24 n_f}$\\
        $\E{7}$&$\Z_2$ & $\frac{1}{2}\operatorname{Pf}(n)$ &$=- {1 \over 2} \; \vec{m}\cdot \vec{k}$ & $ \Z_{36 n_f}$ \\ \hline
    \end{tabu}}
    \caption{Summary of the topological charges $\bmod{1}$ on $\T^4$ for all gauge groups with non-trivial center.  There are two twists, $n_{\mu\nu}^\pm$,  in $Spin(4N)$. The last column shows the discrete chiral symmetry in the theory with $n_f$ massless adjoint Weyl fermions.}
    \label{tab:top_charges1}
\end{table}

The $\Z_p$ center symmetry generators along the $z$-direction are, as before, labeled by $\hat T_3$ (for  $Spin(4N)$, there are two $\Z_2^+ \times \Z_2^-$ generators $\hat T_3^+$ and $\hat T_3^{-}$).
The center-symmetry generators obey the boundary conditions \eqref{eqn:Ti_bcs} with $\vec{k} = (0,0,1)$ (or with $\vec{k}^\pm  = (0,0,1)$ as appropriate). 
The data of table \ref{tab:top_charges1}, the fractional value of the topological charge $Q_\text{top}[\vec{m}, \vec{k}]$, determines the commutation relation of the center symmetry generator with the operator shifting the theta angle by $2 \pi$, eqn.~\eqref{t3vcommutator}. This commutation relation now becomes
\begin{equation}\label{t3vcommutator1}
\hat T_3\;  \hat V_{2\pi} = e^{i 2\pi Q_\text{top}[\text{from table \ref{tab:top_charges1}, with} \;\vec{m}\cdot \vec{k} \rightarrow 1]} \; \hat V_{2\pi} \; \hat T_3,
\end{equation}
where the notation  ``$Q_\text{top}[\text{from table \ref{tab:top_charges1}, with} \;\vec{m}\cdot \vec{k} \rightarrow 1]$'' means that we take it equal to $1/N$ for $SU(N)$ (thus reproducing \eqref{t3vcommutator}), $-N/2$ for $Sp(N)$, $-(1+2N)/4$ for $Spin(4N+2)$, $-1/3$ for $E_6$, and $-1/2$ for $E_7$. 
 The relation \eqref{t3vcommutator1} holds also for each of the center symmetry generators $\hat T_3^{\pm}$ of 
$Spin(4N)$ in our chosen $\vec{m}^+ = \vec{m}^- = (0,0,1)$ background. Here, we simply take $Q_\text{top}[...]=-1/2$ in \eqref{t3vcommutator1}.

\subsection{The parity center-symmetry anomaly}
\label{sec:parityall}

We are now ready to discuss the parity-center algebra at $\theta=0$ and $\theta=\pi$ for all gauge groups.
For all gauge groups, we have the following algebras involving $\hat T_3$ and the parity generators at $\theta=0$ or $\pi$, $\hat P_{0}$ or $\hat P_\pi$:
\begin{eqnarray}
\theta=0:&& \hat P_0 \; \hat T_3\; \hat P_0 = \hat T_3^\dag, \nonumber \\
\theta =\pi: &&\hat{T}_3 \hat P_\pi = e^{i {2 \pi}Q_\text{top}[\text{from table \ref{tab:top_charges1}, with} \;\vec{m}\cdot \vec{k} \rightarrow 1]} \hat P_\pi \hat T_3^\dagger~,
\end{eqnarray}
where we use the  notation explained after \eqref{t3vcommutator1}. We now discuss the implications of these algebras for the various groups in turn (yet again, the $\theta = 0$ algebras are those of appropriate dihedral  groups, and the $\theta = \pi$ ones are their central extensions). We note that the $\T^3$ operator algebra at $\theta=\pi$ for $Spin(2N+1)$ and $Sp(2k) $ is not deformed and there is no anomaly or global inconsistency.
 
 {\flushleft{$\mathbf{Sp(2k-1)}$}}: Now the fractional part of the topological charge in \eqref{t3vcommutator1} is $Q_\text{top} = 1/2$. The energy eigenstates on the torus are labeled by $\Z_2$-electric flux $e_3$. States with $\ket{e_3}$ and $\ket{1-e_3 (\text{mod}\; 2)}$ are degenerate at $\theta = \pi$, implying an anomaly and spontaneous parity breaking, as in the case of $SU(N)$ with even $N$.
 
 {\flushleft{$\mathbf{Spin(4N)}$}}: In the chosen background the algebras for both $\hat T_3^+$ and $\hat T_3^-$ at $\theta=\pi$ have a $\Z_2$ central extension, implying that energy eigenstates labeled by electric fluxes $\ket{e_3^+, e_3^-}$ and their parity partners $\ket{1 - e_3^+(\text{mod}\; 2),1 - e_3^-(\text{mod}\; 2)}$ are degenerate.
 Again, this situation is as in even-$N$ $SU(N)$.
 
 {\flushleft{$\mathbf{Spin(4N+2)}$}}: For even $N$,  the fractional part of the topological charge in \eqref{t3vcommutator1} is $Q_\text{top} = -1/4$, while for odd $N$, $Q_\text{top} = +1/4$. Thus, for even $N$, $\Z_4$ electric fluxes $\ket{e_3}$ and their parity partners $\ket{3 - e_3 (\text{mod} \; 4)}$ are degenerate, while for odd $N$, these are replaced by
 $\ket{e_3}$ and $\ket{1 - e_3 (\text{mod} \; 4)}$. In each case, the deformed algebra implies a double-degeneracy at $\theta =\pi$, absent at $\theta = 0$.
 
 {\flushleft{$\mathbf{E_6}$}}: Here, the $\Z_3$ electric flux states $\ket{e_3}$ and $\ket{2 - e_3 (\text{mod} \; 3)}$ are degenerate. At $\theta = \pi$ the electric flux  $\ket{e_3 = 1}$ state is parity invariant, while at $\theta = 0$ it is the $\ket{e_3 = 0}$ state, implying a global inconsistency, as for odd-$N$ $SU(N)$.
 
 {\flushleft{$\mathbf{E_7}$}}: Here, the situation is that of an anomaly, as  parity maps $\ket{e_3}$ to $\ket{1-e_3 (\text{mod} \;2)}$ energy eigenstates, implying their degeneracy.
 
{\flushleft{ W}}e end with a few comments:
 \begin{enumerate}
 \item Based on our study of the electric flux degeneracies on $\T^3$,  the pattern that emerges is clear:  groups whose  center is of an even order have a parity-center symmetry anomaly at $\theta = \pi$, while groups whose center has an odd order have a global inconsistency. 
 \item There exist almost no semiclassical calculations studying the $\theta=\pi$ behaviour for gauge groups other than $SU(N)$. The   only available semiclassical calculation (known to us) for groups other than $SU(N)$ focusing on $\theta =\pi$ and the implications of the anomaly is that of ref.~\cite{Chen:2020syd}. This work considered
Yang-Mills theories with minimal supersymmetry on $\R^4$, compactified on $\R^3 \times \S^1$ with $\S^1$ of small size $L$ and with supersymmetric boundary conditions. To introduce $\theta$-dependence, a small gaugino mass $m$ was added. The theory can be studied analytically for $L$ and $m$ appropriately small, for details see  \cite{Anber:2014lba}. The target pure Yang-Mills theory on $\R^4$ is obtained in the large $L$, $m$ limits, where semiclassical calculability is lost.
In the small-$m, L$ regime of validity of the semiclassical expansion, spontaneous breaking of parity at $\theta=\pi$ was found for all simple gauge groups, even for the ones without center symmetry. It is not known whether this pattern persists in the $\R^4$ pure gauge theory limit. The parity breaking found at $\theta=\pi$ in the calculable limit appears unrelated to a parity-center anomaly and may be due to the closeness to the supersymmetric theory.   \end{enumerate}

\subsection{The discrete chiral-symmetry/center-symmetry mixed anomaly}
\label{sec:chiralall}

Now we discuss the mixed $\Z_p$-chiral/$1$-form center  anomaly for the theories with $n_f$ Weyl fermions with general gauge groups.
The discussion here will be shorter than in the previous section. 
The chiral-center algebra for $SU(N)$ of eqn.~\eqref{eqn:chiral_algebra2}, generalizes for other gauge groups to
\begin{equation}\label{chiralcentergeneral}
\hat T_3 \; \hat X_{\Z_{p}^{(0)}} = e^{- i {2 \pi} Q_\text{top}[\text{from table \ref{tab:top_charges1}, with} \;\vec{m}\cdot \vec{k} \rightarrow 1]} \; \hat X_{\Z_{p}^{(0)}} \; \hat T_3~,
\end{equation}
where, in addition to the notation introduced after \eqref{t3vcommutator1}, we used  $\hat X_{\Z_{p}^{(0)}}$ to denote the generator of the  appropriate $\Z_p$ chiral symmetry listed in table  \ref{tab:top_charges1}.
As before, for $Spin(4N)$, $\hat T_3$ in (\ref{chiralcentergeneral}) refers to any of the $\hat T_3^\pm$ generators of the $\Z_2^+ \times \Z_2^-$ center. Yet again, for $Sp(2N)$ and $Spin(2N+1)$ the chiral-center algebras on $\T^3$ are not deformed and we do not discuss them further. 

 {\flushleft{$\mathbf{Sp(2k-1)}$}}: Now the fractional part of the topological charge in \eqref{t3vcommutator1} is $Q_\text{top} = 1/2$ and the chiral symmetry is $\Z_{4 n_f k}$. The chiral generator changes maps $\Z_2$ electric flux state $\ket{e_3}$ to $\ket{-1+e_3}$, implying a two-fold degeneracy of the energy eigenstates on $\T^3$.
  
 {\flushleft{$\mathbf{Spin(4N)}$}}: The chiral symmetry is $\Z_{2 n_f (4 N-2)}$ and its generator maps
  energy eigenstates labeled by $\Z_2^+\times \Z_2^-$ electric fluxes $\ket{e_3^+, e_3^-}$ into $\ket{1+ e_3^+(\text{mod}\; 2),1 + e_3^-(\text{mod}\; 2)}$, implying a two-fold degeneracy.  
   
 {\flushleft{$\mathbf{Spin(4N+2)}$}}: The chiral symmetry is $\Z_{8 n_f N}$.  
  For even $N$,  the fractional part of the topological charge in \eqref{t3vcommutator1} is $Q_\text{top} = -1/4$, while for odd $N$, $Q_\text{top} = +1/4$. Thus, for even $N$,  the chiral generator maps $\ket{e_3}$ into $\ket{1 + e_3 (\text{mod} \; 4)}$. 
  On the other hand, 
for odd $N$, 
 $\ket{e_3}$ is mapped to $\ket{-1 + e_3 (\text{mod} \; 4)}$. It is easy to see that in each case, there is  a four-fold degeneracy on $\T^3$.

 {\flushleft{$\mathbf{E_6}$}}:  The chiral symmetry is $\Z_{24 n_f}$ and it maps  states labeled by the $\Z_3$ electric flux  $\ket{e_3}$ into $\ket{1 + e_3 (\text{mod} \; 3)}$. This is a $\Z_3$ orbit,  implying that the states are triply-degenerate.
  
 {\flushleft{$\mathbf{E_7}$}}: The chiral symmetry is $\Z_{36 n_f}$ and maps the electric flux states as $\ket{e_3}$ to $\ket{1+e_3 (\text{mod} \;2)}$ energy eigenstates, implying double degeneracy of the $\T^3$ energy eigenstates.

{\flushleft{A}}gain, we end with some comments and questions for the future:
\begin{enumerate}
\item We see that the central extension of the chiral-center algebra alone implies certain degeneracies. Groups with a $\Z_2$ (or $\Z_2 \times \Z_2$) center have a double degeneracy on $\T^3$, while the groups with $\Z_3$ and $\Z_4$ center have a three-fold and four-fold degeneracy, respectively.
\item
The only case where some aspects of the dynamics are understood is super-Yang-Mills theory, $n_f = 1$. Here, on $\R^4$ the chiral symmetry is known to break, by gaugino condensation, to fermion number $\Z_2$ for each group, implying a large emergent degeneracy in the $\R^4$ limit. 
The same symmetry-breaking pattern is also known to occur, in a semiclassically-calculable manner, in the small-$S^1$ limit of $\R^3 \times S^1$, for super-Yang-Mills with all gauge groups \cite{Davies:2000nw}.

In this respect, we notice that the vacuum degeneracy between the electric flux states implied by the mixed $0$-form/$1$-form anomaly for groups other than $SU(N)$ is very modest, equal to the order of the center-symmetry group. On the other hand, the ``observed'' chiral symmetry breaking pattern suggests a vacuum degeneracy equal to the dual Coxeter number of the gauge group.
The simplest case in point is $SP(N)$, where the dual Coxeter number is equal to $N+1$, while the center symmetry is $Z_2$. There is a mixed chiral-center anomaly only for $N=2k+1$, suggesting two degenerate vacua with  $Z_2$ electric fluxes $0$ and $1$, while the $R^3\times S^1$ analysis \cite{Davies:2000nw} and the Witten index\footnote{See both the early \cite{Witten:1982df} and late \cite{Witten:2000nv} work, especially for groups other than $SU$ and $SP$.} show that there are $2k+2$ vacua. We shall only make two remarks in this regard. First, we note that other 't Hooft anomalies, e.g.~the  mixed anomaly between the discrete chiral symmetry and gravity impose more severe constraints on the chiral symmetry realization, discussed in \cite{Cordova:2019jqi}; these constraints, however, require the validity of dynamical assumptions, namely the existence of a mass gap.
 Second, one might also wonder if there are any other not-yet-identified  symmetries, like the subtle ``noninvertible'' ones of \cite{Komargodski:2020mxz,Nguyen:2021yld,Nguyen:2021naa}, that might also play a role in determining the vacuum degeneracy and symmetry realization.  At the moment, we are not aware of the answer and only note that these are interesting questions to pursue.
 \item 
Not much is known about the dynamics of the nonsupersymmetric versions of these theories with other gauge groups. 
We note that the minimal degeneracies implied by the mixed anomaly on $\T^3$ may be consistent with   symmetry-breaking by  higher-dimensional multi-fermion condensates on $\R^3$, much like the ones argued for in \cite{Anber:2018iof}. Again, we leave this for future work. 
\end{enumerate}

\bigskip

{\flushleft{\bf Acknowledgements:}} We thank Mohamed Anber for comments on the manuscript and for  discussions. This work is supported by an NSERC Discovery Grant. AC was supported by an Ontario Graduate Scholarship.

\bigskip

\appendix

\section{Summary of relevant group theory data}
\label{appx:groupconventions}

We begin by summarizing some known facts about Lie groups, algebras, and representations that we shall use. Our intention here is largely to  set the notation; for more details and proofs, see e.g. \cite{Ramond:2010zz,liegroups}.

\subsection{Notation and conventions}

We consider a general gauge group $G$ with Lie algebra $\lie{g}$, and Cartan subalgebra $\lie{h}$. We use $r$ to denote the rank of the group. We denote the roots by $\vecb{\alpha}$, with $\vecb{\alpha}_i$ for $1\leq i\leq r$ the simple roots, and use $E_\alpha$ to denote the corresponding root vectors. The co-root to the root $\vecb{\alpha}$ is $\vecb{\alpha}^*\equiv 2\frac{\vecb{\alpha}}{\vecb{\alpha}\cdot\vecb{\alpha}}$. Roots live in the root-lattice, $\Lambda_r$, which is spanned by the simple roots, and similarly co-roots live in the co-root lattice, $\Lambda_r^*$, spanned by the co-roots of the simple roots. We denote the set of roots by $\Delta$, and the set of positive roots (with respect to a choice of simple roots) by $\Delta^+$. The fundamental weights are $\vecb{w}_i$ for $i=1,\dots,r$ and satisfy $2\frac{\vecb{w}_i\cdot\vecb{\alpha}_j}{\vecb{\alpha}_j\cdot\vecb{\alpha}_j}=\delta_{ij}$. For each fundamental weight, $\vecb{w}_i$,  the corresponding co-weight is $\vecb{w}_i^*\equiv \frac{2\vecb{w}_i}{\vecb{\alpha}_i\cdot\vecb{\alpha}_i}$. Similar to the roots, weights live in the weigh lattice $\Lambda_w$ spanned by the fundamental weights, and co-weights live in the co-weight lattice, $\Lambda_w^*$, spanned by the co-weights of the fundamental weights. Finally, we take the weights of the defining representation to be $\vecb{\nu}_A$ for $A=1,\dots,\dim{R_{fund.}}$. Weights live in the weight lattice, $\Lambda_w$, spanned by the fundamental weights.

{\flushleft{\bf The Cartan-Weyl basis for $\mathbf{\lie{g}}$.}}
 This basis of the Lie algebra is defined by the following commutation relations
\begin{align}\label{cartancommutators}
    \comm{H^a}{H^b}&=0\\
    \comm{H^a}{E_\alpha}&=(\vecb{\alpha})^aE_\alpha\\
    \comm{E_\alpha}{E_\beta}&=\begin{cases}
    N_{\alpha,\beta}E_{\alpha+\beta} & \vecb{\alpha}+\vecb{\beta}\text{ is a root}\\
    \vecb{\alpha}^*\cdot\vecb{H} & \vecb{\alpha}+\vecb{\beta}=0\\
    0 & \text{otherwise}\end{cases},
\end{align}
where $\{H^a\}_{a=1}^r$ are the Cartan generators, $(H^a)^\dagger = H^a$, which form a basis for $\lie{h}$ with $\vecb{H}=(H^1,H^2,\dots,H^r)$, $E_\alpha$ are the root vectors, $(E_\alpha)^\dagger = E_{- \alpha}$, and $N_{\alpha,\beta}=-N_{\beta,\alpha}$ is some number. We can extend the definition of $N_{\alpha,\beta}$ to include the cases when $\vecb{\alpha}+\vecb{\beta}$ is either not a root or zero, by setting $N_{\alpha,\beta}=0$ in those cases and remembering that $E_{\alpha+\beta}$ really has no meaning when $\vecb{\alpha}+\vecb{\beta}$ is not a root.

We also recall that irreducible representations are specified by their highest weight, $\vecb\lambda \in \Lambda_w$, and that the weights of a given representation are the eigenvalues of the Cartan generators in that representation. For example, in the  defining representation $R_{fund.}$, we  take $H^a$ to be diagonal matrices with components $(H^a)_{AB}=\delta_{AB}\left(\vecb{\nu}_A\right)^a$ where $A,B=1,\dots,\dim{R_{fund.}}$ and $a=1,\dots,r$, with $\vecb\nu_A$---the weights of fundamental representation.

 The  Cartan-Weyl basis and the usual orthogonal basis $\left\{ T^i,i=1,...\dim(\lie{g}) \right\}$ of Hermitean generators are related by \begin{alignat}{2}
    T^a&=H^a&\quad,\quad &a=1,\dots,r\label{eqn:generators_cartan}\\
    T_1^\alpha&=\frac{\abs{\vecb{\alpha}}}{2}\left(E_{\alpha}+E_{-\alpha}\right)&\quad,\quad &\vecb{\alpha}\in\Delta^+\label{eqn:generators_x}\\
    T_2^\alpha&=\frac{\abs{\vecb{\alpha}}}{2i}\left(E_{\alpha}-E_{-\alpha}\right)&\quad,\quad &\vecb{\alpha}\in\Delta^+\label{eqn:generators_y},
\end{alignat}
 where $T^i$ were enumerated as $\left\{T^a, T_1^\alpha, T_2^\alpha\right\}$.
 
{\flushleft{\bf Dynkin index and dimension.}}
The Dynkin index $C(R_\lambda)$ of an irreducible representation $R_\lambda$ of highest weight $\vecb\lambda$ is:
\begin{equation}\Tr_{R_\lambda}(T^iT^j)=C(R_\lambda)\delta^{ij}, ~ {\rm where}~~    C(R_\nu)=\frac{\dim(R_\lambda)}{\dim(\lie{g})}\;\vecb{\lambda}\cdot(\vecb{\lambda}+2\vecb{\rho}),\label{eqn:dynkin}
\end{equation}
where $\vecb\rho =\frac{1}{2}\sum_{\alpha\in\Delta^+}\vecb{\alpha}$ is the Weyl vector.
Note that this may differ by a factor of $\frac{1}{2}$ from definitions seen elsewhere. For use below, notice how $C(R_\lambda)$ scales with a change of normalization of roots: under $\vecb{\alpha}\rightarrow c \vecb{\alpha}$, both $\vecb{\lambda}$ and $\vecb{\rho}$ scale with $c$, so $C(R_\lambda)$ scales with $c^2$. 

Finally, if $R_\lambda$ is an irreducible representation with highest weight $\vecb{\lambda}$, then the dimension of $R$ may be computed from the Weyl dimension formula:
\begin{equation*}
    \dim(R_\lambda)=\frac{\prod_{\alpha\in\Delta^+}\vecb{\alpha}\cdot(\vecb{\lambda}+\vecb{\rho})}{\prod_{\alpha\in\Delta^+}\vecb{\alpha}\cdot\vecb{\rho}},\label{eqn:Weyl-Dimension}
\end{equation*}
where $\vecb{\rho}$ is the Weyl vector, defined after eqn.~(\ref{eqn:dynkin}). 

For the  $Spin(2N)$ groups we will work with a direct sum of two irreducible representations, corresponding to positive- and negative-chirality spinors, for which we cannot directly apply the above.  Suppose we have two irreducible representations $R_{\lambda_1}$ and $R_{\lambda_2}$, with highest weights $\vecb{\lambda}_1$ and $\vecb{\lambda}_2$ respectively, and we are interested in the representation $R_{{\lambda_1}\oplus{\lambda_2}}\equiv R_{\lambda_1}\oplus R_{\lambda_2}$. Suppose that a generator $X\in\lie{g}$ is represented by a $\dim(R_{\lambda_1})\times\dim(R_{\lambda_1})$ matrix $X_{\lambda_1}$ in the representation $R_{\lambda_1}$, and a $\dim(R_{\lambda_2})\times\dim(R_{\lambda_2})$ matrix $X_{\lambda_2}$ in the representation $R_{\lambda_2}$. Then, as a matrix representation for $R_{{\lambda_1}\oplus{\lambda_2}}$ we can simply take $X_{{\lambda_1}\oplus{\lambda_2}}$ to be the block diagonal matrix $\operatorname{diag}(X_{\lambda_1},X_{\lambda_2})$. Thus, we see that traces simply add across the representations, $\Tr_{R_{\lambda_1}\oplus R_{\lambda_2}}=\Tr_{R_{\lambda_1}}+\Tr_{R_{\lambda_2}}$, allowing us to immediately write down an expression for $C(R_{\lambda_1}\oplus R_{\lambda_2})$:
\begin{equation}\label{dynkinproduct}
    C(R_{\lambda_1}\oplus R_{\lambda_2})=C(R_{\lambda_1})+C(R_{\lambda_2})=\frac{\dim(R_{\lambda_1})\vecb{\lambda}_1 \cdot(\vecb{\lambda}_1+2\vecb{\rho})+\dim(R_{\lambda_2})\vecb{\lambda}_2\cdot(\vecb{\lambda}_2+2\vecb{\rho})}{\dim(\lie{g})}.
\end{equation}
This result can be easily generalized to a direct sum of an arbitrary number of irreducible representations.

{\flushleft{\bf The center of the group and the convenient choice of co-weight $\mathbf{\vecb\mu^*}$}}.
A group element $g\in G$ is in the center, $Z(G)$, if and only if $g X g^{-1}=X$ for all generators $X\in \lie{g}$. In the Cartan-Weyl basis, a center element of $G$ is given by
\begin{equation} \label{center1}
g = e^{2\pi i \vecb{\mu}^*\cdot\vecb{H}}\; {\rm  with} \; \vecb{\mu}^*\cdot\vecb{\alpha}\in\Z \;  {\rm for} \; {\rm  all } \; \rm{ roots} \; \vecb{\alpha}, \; {\rm i.e.}\;  \vecb\mu^* \in \Lambda_w^*~,
\end{equation}
or, in words, $\vecb\mu^*$ is an element of the co-weight lattice.\footnote{The commutation relations (\ref{cartancommutators}) imply $g E_\alpha g^{-1} = e^{2\pi i \vecb{\mu}^*\cdot\vecb{\alpha}} E_\alpha$, from which the statement in (\ref{center1}) follows.} Equation~(\ref{center1}) implies that the center of a group is trivial if $\vecb\mu^* \cdot \vecb\nu \in \Z$ for all  weights $\vecb\nu \in \Lambda_w$ since then $g$  is the unit matrix in all representations $R$. If the group has trivial center,\footnote{We shall not prove whether the center is trivial or not for a given group. This can be seen, e.g. by examining the explicit expressions for the roots and weights. A general criterion is to evaluate the determinant of the Cartan matrix relating the root and weight lattices (it equals unity for the groups with trivial center).}  the weights are sums of roots with integer coefficients. Similarly, the roots are integer sums of the weights (thus, neither $\Lambda_r$ or $\Lambda_w$ is finer than the other, and $\Lambda_r/\Lambda_w$ is trivial).

For groups with nontrivial $\Z_k$ centers, in a representation $R_{\vecb\lambda}$ of highest weight $\vecb\lambda$ where (\ref{center1}) is nontrivial, we shall call a choice of co-weight $\vecb{\mu}^*$, such that $\vecb{\mu}^*\cdot\vecb{\lambda}=\frac{1}{k}+\Z$ a {\it convenient choice} of $\vecb\mu^*$. 
In Table \ref{tab:dynkin}, we list all groups with nontrivial centers, the dimensions and Dynkin indices of their corresponding ``convenient'' representations.\footnote{For lack of better terminology, we call the ``convenient'' representation a choice of representation where the center of the group acts faithfully. In each case they are identified by their highest weight, see table~\ref{tab:centers}.}
\begin{table}[h]
    \centering
    {\tabulinesep=1.0mm
    \begin{tabu}{llll}
        \hline
        Group & Rank & Dim($R$) &C($R$)/$\vecb{\alpha}_\text{max}^2$ \\ \hline
        $\SU{N}$ & $N-1$ & $N$ & $\frac{1}{2}$ \\
        $\Sp{N}$ & $N$ & $2N$ & $\frac{1}{2}$ \\
        $\Spin{2N}$ & $N$ & $2^N$ & $2^{N-3}$ \\
        $\Spin{2N+1}$ & $N$ & $2^N$ & $2^{N-3}$ \\
        $\E{6}$ & 6 & 27 & 3 \\
        $\E{7}$ & 7 & 56 & 6 \\
      \hline
    \end{tabu}}
    \caption{Groups with nontrivial centers: their ranks, dimension and Dynkin indices $C(R)$ of the ``convenient'' representation, normalized by the length of the longest root squared.}
    \label{tab:dynkin}
\end{table}

\begin{table}[h]
    \centering
    {\tabulinesep=1.0mm
    \begin{tabu}{llll}
        \hline
        Group & Representation & Center & Convenient Co-Weight \\ \hline
        $\SU{N}$ & $\vecb{w}_1$ & $\Z_N$ & $\vecb{w}_{N-1}^*$ \\
        $\Sp{N}$ & $\vecb{w}_1$ & $\Z_2$ & $\vecb{w}_N^*$ \\
        $\Spin{4N+2}$ & $\vecb{w}_{2N}$ & $\Z_4$ & $\vecb{w}_{2N}^*$ for $N$ even and  $\vecb{w}_{2N+1}^*$ for $N$ odd  \\
        $\Spin{4N+2}$ & $\vecb{w}_{2N+1}$ & $\Z_4$ & $\vecb{w}_{2N+1}^*$ for $N$ even and $\vecb{w}_{2N}^*$ for $N$ odd \\
        $\Spin{8N}$ & $\vecb{w}_{4N-1}$ & $\Z_2^+$ & $\vecb{w}_{4N}^*$ or $\vecb{w}_{2k-1}^*$ for $1\leq k<2N$ \\
        $\Spin{8N}$ & $\vecb{w}_{4N}$ & $\Z_2^-$ & $\vecb{w}_{4N-1}^*$ or $\vecb{w}_{2k-1}^*$ for $1\leq k<2N$ \\
        $\Spin{8N+4}$ & $\vecb{w}_{4N+1}$ & $\Z_2^+$ & $\vecb{w}_{4N+1}^*$ or $\vecb{w}_{2k+1}^*$ for $0\leq k<2N$ \\
        $\Spin{8N+4}$ & $\vecb{w}_{4N+2}$ & $\Z_2^-$ & $\vecb{w}_{4N+2}^*$ or $\vecb{w}_{2k+1}^*$ for $0\leq k<2N$ \\
        $\Spin{2N+1}$ & $\vecb{w}_N$ & $\Z_2$ & $\vecb{w}_{2k+1}^*$ for $0\leq k<(N-1)/2$ \\
        $\E{6}$ & $\vecb{w}_1$ & $\Z_3$ & $\vecb{w}_a^*$ for $a=1,4$ and $2\vecb{w}_b^*$ for $b=2,5$ \\
        $\E{6}$ & $\vecb{w}_5$ & $\Z_3$ & $\vecb{w}_a^*$ for $a=1,4$ and $2\vecb{w}_b^*$ for $b=2,5$ \\
        $\E{7}$ & $\vecb{w}_6$ & $\Z_2$ & $\vecb{w}_a^*$ for $a=4,6,7$ \\ \hline
    \end{tabu}}
    \caption{Centers of irreducible ``convenient'' representations of groups, listed by their highest weights, along with the ``convenient co-weights,'' which correspond to the generators (\ref{center1}) of the centers. These results are obtained in section \ref{appx:convenient}.}
    \label{tab:centers}
\end{table}

\subsection{Groups with nontrivial centers and choice of ``convenient co-weight''}
\label{appx:convenient}

In this section we review the simple Lie groups and their algebras and discuss some of their properties of relevance to us, notably the convenient choice of co-weight $\vecb\mu^*$ to represent the center element (\ref{center1}). The results of this section are conveniently summarized on Table \ref{tab:centers}.

We use $M_n(\mathbb{F})$ to denote the set of $n\times n$ matrices with entries in $\mathbb{F}$ (we take $\mathbb{F}$ to be either $\R$ or $\C$), $U(N)\subset M_N(\C)$ to denote the set of $N\times N$ unitary matrices, and $O(N)\subset M_N(\R)$ to denote the set of $N\times N$ orthogonal matrices.
 For all algebras, we take the roots and weights to be $r$-dimensional vectors where $r$ is the rank of the algebra.\footnote{We note that this is not always the conventional choice, for example  $\su{N}$ roots are easily (and commonly) written down in an $N$-dimensional vector space, even though the rank is $N-1$.}  We use $\vecb{e}_i$ for $i=1,\dots,r$ to denote $r$-dimensional unit vectors, $\vecb{e}_i \cdot \vecb{e}_j = \delta_{ij}$, where $r$ is always assumed to be the rank of the group in question.

\subsubsection{$\mathbf{\SU{N}}$}
The most familiar case, $\SU{N}$, is the group of $N\times N$ unitary matrices with unit determinant. The algebra is $\su{N}$, and the root system is $A_{N-1}$, thus $r = N-1$. Below, we enumerate the defining properties of the group and algebra, the simple roots, fundamental weights and their inner products:
\begin{eqnarray}
    \SU{N}&:=&\left\{U\in U(N)\mid \det(U)=1\right\}\\
    \su{N}&:=&\left\{t\in M_{N}(\C)\mid t=t^\dagger,\ \tr(t)=0\right\}~,
    \end{eqnarray}
    where the simple roots, fundamental weights, and their inner products are
    \begin{eqnarray}
    \vecb{\alpha}_a&=&-\sqrt{\frac{a-1}{2a}}\vecb{e}_{a-1}+\sqrt{\frac{a+1}{2a}}\vecb{e}_a~, ~a = 1,...,N-1, ~\vecb{e}_0\equiv\vecb{0}\\
    \vecb{w}_a&=&a\sum_{j=a}^{N-1}\frac{1}{\sqrt{2j(j+1)}}\vecb{e}_j=\sum_{b=1}^{a-1}\frac{b(N-a)}{N}\vecb{\alpha}_b+\sum_{b=a}^{N-1}\frac{a(N-b)}{N}\vecb{\alpha}_b\\
    \end{eqnarray}
    \begin{eqnarray}
    \vecb{\rho}&=&\frac{1}{2}\sum_{j=1}^{N-1}\sqrt{\frac{j(j+1)}{2}}\vecb{e}_j\\
    \eval{\vecb{w}_a\cdot\vecb{w}_b}_{a\leq b}&=&\frac{\vecb{\alpha}_\text{max}^2}{2}\frac{a(N-b)}{N}\implies \vecb{w}_a^*\cdot\vecb{w}_b=\frac{\min(a,b)(N-\max(a,b))}{N}
\end{eqnarray}

The fundamental representation, $\square$, has highest weight $\vecb{w}_1$. The center of $\SU{N}$ is $\Z_N$, ie the $N^{th}$ roots of unity, which is generated by $e^{2\pi i/N}\identity$. We see from the inner product relation above that $\vecb{w}_{N-1}^*\cdot\vecb{w}_1=\frac{1}{N}$, and thus from the earlier discussion we find $\exp(2\pi i\vecb{w}_{N-1}^*\cdot\vecb{H})=e^{2\pi i/N}\identity$, so we have found the generator of the $\Z_N$ center. Thus, $\vecb{\mu}^*=\vecb{w}_{N-1}^*$ is a convenient choice. Then, an arbitrary center element can be written as
\begin{equation}
    e^{2\pi i x/N}\identity=e^{2\pi ix\vecb{w}_{N-1}^*\cdot\vecb{H}}~, ~ x \in \Z\; ({\rm mod} N)
\end{equation}

The weights of the fundamental representation are given by $\vecb{\nu}_A=\vecb{w}_1-\sum_{a=1}^{A-1}\vecb{\alpha}_a$. Plugging in our expressions for the simple roots and $\vecb{w}_1$ we find an expression for $\vecb{\nu}_A$:
\begin{equation}
    \vecb{\nu}_A=-\sqrt{\frac{A-1}{2A}}\vecb{e}_{A-1}+\sum_{j=A}^{N-1}\frac{1}{\sqrt{2j(j+1)}}\vecb{e}_j.
\end{equation}
The positive roots are $\vecb{\alpha}_{ab}=\sum_{c=a}^b\vecb{\alpha}_c$, for $1\leq a\leq b\leq N-1$, where the simple roots are $\vecb{\alpha}_{aa}$, and it is easily seen that there are $N(N-1)/2$ positive roots. As a quick check, we know that the dimension of any $\lie{g}$ is twice the number of positive roots plus the rank of $\lie{g}$, so here we have $N(N-1)+(N-1)=N^2-1$ as expected. The Weyl vector is $\vecb{\rho}=\frac{1}{2}\sum_{a=1}^{N-1}\sum_{b=a}^{N-1}\vecb{\alpha}_{ab}$
Using equation \eqref{eqn:dynkin} we can calculate the Dynkin index of the fundamental\begin{equation}
    C(\square)=\frac{\vecb{\alpha}_\text{max}^2}{2},
\end{equation}
where $\vecb{\alpha}_\text{max}^2$ is the length squared of any root, which we have taken above to be 1.

\subsubsection{$\mathbf{\Sp{N}}$}
$\Sp{N}$, sometimes written as $USp(2N)$, is the compact symplectic group, defined as the subgroup of $\SU{2N}$ which preserves the symplectic form $J=\begin{pmatrix}0 & \identity_N \\ -\identity_N & 0\end{pmatrix}$:
\begin{eqnarray}
    \Sp{N}&:=&\left\{U\in \SU{2N}\mid U^TJU=J\right\}\\
    \sym{N}&:=&\left\{t\in M_{2N}(\C)\mid t=t^\dagger,\ t^TJ+Jt=0\right\}
    \end{eqnarray}
The algebra is denoted by $\sym{N}$, the root system is $C_N$, thus $r=N$. The simple roots and fundamental weights are
\begin{eqnarray}
        \vecb{\alpha}_{a<N}&=&\vecb{e}_a-\vecb{e}_{a+1},\qquad \vecb{\alpha}_N=2\vecb{e}_N\\
    \vecb{w}_a&=&\sum_{j=1}^a\vecb{e}_j\\
    \vecb{\rho}&=&\sum_{j=1}^N(N-j+1)\vecb{e}_j
    \end{eqnarray}
    and the inner products of the (co-) weights are:
    \begin{eqnarray}
    \vecb{w}_a\cdot\vecb{w}_b&=&\frac{\vecb{\alpha}_\text{max}^2}{4}\min(a,b)\\
    \vecb{w}_a^*\cdot\vecb{w}_b&=&\begin{cases}\min(a,b) & a<N \\ \frac{1}{2}\min(a,b)=\frac{b}{2} & a=N\end{cases}
\end{eqnarray}
The positive roots come in four types:
\begin{align}
    \vecb{e}_a-\vecb{e}_{b+1}&=\sum_{c=a}^b\vecb{\alpha}_c,\qquad 1\leq a\leq b<N\\
    \vecb{e}_a+\vecb{e}_N&=\sum_{c=a}^{N}\vecb{\alpha}_c,\qquad 1\leq a\leq N\\
    \vecb{e}_a+\vecb{e}_b&=\sum_{c=a}^N\vecb{\alpha}_c+\sum_{c=b}^{N-1}\vecb{\alpha}_c,\qquad 1\leq a<b<N\\
    2\vecb{e}_a&=2\sum_{c=a}^{N-1}\vecb{\alpha}_c+\vecb{\alpha}_N,\qquad 1\leq a< N.
\end{align}
There are $N(N-1)/2$ positive roots of the first type, $N$ of the second, $(N-1)(N-2)/2$ of the third type, and $N-1$ of the fourth type, giving us a total of $N^2$ positive roots. Thus, the dimension of $\sym{N}$ is $2N^2+N=N(2N+1)$.

The fundamental representation has highest weight $\vecb{w}_1$. $\Sp{N}$ has a $\Z_2$ center, so we just need to find a co-weight which gives an odd integer when dotted with $\vecb{w}_1$. From the inner product relation above we see that only $(2k+1)\vecb{w}_N^*$ for $k\in\Z$ works, and thus the $\Z_2$ center is generated by $\exp(2\pi i\vecb{w}_N^*\cdot\vecb{H})=-\identity$, thus $\vecb{\mu}^*=\vecb{w}_N^*$ is a convenient choice. Then an arbitrary center element can be written as
\begin{equation}
    e^{2\pi ix/2}\identity=e^{2\pi i x \vecb{w}_N^*\cdot\vecb{H}}~,~~ x \in \Z \; ({\rm mod}\; 2).
\end{equation}

We can calculate the Dynkin index quite easily. It is clear that $\vecb{w}_1\cdot\vecb{w}_1=1$ and $\vecb{w}_1\cdot\vecb{\rho}=N$, and we know $\dim(R_{w_1})=2N$ while $\dim(\sym{N})=N(2N+1)$, so we find $C(R_{w_1})=2$, where the longest root has length 2. With an arbitrary normalization of roots, where the longest root, $\vecb{\alpha}_N$ in this case, has length squared $\vecb{\alpha}_\text{max}^2$ we find
\begin{equation}
    C(R_{\vecb{w}_1})=\frac{\vecb{\alpha}_\text{max}^2}{2}.
\end{equation}

\subsubsection{$\mathbf{\Spin{2N}}$}
$\Spin{k}$ is defined as the universal cover of the special orthogonal group, $SO(k)$, and as such has the same algebra, $\mathfrak{so}(k)$. The root system depends on whether $k$ is even or odd.
For $\Spin{2N}$ the root system is $D_N$ and the rank is $r=N$. The simple roots and fundamental weights are:
 \begin{eqnarray}
    \vecb{\alpha}_{a<N}&=&\vecb{e}_a-\vecb{e}_{a+1},\qquad \vecb{\alpha}_N=\vecb{e}_{N-1}+\vecb{e}_N\\
    \vecb{w}_{a\leq N-2}&=&\sum_{j=1}^a\vecb{e}_j,\qquad \vecb{w}_{N-1}=\frac{1}{2}\left(\sum_{j=1}^{N-1}\vecb{e}_j-\vecb{e}_N\right),\qquad \vecb{w}_N=\frac{1}{2}\sum_{j=1}^N\vecb{e}_j\\
    \vecb{\rho}&=&\sum_{j=1}^{N-1}(N-j)\vecb{e}_j
  \end{eqnarray}
  The inner products of the weights and co-weights are:
  \begin{eqnarray}\label{spin_inner}
    \eval{\vecb{w}_a\cdot\vecb{w}_b}_{a\leq b}&=&\frac{\vecb{\alpha}_\text{max}^2}{2}\begin{cases}a & a,b\leq N-2 \\ \frac{\min(a,b)}{2} & \min(a,b)<N-1\leq \max(a,b) \\ \frac{N}{4}-\frac{\abs{b-a}}{2} & a,b\geq N-1\end{cases}\\
    \vecb{w}_a^*\cdot\vecb{w}_b&=&\vecb{w}_a\cdot\vecb{w}_b^*=\begin{cases}\min(a,b) & a,b\leq N-2 \\ \frac{a}{2} & a<N-1\leq b \\ \frac{N}{4}-\frac{\abs{b-a}}{2} & a,b\geq N-1\end{cases}~.
\end{eqnarray}

$\Spin{2N}$ has two irreducible fundamental representations, corresponding to left- and right-chirality spinors, which have highest weights $\vecb{w}_{N-1}$ and $\vecb{w}_N$, each of dimension $2^{N-1}$. Since all the simple roots have length 2, we identify weights with co-weights. As a convention, we call the representation with highest weight $\vecb{w}_{N-1}$ the positive chirality representation, $S^+$, and the representation with highest weight $\vecb{w}_N$ the negative chirality representation, $S^-$. The center of $\Spin{2N}$ is $\Z_4$ if $N$ is odd, and $\Z_2^{+}\times\Z_2^-$ if $N$ is even, and will be discussed in more detail below.

The weights of the positive chirality representation, $S^+$, are the $2^{N-1}$ vectors with entries of $\pm\frac{1}{2}$ where there are an odd number of $-\frac{1}{2}$ entries. Similarly, the weights of $S^-$ are those with an even number of $- \frac{1}{2}$ entries.

Both of the highest weights have the same length squared, $\vecb{w}_{N-1}^2=\vecb{w}_N^2=\frac{N}{4}$. Further, they both have the same first $N-1$ components, so they will have the same inner product with $\vecb{\rho}$, namely $\vecb{w}_{N-1}\cdot\vecb{\rho}=\vecb{w}_{N}\cdot\vecb{\rho}=\frac{N(N-1)}{4}$. As noted above both spinor representations have dimension $2^{N-1}$, and the dimension of the algebra is $N(2N-1)$. Thus, we compute the Dynkin index of the two spinor representations as $C(S^-)=C(S^+)=2^{N-3}$ for our normalization of roots, and for an arbitrary normalization as
\begin{equation}
    C(S^-)=C(S^+)=\vecb{\alpha}_\text{max}^2 2^{N-4},
\end{equation}
where $\vecb{\alpha}_\text{max}^2$ above is taken to be 2. As per (\ref{dynkinproduct}), the Dynkin index of the direct sum representation $S^+\oplus S^-$ is simply the sum of the two Dynkin indices above,
\begin{equation}\label{dynkinspdirect}
    C(S^+\oplus S^-)=\vecb{\alpha}_\text{max}^2 2^{N-3}.
\end{equation}
We now discuss the odd- and even-$N$ cases of $Spin(2N)$ in turn:

\paragraph{$\mathbf{\Spin{4N+2}}$:}
In this case we have $\vecb{w}_{2N+1}^*\cdot\vecb{w}_{2N+1}=\vecb{w}_{2N}^*\cdot\vecb{w}_{2N}=\frac{2N+1}{4}=\frac{1}{4}+\frac{N}{2}$ and $\vecb{w}_{2N+1}^*\cdot\vecb{w}_{2N}=\vecb{w}_{2N}^*\cdot\vecb{w}_{2N+1}=\frac{2N-1}{4}=-\frac{1}{4}+\frac{N}{2}$, so either $\vecb{w}_{2N+1}^*$ or $\vecb{w}_{2N}^*$ can work for generating the $\Z_4$ center. Thus we find that both $\exp(2\pi i\vecb{w}_{2N+1}^*\cdot\vecb{H})$ and $\exp(2\pi i\vecb{w}_{2N}^*\cdot\vecb{H})$ generate the $\Z_4$ center, for both representations. In particular, on the positive chirality representation, $\exp(2\pi i\vecb{w}_{2N}^*\cdot\vecb{H})=e^{2\pi i(\frac{1}{4}+\frac{N}{2})}\identity=(-1)^Ne^{2\pi i/4}$ and $\exp(2\pi i\vecb{w}_{2N+1}^*\cdot\vecb{H})=e^{2\pi i(-\frac{1}{4}+\frac{N}{2})}\identity=(-1)^{N}e^{-2\pi i /4}$. When $N$ is even, we have $\exp(2\pi i\vecb{w}_{2N}^*\cdot\vecb{H})=e^{2\pi i/4}\identity$ so it is most convenient to take $\vecb{\mu}^*=\vecb{w}_{2N}^*$. Similarly, when $N$ is odd, it is most convenient to take $\vecb{\mu}^*=\vecb{w}_{2N+1}^*$. For the negative chirality representation we find that it is most convenient to take $\vecb{\mu}^*=\vecb{w}_{2N+1}^*$ when $N$ is even, and $\vecb{\mu}^*=\vecb{w}_{2N}^*$ when $N$ is odd. In practice however, we can only choose one of these, for instance if we take $\vecb{\mu}^*$ to be the convenient choice for the positive chirality representation, then in the direct sum representation we will have $\exp(2\pi i\vecb{\mu}^*\cdot\vecb{H})=e^{2\pi i/4}\identity_+\oplus e^{-2\pi i/4}\identity_-$. In general the action of a $\Z_4$ element on the one representation will be the conjugate of that on the other, and we can write it in the most general way as
\begin{equation}
    e^{2\pi i x/4}\identity_+\oplus e^{-2\pi i x/4}\identity_-=\begin{cases}
    e^{2\pi i x\vecb{w}_{2N}^*\cdot\vecb{H}} & N\text{ even}\\
    e^{2\pi i x\vecb{w}_{2N+1}^*\cdot\vecb{H}} & N\text{ odd}
    \end{cases}, ~~x \in \Z \; ({\rm mod} \; 4)
\end{equation}
where $\vecb{H}$ is understood to be in the direct sum representation $S^+\oplus S^-$ of $Spin(4N+2)$.

\paragraph{$\mathbf{\Spin{4N}}$:}
For this case we have to worry about each of the two chiral representations separately. For the positive chirality representation, $S^+$ with highest weight $\vecb{w}_{2N-1}$, we want to find a co-weight, $\vecb{\mu}^*$, which satisfies $\vecb{\mu}^*\cdot\vecb{w}_{2N-1}=\frac{1}{2}+\Z$. Consider $\vecb{\mu}^*=\vecb{w}_{2N}^*=\vecb{w}_{2N}$: $\vecb{w}_{2N}^*\cdot\vecb{w}_{2N-1}=\frac{(2N-1)-1}{4}=\frac{N-1}{2}$ which will be half-integer when $N$ is even, so $\exp(2\pi i\vecb{w}_{2N}^*\cdot\vecb{H}_+)$ generates $\Z_2^+$ when $N$ is even.\footnote{We use $\vecb{H}_{+}$ to denote the Cartan generators in the $S^{+}$ representation (and $\vecb{H}_-$ for  $S^-$).}
When $N$ is odd we can instead take $\vecb{\mu}^*=\vecb{w}_{2N-1}^*$: $\vecb{w}_{2N-1}^*\cdot\vecb{w}_{2N-1}=\frac{2N}{4}=\frac{N}{2}$, so $\exp(2\pi i\vecb{w}_{2N-1}^*\cdot\vecb{H}_+)$ generates $\Z_2^+$ when $N$ is odd. It is clear that $\vecb{\mu}^*=\vecb{w}_{2N}^*$ and $\vecb{\mu}^*=\vecb{w}_{2N-1}^*$ are convenient choices, for $N$ even and odd respectively. We also stress that the $\Z_2^+$ part of the center acts trivially on $S^-$, as follows upon inspection by replacing $\vecb{H}_+$ with $\vecb{H}_-$ and using the inner products of weights \eqref{spin_inner}.\footnote{\label{pitfallnote}The reader is warned to avoid a notational pitfall while using the formulae given in this section. This is due to our choice of notation and should be self-explanatory, but  is nonetheless worth pointing out. For example, in the expression for the center elements given in (\ref{direct1},\ref{direct2}), $N$ refers to the group $Spin(4N)$, while in the Dynkin index formula for $S^+\oplus S^-$ given in (\ref{dynkinspdirect}) as well as in the inner product relations \eqref{spin_inner}, $N$ refers to $Spin(2N)$.}  

For the negative chirality representation, $S^-$ with highest weight $\vecb{w}_{2N}$,  only $\Z_2^{-}$ acts nontrivially. Essentially we just need to swap $\vecb{w}_{2N}$ and $\vecb{w}_{2N-1}$, since we are looking for a co-weight $\vecb{\mu}^*$ such that $\vecb{\mu}^*\cdot\vecb{w}_{2N}=\frac{1}{2}+\Z$ and we identify weights with co-weights for $Spin(2N)$. We found above that $\vecb{w}_{2N}^*\cdot\vecb{w}_{2N-1}=\frac{1}{2}+\Z$ when $N$ is even, and since we identify weights with co-weights we immediately see that $\vecb{w}_{2N-1}^*\cdot\vecb{w}_{2N}=\frac{1}{2}+\Z$. Thus we find that $\exp(2\pi i\vecb{w}_{2N-1}^*\cdot\vecb{H}_-)$ generates $\Z_2^-$ when $N$ is even. Similarly, we find that $\exp(2\pi i\vecb{w}_{2N}^*\cdot\vecb{H}_-)$ generates $\Z_2^-$ when $N$ is odd. It is clear that $\vecb{\mu}^*=\vecb{w}_{2N-1}^*$ and $\vecb{\mu}^*=\vecb{w}_{2N}^*$ are convenient choices for $N$ even and odd respectively.

On the direct sum representation $S^+\oplus S^-$ of $Spin(4N)$, we can then write arbitrary center elements as
\begin{equation} \label{direct1}
    e^{2\pi i x^+/2}\identity_+\oplus e^{2\pi i x^-/2}\identity_-=\begin{cases}e^{2\pi i x^+\vecb{w}_{2N}^*\cdot\vecb{H}_+}\oplus e^{2\pi i x^-\vecb{w}_{2N-1}^*\cdot\vecb{H}_-} & $N$\text{ even}\\
    e^{2\pi i x^+\vecb{w}_{2N-1}^*\cdot\vecb{H}_+}\oplus e^{2\pi i x^-\vecb{w}_{2N}^*\cdot\vecb{H}_-} & $N$\text{ odd}\end{cases},~x^{+ (-)} \in \{0,1\}.
\end{equation}
Conveniently, we can write the right hand side in terms of the direct sum generators, which we write explicitly as $\vecb{H} = {\rm diag}( \vecb{H}_+, \vecb{H}_-)$,
\begin{equation}\label{direct2}
    e^{2\pi i x^+/2}\identity_+\oplus e^{2\pi i x^-/2}\identity_-=\begin{cases}e^{2\pi i (x^+\vecb{w}_{2N}^*+x^-\vecb{w}_{2N-1}^*)\cdot\vecb{H}} & $N$\text{ even}\\
    e^{2\pi i (x^+\vecb{w}_{2N-1}^*+x^-\vecb{w}_{2N})\cdot\vecb{H}} & $N$\text{ odd}\end{cases},~x^{+ (-)}  \in \{0,1\}.
\end{equation}
since whenever $e^{2\pi i\vecb{\mu}^*\cdot\vecb{H}}=-\identity$ on one representation, it is the identity on the other.

Also note that $\vecb{\mu}^*=\vecb{w}_{2k+1}^*$ for $1\leq 2k+1<2N-1$ is a convenient choice for both representations, but it isn't much help to us since it treats the two centers the same, that is, $\exp(2\pi i x\vecb{w}_{2k+1}^*\cdot\vecb{H})=\exp(2\pi i x/2)(\identity_{+}\oplus\identity_-)$. Thus, when working in the direct sum representation, we can't separate the two centers if we use $\vecb{w}_{2k+1}^*$, so we opt to use the others described above.

\subsubsection{$\mathbf{\Spin{2N+1}}$}
For $\Spin{2N+1}$ the root system is $B_N$, $r=N$, with simple roots and fundamental weights given by:
\begin{eqnarray}
    \vecb{\alpha}_{a<N}&=&\vecb{e}_a-\vecb{e}_{a+1},\qquad \vecb{\alpha}_N=\vecb{e}_N\\
    \vecb{w}_{a\leq N-1}&=&\sum_{j=1}^a\vecb{e}_j,\qquad \vecb{w}_N=\frac{1}{2}\sum_{j=1}^N\vecb{e}_j
    \end{eqnarray}
    \begin{eqnarray}
    \vecb{\rho}&=&\sum_{j=1}^N\left(N-j+\frac{1}{2}\right)\vecb{e}_j
    \end{eqnarray}
    and inner products of (co-) weights
    \begin{eqnarray}
    \eval{\vecb{w}_a\cdot\vecb{w}_b}_{a\leq b}&=&\frac{\vecb{\alpha}_\text{max}^2}{2}\begin{cases}a & a,b\leq N-1 \\ \frac{a}{2} & a<N=b \\ \frac{N}{4} & a=b=N\end{cases}\\
    \vecb{w}_a^*\cdot\vecb{w}_b&=&\begin{cases}\min(a,b) & b<N \\ \frac{1}{2}\min(a,b)=\frac{a}{2} & b=N\end{cases}~.
\end{eqnarray}

For $\Spin{2N+1}$ there is just one spinor representation, with highest weight $\vecb{w}_N$. We consider $N\geq2$ since $\Spin{3}\cong\SU{2}$. The center is $\Z_2$, so we want to find a co-weight, $\vecb{\mu}^*$, such that $\vecb{\mu}^*\cdot\vecb{w}_N=\frac{1}{2}+\Z$. From the inner product above we see that $\vecb{w}_{2k+1}^*\cdot\vecb{w}_N=\frac{1}{2}+\Z$ for $0\leq k<(N-1)/2$, giving us a whole set of equivalent generators. Each of these choices of $\vecb{\mu}^*$ are convenient choices, and thus an arbitrary center element can be written as
\begin{equation}
    e^{2\pi i x/2}\identity=e^{2\pi i x \vecb{w}_{2k+1}^*\cdot\vecb{H}},\ 1\leq k<(N-1)/2, ~ x \in \Z \; ({\rm mod} \; 2)~.
\end{equation}

The weights of the spinor representation are the $2^N$ vectors with entries of $\pm\frac{1}{2}$. Thus, we find $\dim(R_{w_N})=2^N$. We see that $\vecb{w}_N^2=\frac{N}{4}$, and $\vecb{w}_N\cdot\vecb{\rho}=\left(\frac{N}{2}\right)^2$. The dimension of the algebra is $N(2N+1)$, and the dimension of the spinor representation is $2^N$, so the Dynkin index is $C(R_{w_N})=2^{N-2}$, where the longest roots have length squared 2. In arbitrary root normalization we get
\begin{equation}
    C(R_{w_N})=\vecb{\alpha}_\text{max}^2 2^{N-3},
\end{equation}
similar to the even spin groups.

\subsubsection{$\mathbf{\E{6}}$}
$\E{6}$ and its root system share the same name, and the same occurs for all the other exceptional algebras. Here, $r=6$ and the list of simple roots and fundamental weights is as follows:
\begin{eqnarray}
    \vecb{\alpha}_{a<4}&=&\vecb{e}_a-\vecb{e}_{a+1},\;\;  \vecb{\alpha}_4=\vecb{e}_4+\vecb{e}_5,\;\; \vecb{\alpha}_5=\frac{1}{2}\left(\sqrt{3}\vecb{e}_6-\sum_{i=1}^{5}\vecb{e}_i\right),\;\; \vecb{\alpha}_6=\vecb{e}_4-\vecb{e}_5\\
    \vecb{w}_{a<4}&=&\sum_{j=1}^a\vecb{e}_j+\frac{a}{\sqrt{3}}\vecb{e}_6,\;\; \vecb{w}_4=\frac{1}{2}\left(\sum_{j=1}^5\vecb{e}_j+\frac{5}{\sqrt{3}}\vecb{e}_6\right),\;\; \vecb{w}_5=\frac{2}{\sqrt{3}}\vecb{e}_6,\;\; \vecb{w}_6=\frac{1}{2}\left(\sum_{j=1}^4\vecb{e}_j-\vecb{e}_5+\sqrt{3}\vecb{e}_6\right) \nonumber\\
    \vecb{\rho}&=&\sum_{j=1}^4(5-j)\vecb{e}_j+4\sqrt{3}\vecb{e}_6~.
    \end{eqnarray}
The inner product of weights are 
    \begin{eqnarray}
    \left[\vecb{w}_a\cdot\vecb{w}_b\right]&=&\frac{\vecb{\alpha}_\text{max}^2}{2}\begin{pmatrix}4/3 & 5/3 & 2 & 4/3 & 2/3 & 1 \\
 5/3 & 10/3 & 4 & 8/3 & 4/3 & 2 \\
 2 & 4 & 6 & 4 & 2 & 3 \\
 4/3 & 8/3 & 4 & 10/3 & 5/3 & 2 \\
 2/3 & 4/3 & 2 & 5/3 & 4/3 & 1 \\
 1 & 2 & 3 & 2 & 1 & 2 \end{pmatrix}\\
    \left[\vecb{w}_a^*\cdot\vecb{w}_b\right]&=&\begin{pmatrix}4/3 & 5/3 & 2 & 4/3 & 2/3 & 1 \\
 5/3 & 10/3 & 4 & 8/3 & 4/3 & 2 \\
 2 & 4 & 6 & 4 & 2 & 3 \\
 4/3 & 8/3 & 4 & 10/3 & 5/3 & 2 \\
 2/3 & 4/3 & 2 & 5/3 & 4/3 & 1 \\
 1 & 2 & 3 & 2 & 1 & 2 \end{pmatrix}
\end{eqnarray}

There are two equivalent fundamental representations of $\E{6}$, with highest weight $\vecb{w}_1$ and $\vecb{w}_5$ respectively. We start off by using the representation with highest weight $\vecb{w}_1$. The center is $\Z_3$, so we want to find a co-weight $\vecb{\mu}^*$ such that $\vecb{\mu}^*\cdot\vecb{w}_1=\frac{l}{3}+\Z$ for $l=1$ or $l=2$. Again here all the roots have length $\sqrt{2}$, so we identify weights with co-weights. We find that $\vecb{w}_1^*\cdot\vecb{w}_1=\vecb{w}_4^*\cdot\vecb{w}_1=\frac{1}{3}+1$, $\vecb{w}_2^*\cdot\vecb{w}_1=\frac{2}{3}+1$, and $\vecb{w}_5^*\cdot\vecb{w}_1=\frac{2}{3}$, so we conclude that the $\Z_3$ center of the representation with highest weight $\vecb{w}_1$ is generated by $\exp(2\pi i\vecb{w}_a^*\cdot\vecb{H})$ for $a=1,2,4,5$. Of these choices of $\vecb{\mu}^*$, only $a=1,4$ are convenient choices, while for $a=2,5$ we must scale $\vecb{\mu}^*$ by 2 to make them convenient choices. We can then write an arbitrary center element in the following ways
\begin{equation}
    e^{2\pi i x/3}\identity=e^{2\pi i x\vecb{w}_{1,4}^*\cdot\vecb{H}}=e^{2\pi i 2x\vecb{w}_{2,5}^*\cdot\vecb{H}},~  x \in \Z \; ({\rm mod} \;3)~,
\end{equation}
where $\vecb{w}_{a,b}^*$ could be either $\vecb{w}_a^*$ or $\vecb{w}_b^*$.

Similarly, for the representation with highest weight $\vecb{w}_5$ we find that the $\Z_3$ center is generated by $\exp(2\pi i\vecb{w}_a^*\cdot\vecb{H})$ for $a=1,2,4,5$, exactly the same as before, where now $a=2,5$ are convenient, and $a=1,4$ must be scaled by 2 to be convenient.

The weights of the fundamental representation with highest weight $\vecb{w}_1$ are given below (in no particular order, except that $\vecb{\nu}_1=\vecb{w}_1$)
\begin{equation}
    \vecb{\nu}_{1\leq A\leq 5}=\vecb{e}_A+\frac{1}{\sqrt{3}}\vecb{e}_6,\;\; \vecb{\nu}_{6\leq A\leq 10}=-\vecb{e}_{A-5}+\frac{1}{\sqrt{3}}\vecb{e}_6,\;\;\vecb{\nu}_{11}=-\frac{2}{\sqrt{3}}\vecb{e}_6,\;\;  \vecb{\nu}_{A\geq 12}=\frac{1}{2}\left(\sum_{a=1}^5(-1)^{q_a}\vecb{e}_a-\frac{1}{\sqrt{3}}\vecb{e}_6\right),\nonumber
\end{equation}
where $\sum_a q_a$ is odd. There are 27 weights, each with multiplicity one, so the dimension of $R_{w_1}$ is 27. We see that $\vecb{w}_1^2=\frac{4}{3}$, while $\vecb{w}_1\cdot\vecb{\rho}=8$. The dimension of the algebra is 78, and thus the Dynkin index is $C(R_{w_1})=6$ with $\vecb{\alpha}_\text{max}^2=2$, so in general we get
\begin{equation}
    C(R_{w_1})=3\vecb{\alpha}_\text{max}^2
\end{equation}
We get the same result if we use the representation with highest weight $\vecb{w}_5$.

\subsubsection{$\mathbf{\E{7}}$}
The rank of this group is $7$ and the simple roots and fundamental weights are
 \begin{eqnarray}
    \vecb{\alpha}_1&=&\frac{1}{2}\left(\sqrt{2}\vecb{e}_7-\sum_{i=1}^6\vecb{e}_i\right),\;\; \vecb{\alpha}_2=\vecb{e}_5+\vecb{e}_6,\qquad \vecb{\alpha}_{3\leq a\leq 6}=\vecb{e}_{7-a}-\vecb{e}_{7-a+1},\;\; \vecb{\alpha}_7=\vecb{e}_5-\vecb{e}_6\\
    \vecb{w}_1&=&\sqrt{2}\vecb{e}_7,\;\; \vecb{w}_2=\frac{1}{2}\left(\sum_{j=1}^6\vecb{e}_j+3\sqrt{2}\vecb{e}_7\right),\;\; \vecb{w}_{3\leq a\leq 6}=\sum_{j=1}^{7-a}\vecb{e}_j+\frac{7-a}{\sqrt{2}}\vecb{e}_7,\;\; \vecb{w}_7=\frac{1}{2}\left(\sum_{j=1}^{5}\vecb{e}_j-\vecb{e}_6+2\sqrt{2}\vecb{e}_7\right)\nonumber \\
    \vecb{\rho}&=&\sum_{j=1}^5(6-j)\vecb{e}_j+\frac{17}{\sqrt{2}}\vecb{e}_7  
    \end{eqnarray}
    \begin{eqnarray}
      \left[\vecb{w}_a \cdot \vecb{w}_b\right]&=&\frac{\vecb{\alpha}_\text{max}^2}{2}\begin{pmatrix}
 2 & 3 & 4 & 3 & 2 & 1 & 2 \\
 3 & 6 & 8 & 6 & 4 & 2 & 4 \\
 4 & 8 & 12 & 9 & 6 & 3 & 6 \\
 3 & 6 & 9 & 15/2 & 5 & 5/2 & 9/2 \\
 2 & 4 & 6 & 5 & 4 & 2 & 3 \\
 1 & 2 & 3 & 5/2 & 2 & 3/2 & 3/2 \\
 2 & 4 & 6 & 9/2 & 3 & 3/2 & 7/2
 \end{pmatrix}
     \end{eqnarray}
    \begin{eqnarray}
    \left[\vecb{w}_a^*\cdot\vecb{w}_b\right]&=&\begin{pmatrix}2 & 3 & 4 & 3 & 2 & 1 & 2 \\
 3 & 6 & 8 & 6 & 4 & 2 & 4 \\
 4 & 8 & 12 & 9 & 6 & 3 & 6 \\
 3 & 6 & 9 & 15/2 & 5 & 5/2 & 9/2 \\
 2 & 4 & 6 & 5 & 4 & 2 & 3 \\
 1 & 2 & 3 & 5/2 & 2 & 3/2 & 3/2 \\
 2 & 4 & 6 & 9/2 & 3 & 3/2 & 7/2\end{pmatrix}
    \end{eqnarray}
    
There is a single fundamental representation of $\E{7}$ which has highest weight $\vecb{w}_6$. The center is $\Z_2$, so we want to find a co-weight $\vecb{\mu}^*$ such that $\vecb{\mu}^*\cdot\vecb{w}_6=\frac{1}{2}+\Z$. Again, we can identify weights with co-weights here. We find that $\vecb{w}_6^*\cdot\vecb{w}_6=\vecb{w}_7^*\cdot\vecb{w}_6=\frac{1}{2}+1$ and that $\vecb{w}_4^*\cdot\vecb{w}_6=\frac{1}{2}+2$, so we conclude that the $\Z_2$ center is generated by $\exp(2\pi i\vecb{w}_a^*\cdot\vecb{H})$ for $a=4,6,7$, all of which are convenient choices. Then, an arbitrary center element can be written as
\begin{equation}
    e^{2\pi ix/2}\identity=e^{2\pi ix\vecb{w}_a^*\cdot\vecb{H}},\ a=4,6,7,~  x \in \Z \; ({\rm mod} \; 2)~.
     \end{equation}

The 56 weights of the fundamental representation are of the form $\pm \vecb{e}_i\pm\frac{1}{\sqrt{2}}\vecb{e}_7$ for $1\leq i\leq 6$, as well as $\frac{1}{2}\sum_{i=1}^6(-1)^{q_i}\vecb{e}_i$ where $\sum_i q_i$ is odd. We see that $\vecb{w}_6^2=\frac{3}{2}$ and $\vecb{w}_6\cdot\vecb{\rho}=\frac{27}{2}$. The dimension of the algebra is 133, so the Dynkin index is $C(R_{w_6})=12$, or in an arbitary normalization,
\begin{equation}
    C(R_{w_6})=6\vecb{\alpha}_\text{max}^2.
\end{equation}

\section{'t Hooft twists  for all gauge groups}
\label{appx:fractionalcharge}

In this appendix, we describe in detail the introduction of twisted boundary conditions on $\T^4$ (and, by restriction,  $\T^3$) for all  compact simple Lie groups with nontrivial center.
While the results are not new and have been already given\footnote{More recently, these were used in ref.~\cite{Cordova:2019uob}, also in the framework of generalized anomalies. This reference also considered non-spin manifolds. For completeness, we use discussion of \cite{Anber:2020gig} to study the fractional topological charge for all groups on the non-spin manifold $\mathbb{CP}^2$ in appx.~\ref{appx:nonspin}.}  in \cite{Witten:2000nv}, our derivation using transition functions and co-cycle conditions on $\T^4$ is  quite explicit and physicist-friendly.

 To the best of our knowledge a discussion along the lines of \cite{tHooft:1981sps,vanBaal:1982ag} for general gauge groups has not previously appeared in the literature.  
 The formulae of this appendix may also be helpful in the studies of other types of generalized anomalies and we hope they will be of use to physicists.

\subsection{Normalizing the topological charge: the BPST instanton}
\label{appx:normalizingQ}

Here, we shall  properly normalize the topological charge  $Q_\text{top}$ in any representation. We need to dwell on this detail, because our explicit description of the $\T^4$ bundle and the 't Hooft twists on $\T^3$  requires us to study the gauge field using the generators of the ``convenient'' representation of $G$, where the center of $G$ acts nontrivially.  We begin by the expression for the topological charge in a general representation $R$ 
\begin{equation} \label{a4}
    Q_\text{top}=\frac{N(G)}{32\pi^2}\int\dd[4]{x}\Tr_R \left(F_{\mu\nu}\tilde{F}^{\mu\nu}\right),
\end{equation}
where $N(G)$ is normalization factor that we want to determine. 
The above expression for $Q_\text{top}$ is valid on $\T^4$ as well as in the $\R^4$ limit. Being an integral of a total divergence, $Q_\text{top}$ only depends on appropriate transition functions, a fact that we  explicitly use below, see (\ref{vanbaaltwo}). 

  We shall determine the 
 normalization factor $N(G)$, such that upon embedding an $\R^4$ BPST instanton solution into an $\SU{2}$ subgroup of the gauge group $G$, the minimum topological charge we obtain is $Q_\text{top} = 1$.  
 For the reader interested only in the results,  in Table \ref{tab:top-normalization} we give the root lengths and, most importantly, the result for the normalization factor $N(G)$ for all groups, as determined in the rest of this section.\footnote{We also stress that the topological charge for the twisted bundles on $\T^4$ that we calculate in section \ref{appx:Qforallgroups} is independent on the normalization of roots, see the discussion after eqn.~(\ref{a19}). }

\begin{table}[h]
    \centering
    {\tabulinesep=1.0mm
    \begin{tabu}{llll}
        \hline
        Group & Root Lengths & $C(R)$/$\vecb{\alpha}_\text{max}^2$ & $N(G) = \frac{\vecb{\alpha}_\text{max}^2}{C(R)}$ \\ \hline
        $\SU{N}$ & 1 & $\frac{1}{2}$ & 2 \\
        $\Sp{N}$ & $\sqrt{2},\ 2$ & $\frac{1}{2}$ & $2$ \\
        $\Spin{2N}$ & $\sqrt{2}$ & $2^{N-3}$ & $2^{3-N}$ \\
        $\Spin{2N+1}$ & $1,\ \sqrt{2}$ & $2^{N-3}$ & $2^{3-N}$ \\
        $\E{6}$ & $\sqrt{2}$ & 3 & $\frac{1}{3}$ \\
        $\E{7}$ & $\sqrt{2}$ & 6 & $\frac{1}{6}$ \\ \hline
    \end{tabu}}
    \caption{Groups with their root lengths, ``convenient'' representation Dynkin indices $C(R)$, as well as the normalization $N(G)$ of the topological charge (\ref{a4}).}
    \label{tab:top-normalization}
\end{table}

 For those interested in the details, we begin by noting that 
given a positive root $\vecb{\alpha}$ we can construct $\su{2}$ generators $\tau^a$:
\begin{equation}\label{a2}
    \tau^1=\frac{1}{2}(E_\alpha+E_{-\alpha}),\qquad \tau^2=\frac{1}{2i}(E_\alpha-E_{-\alpha}),\qquad \tau^3=\frac{1}{2}\vecb{\alpha}^*\cdot\vecb{H},
\end{equation}
where $\vecb{\alpha}^*=\frac{2}{\vecb{\alpha}\cdot\vecb{\alpha}} \vecb{\alpha}$ is the co-root associated to the root $\alpha$. These generators will satisfy the $\su{2}$ algebra provided that $E_{\pm\alpha}$ are normalized properly:
\begin{equation*}
    \comm{E_\alpha}{E_{-\alpha}}=\vecb{\alpha}^*\cdot\vecb{H}\implies \comm{\tau^a}{\tau^b}=i\varepsilon^{abc}\tau^c.
\end{equation*}
If the root vectors are not normalized as above, then we will have $\comm{\tau^a}{\tau^b}\propto i\varepsilon^{abc}\tau^c$, where the value of the proportionality is different for the different values of $a,b$. Further, the $\su{2}$ commutation relations guarantee that $\Tr(\tau^a\tau^a)$ is the same for all choices of $a$. To see this, we consider $\Tr(\tau^3\tau^3)$. Then, use $\tau^3=-i\comm{\tau^1}{\tau^2}$ to get $\Tr(\tau^3\tau^3)=\Tr(-i\comm{\tau^1}{\tau^2}\tau^3)=-i\Tr(\tau^1\comm{\tau^2}{\tau^3})=-i\Tr(\tau^1i\tau^1)=\Tr(\tau^1\tau^1)$. In the same way we can show that $\Tr(\tau^3\tau^3)=\Tr(\tau^2\tau^2)$. Note that this would still be true if $\comm{\tau^a}{\tau^b}=ix\varepsilon^{abc}\tau^c$ for some constant $x$. Then, for the embedding corresponding to the root $\vecb{\alpha}$, following from our definition of the Dynkin index, $C(R)$, we have
\begin{equation}
\label{a3}
    \Tr(\tau^a\tau^b)=\Tr(\tau^3\tau^3)\delta^{ab}=\delta^{ab}C(R)\left(\frac{1}{2}\vecb{\alpha}^*\right)^2=\frac{C(R)}{\vecb{\alpha}^2}\delta^{ab}~.
\end{equation}
In table \ref{tab:dynkin}, we give the relevant (fundamental) Dynkin indices for all groups.

\paragraph{}
Now, we consider the $SU(2)$ BPST instanton solution  with field strength $F_{\mu\nu}^a$, $a=1,2,3$, embedded into $G$ via (\ref{a2}), 
\begin{equation}\label{a33}
    F_{\mu\nu}=\tau^a F_{\mu\nu}^a = -4 \tau^a \eta_{a\mu\nu}\frac{\rho^2}{\left[(x-x_0)^2+\rho^2\right]^2}, ~ {\rm with }~ \int\dd[4]{x}F^a_{\mu\nu}\tilde{F}^{a,\mu\nu}= 32\pi^2,
\end{equation}
where $\rho$ is the size of the instanton, $x_0$ is its position, and $\eta_{a\mu\nu}$ are the 't Hooft symbols (e.g. \cite{Vandoren:2008xg}). For the embedding (\ref{a2}) according to the root $\vecb{\alpha}$, the topological charge is then given by
\begin{equation}
    Q_\text{top}=\frac{N(G)}{32\pi^2}\int\dd[4]{x}\Tr_R\left(F_{\mu\nu}\tilde{F}^{\mu\nu}\right)=N(G)\frac{C(R)}{\vecb{\alpha}^2},
\end{equation}
where  we used (\ref{a33}) and (\ref{a3}). 

As already discussed, in our study of $\T^4$ bundle, we want to define $Q_\text{top}$ so that the \textit{minimum} possible charge for an embedding of the BPST instanton is $Q_\text{top}=1$, so we must set $N(G)$ as
\begin{equation}
    N(G)=\frac{\vecb{\alpha}_\text{max}^2}{C(R)},
\end{equation}
where $\vecb{\alpha}_\text{max}^2$ is the length squared of the longest root. Embeddings with shorter roots correspond to multi-instantons, see e.g. \cite{Vandoren:2008xg}. Finally, notice that $N(G)$ is independent of the normalization of the roots, as required. The results for the root lengths and $N(G)$ for the various groups are summarized in Table \ref{tab:top-normalization}.

\subsection{Fractional topological charge on $\mathbf{\T^4}$}
\label{appx:Qforallgroups}
\paragraph{}
In this section, we calculate the topological charge for a $\T^4$-bundle twisted by the center, for general simple gauge group. We assume that the center is a cyclic group $\Z_k$ for some $k$, and  follow van Baal's work for $\SU{N}$ \cite{vanBaal:1982ag}.\footnote{Ref.~\cite{vanBaal:1982ag} calculates the $SU(N)$ topological charge with  $N(G)=2$, consistent with our Table \ref{tab:top-normalization}.} The discussion here holds for all gauge groups, except $\Spin{4N}$ which gets a similar, but notably different, treatment in the relevant section below. 

We take the side lengths of $\T^4$ to be $L_\mu$ for $\mu=1,2,3,4$. We use $\Omega_\mu$ to denote the transition function relating the gauge field at $x_\mu=L_\mu$ with the field at $x_\mu=0$:
\begin{equation}\label{match1}
    A(x_\mu=L_\mu)=\Omega_\mu \circ \ A(x_\mu=0)~,
\end{equation}
where in accordance with usual notation, we do not display the arguments of $\Omega_\mu$ (noting only that, obviously, $\Omega_\mu$ does not depend on $x^\mu$).
With a  co-cycle condition relaxed by a center element, transition functions in the corners must commute up to a center element. In the $\mu$-$\nu$ plane we call this center element $Z_{\mu\nu}\in\Z_k$:
\begin{equation}\label{twist1}
    Z_{\mu\nu}=\Omega_\mu(x_\nu=L_\nu)\Omega_\nu(x_\mu=0)\Omega_\mu^{-1}(x_\nu=0)\Omega_\nu^{-1}(x_\mu=L_\mu)\equiv e^{2\pi i n_{\mu\nu}/k}\identity,
\end{equation}
which defines the integer $n_{\mu\nu}=-n_{\nu\mu}$. In  Theorem 3.1 of \cite{vanBaal:1982ag}, whose proof  holds for any simple Lie group with trivial $\pi_2(G)$, it was shown that for the purposes of calculating the non-integer part of the topological charge it suffices to take the transition functions to lie in the maximal torus. Thus, we take 
\begin{equation}\label{omega1}
\Omega_\mu=\exp{2\pi i\vecb{f}_\mu(x)\cdot\vecb{H}},
\end{equation}
 where $\vecb{H}$ are the Cartan generators in the appropriate ``convenient'' representation $R$. Define $\vecb{n}_{\mu\nu}$ as follows
\begin{equation}\label{endef}
    \vecb{n}_{\mu\nu}\equiv \vecb{f}_\mu(x_\nu=L_\nu)+\vecb{f}_\nu(x_\mu=0)- \vecb{f}_\mu(x_\nu=0)-\vecb{f}_\nu(x_\mu=L_\mu).
\end{equation}
Notice that, by continuity, $\vecb{n}_{\mu\nu} = - \vecb{n}_{\nu\mu}$ does not depend on the transverse coordinates. In fact, as in Theorem 3.1 of \cite{vanBaal:1982ag},   it suffices to take 
\begin{equation}\label{fmu}
\vecb{f}_\mu =  \sum\limits_\nu \vecb{n}_{\mu\nu} \; {x^\nu \over 2 L^\nu} . 
\end{equation}
Then, we have for $Z_{\mu\nu}$:
\begin{equation}
    Z_{\mu\nu}=\exp(2\pi i \vecb{n}_{\mu\nu}\cdot\vecb{H}).
\end{equation}
To be consistent with the definition of $n_{\mu\nu}$ we require that $\vecb{n}_{\mu\nu}=n_{\mu\nu}\vecb{\mu}^*+\vecb{\alpha}^*_{\mu\nu}$ where $\vecb{\alpha}^*_{\mu\nu}$ is an arbitrary vector in the co-root lattice and can be different for each $\mu-\nu$ plane, and $\vecb{\mu}^*$ is the co-weight which generates the center. We have assumed that for the representation $R_{\vecb\nu}$ with highest weight $\vecb{\nu}$, $\vecb{\mu}^*$ is a convenient choice - which can always be done, as discussed in the previous section.

\paragraph{}
From \cite{vanBaal:1982ag}, see Lemma 3.1 there,\footnote{Eqn.~(\ref{vanbaaltwo})  follows from (\ref{a4}) upon integrating by parts on $\T^4$ and repeated use of the co-cycle conditions.}  we find that the topological charge, for $\Omega_\mu$ in the maximal torus, is given by
\begin{equation}\label{vanbaaltwo}
    Q_\text{top}=\frac{N(G)}{2}\frac{1}{8\pi^2}\sum_{\mu,\nu}\int\dd[2]S_{\mu\nu}\varepsilon_{\mu\nu\alpha\beta}\Tr\left[\left(\Omega_\nu^{-1}\partial_\alpha\Omega_\nu\right)_{x_\mu=L_\mu}\left(\Omega_\mu\partial_\beta\Omega_\mu^{-1}\right)_{x_\nu=0}\right],
\end{equation}
where $\int\dd[2]S_{12}=\int\limits_{0}^{L_3} \dd{x_3}\int\limits_{0}^{L_4}\dd{x_4}$, etc. We can plug in $\Omega$ from (\ref{omega1}):
\begin{align}
    \Omega_\nu^{-1}\partial_\alpha\Omega_\nu=2\pi i\partial_\alpha \vecb{f}_\nu(x)\cdot\vecb{H}~~,~~   \Omega_\mu\partial_\beta\Omega_\mu^{-1}=-2\pi i\partial_\beta \vecb{f}_\mu(x)\cdot\vecb{H} ~,
\end{align}
and obtain
\begin{align}
    Q_\text{top}&=\frac{N(G)}{2}\frac{1}{8\pi^2}\sum_{\mu,\nu}\int\dd[2]S_{\mu\nu}\varepsilon_{\mu\nu\alpha\beta}\Tr\left[\left(2\pi i\partial_\alpha \vecb{f}_\nu(x_\mu=L_\mu)\cdot\vecb{H}\right)\left(-2\pi i\partial_\beta \vecb{f}_\mu(x_\nu=0)\cdot\vecb{H}\right)\right] \nonumber \\
    &=\frac{N(G)}{2}\frac{C(R)}{2}\sum_{\mu,\nu}\int\dd[2]S_{\mu\nu}\varepsilon_{\mu\nu\alpha\beta} \partial_\alpha \vecb{f}_\nu(x_\mu=L_\mu)\cdot \partial_\beta \vecb{f}_\mu(x_\nu=0) \nonumber \\
    &=\frac{\vecb{\alpha}_\text{max}^2}{2C(R)}\frac{C(R)}{2}\varepsilon_{\mu\nu\alpha\beta}\frac{\vecb{n}_{\nu\alpha}}{2}\cdot\frac{\vecb{n}_{\mu\beta}}{2} =\frac{\vecb{\alpha}_\text{max}^2}{4}\varepsilon_{\mu\nu\alpha\beta}\frac{n_{\nu\alpha}\vecb{\mu}^*+\vecb{\alpha}^*_{\nu\alpha}}{2}\cdot\frac{n_{\mu\beta}\vecb{\mu}^*+\vecb{\alpha}^*_{\mu\beta}}{2}\nonumber \\
    &=\frac{\vecb{\alpha}_\text{max}^2}{4}\left[(\vecb{\mu}^*)^2\; 2\operatorname{Pf}(n)+\vecb{\mu}^*\cdot\varepsilon_{\mu\nu\alpha\beta}\;\vecb{\alpha}_{\mu\beta}^*\frac{n_{\nu\alpha}}{2}+\varepsilon_{\mu\nu\alpha\beta}\frac{\vecb{\alpha}_{\nu\alpha}^*}{2}\cdot\frac{\vecb{\alpha}_{\mu\beta}^*}{2}\right] \label{eqn:top_unsimple},
\end{align}
where $\operatorname{Pf}(n)$ is the Pfaffian of $n$, defined as \begin{equation}
\operatorname{Pf}(n) \equiv \frac{1}{8}\varepsilon_{\mu\nu\alpha\beta}n_{\mu\nu}n_{\alpha\beta}\label{pfaffiandefinition}
\end{equation}
 in four dimensions.\footnote{For an antisymmetric matrix with  integer-valued entries $n_{\mu\nu}$, 
$\operatorname{Pf}(n)$ is an integer. The simplest example is the matrix with all entries zero but $n_{12} = - n_{21} = n_{34} = - n_{43} = 1$ which has $\operatorname{Pf}(n) = 1$.}

 Now, we examine the various terms above. 
 Examining the second term in (\ref{eqn:top_unsimple}) closer,  consider $\frac{1}{8}\vecb{\alpha}_\text{max}^2\vecb{\mu}^*\cdot\vecb{\alpha}_{\mu\beta}^*$. We know that $\vecb{\mu}^*\in\Lambda_w^*$ so $\vecb{\mu}^*=\sum_i\mu_i\vecb{w}_i^*$ for $\mu_i\in\Z$, and similarly $\vecb{\alpha}_{\mu\beta}^*\in\Lambda_r^*$ so $\vecb{\alpha}_{\mu\beta}^*=\sum_i (\alpha_{\mu\beta})_i\vecb{\alpha}_i^*$ for $(\alpha_{\mu\beta})_i\in\Z$. We then have $\vecb{\mu}^*\cdot\vecb{\alpha}_{\mu\beta}^*=\sum_{i,j}\mu_i(\alpha_{\mu\beta})_j\vecb{w}_i^*\cdot\vecb{\alpha}_j^*$, but by definition we have $\vecb{w}_i^*\cdot\vecb{\alpha}_j^*=\frac{2}{\vecb{\alpha}_i^2}\delta_{ij}$, so we find $\frac{1}{8}\vecb{\alpha}_\text{max}^2\vecb{\mu}^*\cdot\vecb{\alpha}_{\mu\beta}^*=\frac{1}{8}\vecb{\alpha}_\text{max}^2\sum_i\mu_i(\alpha_{\mu\beta})_i\frac{2}{\vecb{\alpha}_i^2}=\frac{1}{4}\sum_i\mu_i(\alpha_{\mu\beta})_i\frac{\vecb{\alpha}_\text{max}^2}{\vecb{\alpha}_i^2}$. Finally, recall that the ratio of the lengths of any two roots (with the longer root in the numerator) is one of $1, \sqrt{2}, \sqrt{3}$, and thus $\frac{\vecb{\alpha}_\text{max}^2}{\vecb{\alpha}_i^2}\in\Z$. Now we define $\xi_{\mu\beta}\equiv \sum_i\mu_i(\alpha_{\mu\beta})_i\frac{\vecb{\alpha}_\text{max}^2}{\vecb{\alpha}_i^2}$, which must be an integer, as just argued, and is antisymmetric since $\vecb{\alpha}^*_{\mu\beta}$ is antisymmetric. Including the Levi-Cevita symbol and $n_{\nu\alpha}$ we find the total second term to be $\frac{1}{4}\varepsilon_{\mu\nu\alpha\beta}n_{\nu\beta}\xi_{\mu\alpha}$. It is not   hard to see that this must be an integer, since we get a factor of 4 coming from the antisymmetry of both $n$ and $\xi$, canceling the overall factor of $\frac{1}{4}$.

Looking at the last term in (\ref{eqn:top_unsimple}), consider $\vecb{\alpha}_{\nu\alpha}^*\cdot\vecb{\alpha}_{\mu\beta}^*$. Recalling that the co-weight lattice spans the co-root lattice, we can directly import our previous work with $\vecb{\mu}^*\cdot\vecb{\alpha}_{\mu\beta}^*$ to find that $(\vecb{\alpha}_\text{max}^2)\vecb{\alpha}_{\nu\alpha}^*\cdot\vecb{\alpha}_{\mu\beta}^*\in 2\Z$. Defining $\zeta_{\nu\alpha\mu\beta}\equiv \frac{1}{2}(\vecb{\alpha}_\text{max}^2)\vecb{\alpha}_{\nu\alpha}^*\cdot\vecb{\alpha}_{\mu\beta}^*\in\Z$, it is clear that $\zeta$ is antisymmetric in its first two indices and its last two indices, and is symmetric with respect to swapping the first two indices with the second two ($\nu\alpha\leftrightarrow \mu\beta$). We then have $\frac{1}{8}\varepsilon_{\mu\nu\alpha\beta}\zeta_{\nu\alpha\mu\beta}$. Again, it is not too difficult to see that this must also be an integer, since we can swap the first two indices of $\zeta$, the last two, and the first two with the last two, each of which contributes a factor of 2.

In conclusion, we see that the second and third term in equation \eqref{eqn:top_unsimple} are integers, and hence we find
\begin{equation}\label{a19}
    Q_\text{top}=\frac{\vecb{\alpha}_\text{max}^2\; (\vecb{\mu}^*)^2}{2}\operatorname{Pf}(n)+\Z, 
\end{equation}
with $\operatorname{Pf}(n)$ defined in (\ref{pfaffiandefinition}). Now we  ask if this $Q_\text{top}$ is invariant under changes of normalization of roots. Recall that if we rescale our roots $\vecb{\alpha}\rightarrow c \vecb{\alpha}$, then the weights must also scale with $c$, while co-roots and co-weights scale with $\frac{1}{c}$. Thus, in our above expression $\vecb{\alpha}_\text{max}^2\rightarrow c^2\vecb{\alpha}_\text{max}^2$ will be compensated by $(\vecb{\mu}^*)^2\rightarrow \frac{1}{c^2}(\vecb{\mu}^*)^2$. We then find that the topological charge is invariant, as it should be.

\begin{table}[h]
    \centering
    {\tabulinesep=1.0mm
    \begin{tabu}{llll}
        \hline
        Group &Center  & $Q_\text{top}\pmod{1}$ &$Q_\text{top}^{\mathbb{CP}^2}  \pmod{1}$ \\ \hline
        $\SU{N}$  & $\Z_N$ &$-\frac{1}{N}\operatorname{Pf}(n)$ &$-{ 1 \over 2 N} \;n^2$ \\
        $\Sp{N}$ & $\Z_2$&$\frac{N}{2}\operatorname{Pf}(n)$ &${N\over 4} \;n^2$ \\
        $\Spin{8N}$& $\Z_2 \times \Z_2$ &$\frac{1}{2} \left({1\over 4} {\epsilon_{\mu\nu\lambda\sigma} n^+_{\mu\nu} n^-_{\lambda\sigma}} \right)$ &${N\over 2} \;(n_+^2 + n_-^2) - {1 \over 2}\; n_+ n_-$ \\
        $\Spin{8N+4}$&$\Z_2 \times \Z_2$  &$\frac{1}{2}\left(\operatorname{Pf}(n^+)+\operatorname{Pf}(n^-)\right)$ &$({N\over 2} + {1 \over 4})\; (n_+^2 + n_-^2)$ \\
        $\Spin{4N+2}$ &$\Z_4$ &$\frac{1+2N}{4}\operatorname{Pf}(n)$ & ${1+2N \over 8} \; n^2$\\
        $\Spin{2N+1}$& $\Z_2$ &0 & ${1 \over 2}\; n^2$\\
        $\E{6}$& $\Z_3$ & $\frac{1}{3}\operatorname{Pf}(n)$&${2 \over 3} \; n^2$\\
        $\E{7}$&$\Z_2$ & $\frac{1}{2}\operatorname{Pf}(n)$ & ${1 \over 4} \; n^2$\\ \hline
    \end{tabu}}
    \caption{Summary of the topological charges $\bmod{1}$ on $\T^4$ for all gauge groups with non-trivial center, derived in section \ref{appx:Qforallgroups}. The third column shows the result of our calculation of the topological charge on the non-spin manifold $\mathbb{CP}^2$, where  $n, n^\pm$ are the corresponding integer twists, for derivation and explanation, see appx.~\ref{appx:nonspin}}
    \label{tab:top_charges}
\end{table}

Table \ref{tab:top_charges} summarizes our results for the various groups. The numbers given in the table follow from eqn.~(\ref{a19}) and are obtained in what  follows, beginning with the groups with cyclic center.

\subsubsection{Groups with cyclic center}

\paragraph{$\mathbf{\SU{N}}$:}
For $\SU{N}$ we have $\vecb{\mu}^*=\vecb{w}_{N-1}^*=\sqrt{\frac{2(N-1)}{N}}\vecb{e}_{N-1}$ giving us $(\vecb{\mu}^*)^2=\frac{2(N-1)}{N}$, and $\vecb{\alpha}_\text{max}^2=1$, so we find $Q_\text{top}=\frac{N-1}{N}\operatorname{Pf}(n)+\Z$, as expected.

\paragraph{$\mathbf{\Sp{N}}$:}
For $\Sp{N}$ we have $\vecb{\mu}^*=\vecb{w}_{N}^*=\frac{1}{2}\sum_{j=1}^N\vecb{e}_j$ giving us $(\vecb{\mu}^*)^2=\frac{N}{4}$, and $\vecb{\alpha}_\text{max}^2=4$, so we find $Q_\text{top}=\frac{N}{2}\operatorname{Pf}(n)+\Z$.

\paragraph{$\mathbf{\Spin{4N+2}}$:}
For $\Spin{4N+2}$ since the two chiral representations share the same center, we can compute the topological charge in the direct sum representation. Indeed, the normalizations were computed with this in mind. We have $\vecb{\mu}^*=\vecb{w}_{2N}^*$ if $N$ is even and $\vecb{\mu}^*=\vecb{w}_{2N+1}^*$ if $N$ is odd. In either case we find $(\vecb{\mu}^*)^2=\frac{2N+1}{4}=\frac{1}{4}+\frac{N}{2}$. We have $\vecb{\alpha}_\text{max}^2=2$, so the topological charge is $Q_\text{top}=\left(\frac{1}{4}+\frac{N}{2}\right)\operatorname{Pf}(n)+\Z=\frac{1+2N}{4}\operatorname{Pf}(n)+\Z$.

\paragraph{$\mathbf{\Spin{2N+1}}$:}
For $\Spin{2N+1}$ we have $\vecb{\mu}^*=\vecb{w}_{2k+1}^*$ for $1\leq k<(N-1)/2$ giving us $(\vecb{\mu}^*)^2=2k+1$, and $\vecb{\alpha}_\text{max}^2=2$, so we find $Q_\text{top}=(2k+1)\operatorname{Pf}(n)+\Z\in\Z$.

\paragraph{$\mathbf{\E{6}}$:}
For $\E{6}$ we have $\vecb{\mu}^*=\vecb{w}_{a}^*$ for $a=1,4$ or $\vecb{\mu}^*=2\vecb{w}_b^*$ for $b=2,5$, each of these gives us $(\vecb{\mu}^*)^2=\frac{1}{3}+\Z$, and $\vecb{\alpha}_\text{max}^2=2$, so we find $Q_\text{top}=\frac{1}{3}\operatorname{Pf}(n)+\Z$.

\paragraph{$\mathbf{\E{7}}$:}
For $\E{7}$ we have $\vecb{\mu}^*=\vecb{w}_{a}^*$ for $a=4,6,7$ giving us $(\vecb{\mu}^*)^2=\frac{1}{2}+\Z$, and $\vecb{\alpha}_\text{max}^2=2$, so we find $Q_\text{top}=\frac{1}{2}\operatorname{Pf}(n)+\Z$.

\subsubsection{$\mathbf{\Spin{4N}}$}
For $\Spin{4N}$ we have to treat the two chiral representations separately in $Z_{\mu\nu}$, where now we have $\vecb{n}_{\mu\nu}=n_{\mu\nu}^+\vecb{\mu}^*_++n_{\mu\nu}^-\vecb{\mu}^*_-+\vecb{\alpha}_{\mu\nu}^*$, where the $\Z_2^{+} \times \Z_2^{-}$ twists $n^{\pm}_{\mu\nu}$ are mod 2 integers. We take $\vecb{\mu}^*_+=\vecb{w}_{2N}^*$ for even $N$, and $\vecb{\mu}^*_+=\vecb{w}_{2N-1}^*$ for odd $N$, while we take $\vecb{\mu}^*_-=\vecb{w}_{2N-1}^*$ for even $N$, and $\vecb{\mu}^*_-=\vecb{w}_{2N}^*$ for odd $N$.  

As $Spin(4N)$ is a special case, let us be more explicit. 
Each of the transition functions \eqref{omega1} is periodic up to a center element, eqn.~\eqref{direct2}, in the  $S^+\oplus S^-$ representation with generators $\vecb{H} = {\rm diag}( \vecb{H}_+, \vecb{H}_-)$. Explicitly, for even $N$, we take
\begin{equation}\label{spin_trans}
\Omega_\mu = \exp(2 \pi i \sum\limits_\nu( n_{\mu\nu}^+ {x^\nu \over 2 L_\nu}\vecb{w}_{2N}^*+n_{\mu\nu}^- {x^\nu \over 2 L_\nu}\vecb{w}_{2N-1}^*)\cdot\vecb{H})~,
\end{equation} 
where, for brevity, we ignored $\vecb{\alpha}_{\mu\nu}^*$ (restored below). We then  evaluate (\ref{vanbaaltwo}) as in deriving \eqref{eqn:top_unsimple}, using  \eqref{dynkinspdirect} and keeping in mind footnote \ref{pitfallnote}. For even $N$, we find
\begin{eqnarray}
    Q_\text{top}&=&\frac{\vecb{\alpha}_\text{max}^2}{4}\varepsilon_{\mu\nu\alpha\beta}\frac{n_{\nu\alpha}^+\vecb{w}_{2N}^* + n_{\nu\alpha}^-\vecb{w}_{2N-1}^* + \vecb{\alpha}_{\nu\alpha}^*}{2}\cdot\frac{n_{\mu\beta}^+\vecb{w}_{2N}^* + n_{\mu\beta}^-\vecb{w}_{2N-1}^* + \vecb{\alpha}_{\mu\beta}^*}{2}\nonumber \\
    &=&\frac{\vecb{\alpha}_\text{max}^2}{2}\left((\vecb{w}_{2N}^*)^2\operatorname{Pf}(n^+)+(\vecb{w}_{2N-1}^*)^2\operatorname{Pf}(n^-)+\frac{\vecb{w}_{2N}^*\cdot\vecb{w}_{2N-1}^*}{8}\varepsilon_{\mu\nu\alpha\beta}(n_{\nu\alpha}^+n_{\mu\beta}^-+n_{\nu\alpha}^+n_{\mu\beta}^-)\right)+\Z\\
    &=&\frac{\vecb{\alpha}_\text{max}^2}{2}\left((\vecb{w}_{2N}^*)^2\operatorname{Pf}(n^+)+(\vecb{w}_{2N-1}^*)^2\operatorname{Pf}(n^-)+\vecb{w}_{2N}^*\cdot\vecb{w}_{2N-1}^*(\operatorname{Pf}(n^++n^-)-\operatorname{Pf}(n^+)-\operatorname{Pf}(n^-))\right)+\Z, \nonumber
\end{eqnarray}
where we made use of the identity $\operatorname{Pf}(a+b)=\operatorname{Pf}(a)+\operatorname{Pf}(b)+\frac{1}{4}\varepsilon_{\mu\nu\alpha\beta}a_{\mu\nu}b_{\alpha\beta}$. When $N$ is odd we simply swap $n^+$ for $n^-$, and we can see that $Q_\text{top}$ will still be of the same form since $(\vecb{w}_{2N-1}^*)^2=(\vecb{w}_{2N}^*)^2$. In particular, $(\vecb{w}_{2N-1}^*)^2=\frac{2N}{4}=\frac{N}{2}$ and $\vecb{w}_{2N}^*\cdot\vecb{w}_{2N-1}^*=\frac{(2N-1)-1}{4}=\frac{N-1}{2}$. Plugging these in, and using the fact that $\vecb{\alpha}_\text{max}^2=2$, we find the following expression for $Q_\text{top}$,
\begin{align}
    Q_\text{top}&=\frac{N}{2}\left(\operatorname{Pf}(n^+)+\operatorname{Pf}(n^-)\right)+\frac{N-1}{2}\left(\operatorname{Pf}(n^++n^-)-\operatorname{Pf}(n^+)-\operatorname{Pf}(n^-)\right)+\Z\\
    &=\begin{cases}
    \frac{\operatorname{Pf}(n^++n^-)-\operatorname{Pf}(n^+)-\operatorname{Pf}(n^-)}{2}+\Z & N\text{ even}\\
    \frac{\operatorname{Pf}(n^+)+\operatorname{Pf}(n^-)}{2}+\Z & N\text{ odd}
    \end{cases}.
\end{align}
We note the difference between $N$ even and $N$ odd: when $N$ is even we can get fractional topological charge only when we turn on 't Hooft fluxes for both representations, while when $N$ is odd we can get fractional topological charge only when we turn on just one 't Hooft flux out of the two representations.

\subsection{Fractional topological charge on  $\mathbb{CP}^2$ } 
\label{appx:nonspin}

This appendix is included here merely for completeness, due to its close resemblance of the calculations already done on $\T^4$. At present, we are not aware of any relation to the Hamiltonian framework which is our main interest in this paper. Nonetheless, we note that ref.~\cite{Cordova:2019uob} quoted the fractional topological charges due to backgrounds gauging the $1$-form symmetry on non-spin manifolds. In particular, their results imply the existence of a ``$\theta$-periodicity anomaly'' on such manifolds in cases when  no anomaly is present on spin manifolds, as in the $Spin(2N+1)$ case on $\T^4$, as per  Table \ref{tab:top_charges}. 

We feel that for future applications,  it may be useful to have a more pedestrian derivation of the fractional topological charge on non-spin manifolds as well, akin to our $\T^4$  calculation. The main
  point of this appendix is that the results quoted in \cite{Cordova:2019uob} can be understood in the explicit framework of \cite{Anber:2020gig}. It is based on considering ``'t Hooft flux'' backgrounds proportional to the K\" ahler 2-form of $\mathbb{CP}^2$,  the well-known explicit example of a compact non-spin manifold. In \cite{Anber:2020gig}, only $SU(N)$ gauge groups were considered. Here, we generalize the computation of the fractional $Q_\text{top}$ in 
  't Hooft flux backgrounds to the other gauge groups. 
  
  To set the stage, let us return to $\T^4$ and note that our calculation of $Q_\text{top}$ relied on using  transition functions $\Omega_\mu$ (\ref{omega1}) which obey a co-cycle condition twisted by center elements, as in \eqref{twist1}. The fractional part  of the topological charge, naturally, only depends on the twists $n_{\mu\nu}$. Thus, the calculation of $Q_\text{top}$ can be made using any particular gauge background on $\T^4$, periodic up to transition function $\Omega_\mu$ which obey the same co-cycle conditions. For example,  we can take the following background, switching to form notation to be used later:
  \begin{equation}\label{tflux1}
  A = A_\lambda dx^\lambda =   \vecb\mu^* \cdot \vecb{H}\; \sum\limits_{\nu,\mu}{\pi x^\nu d x^\mu  {n}_{\mu\nu} \over   L_\mu L_\nu} \; ~, 
  \end{equation}
  which obeys
  \begin{equation}~ A (x_\nu=L_\nu) = \Omega_\nu \circ A (x_\nu=0)~,~ {\rm with} ~ \Omega_\nu = e^{i 2 \pi  {x^\mu\over 2 L_\mu} {n}_{\nu\mu} \vecb\mu^* \cdot \vecb{H}} ~,
  \end{equation}
exactly as in (\ref{match1},\ref{omega1},\ref{fmu}) (with the convenient co-weight $\vecb\mu^*$ inserted in $\vecb{n}_{\mu\nu}$) showing that this background obeys the co-cycle conditions with the chosen twists.  To calculate the topological charge, we can then use the constant field strength of (\ref{tflux1}) \begin{equation}\label{tflux2}
  F =  \vecb\mu^* \cdot \vecb{H}  \sum\limits_{\nu,\mu}{ \pi {n}_{\mu\nu}  \over L_\mu  L_\nu}    \; dx^\nu \wedge dx^\mu  ~,
  \end{equation}
  and use (\ref{a4}), rewritten in form notation, to obtain \eqref{a19}:
  \begin{equation} \label{qtopt4flux}
  Q_\text{top} = {\vecb\alpha_{\text{max}}^2 \over C(R)} \int\limits_{\T^4} Tr_R \;{ F \wedge F \over 16 \pi^2} = {\vecb\alpha_{\text{max}}^2 \; (\vecb\mu^*)^2\over 2} \;\operatorname{Pf}(n)~,
  \end{equation} 
the result obtained earlier.\footnote{For $Spin(4N)$ we need to simply replace $\vecb{\mu}^* n_{\mu\nu}$ by
$\vecb{w}^*_{2N} n_{\mu\nu}^+ + \vecb{w}^*_{2N-1} n_{\mu\nu}^-$ for even $N$ (and the identical expression with $n_{\mu\nu}^+$ and $n_{\mu\nu}^-$ interchanged for odd $N$).}

Next, following \cite{Anber:2020gig}, we generalize the background flux (\ref{tflux2}) to one appropriate to $\mathbb{CP}^2$. 
$\mathbb{CP}^2$ is a compact manifold, the set of  lines in the three-dimensional complex space, $\mathbb C^3$, passing through the origin. $\mathbb{CP}^2$ can be described by the complex coordinates $\Xi=(\xi_1,\xi_2, \xi_3)\neq (0,0,0)$ (here $\xi_{1,2,3}\in \mathbb C$) modulo the identification $\Xi\equiv \lambda \Xi$ for any complex number $\lambda \neq 0$. 

We now quickly review some facts about $\mathbb{CP}^2$ that we shall need, see \cite{Gibbons:1978zy,Eguchi:1980jx} or the appendix of \cite{Anber:2020gig} for details and derivations.  
One can cover $\mathbb{CP}^2$ with three patches $U_i$  ($i=1,2,3$, where $U_i$ covers $\xi_i \ne 0$)  such that the transition functions on the overlap $U_i\cap U_j$ are holomorphic. In our discussion below, we shall consider one patch, the $U_3$ patch with $\xi_3 \ne 0$. Thus, we take $z^1 \equiv \xi^1/\xi^3, z^2 \equiv \xi^2/\xi^3$. At the points $\xi^3=0$ in $\mathbb{CP}^2$, we have $(\xi^1, \xi^2) \equiv \lambda (\xi^1, \xi^2)$, i.e. a two-sphere  $\S^2=\mathbb{CP}$$^1$. We now introduce   polar coordinates $r,\theta,\phi,\psi$ 
\begin{eqnarray}
  \label{polar1} 
  z_1 =   r\cos\frac{\theta}{2} \; e^{i {\psi + \phi \over 2}} ~,~~ 
  z_2 =   r\sin\frac{\theta}{2} \; e^{i {\psi - \phi \over 2}}~,
\end{eqnarray}
where $0\leq r<\infty$, $0\leq \theta<\pi$, $0\leq \phi<2\pi$, $0\leq \psi<4\pi$, and note that the $\mathbb{S}^2$ is at $r \rightarrow \infty$.
In these coordinates, the
 Fubini-Study metric on $\mathbb{CP}^2$ is
\begin{equation}\label{metricpolar}
ds^2 = {d r^2 \over (1+r^2)^2} + {r^2\over 4(1+r^2)^2} (d \psi + \cos\theta d\phi)^2 + {r^2 \over 4 (1 + r^2)} (d \theta^2 + \sin^2\theta d \phi^2)~.
\end{equation}
To study the points at $r \rightarrow \infty$, one can introduce a new coordinate $u=1/r$ and observe that at $u=0$ there is a $\S^2$ (or $\mathbb{CP}^1$) of area $\pi$ (the metric is well behaved at $u=0$ and the singularity apparent in the first two terms of (\ref{metricpolar}) at $1/r = u \rightarrow 0$ is only a coordinate one). We also note that we have scaled to dimensionless coordinates, where the Ricci tensor of the metric (\ref{metricpolar}) is $R_{ab} = 6 \delta_{ab}$ and that $\mathbb{CP}^2$ is a solution of the Euclidean vacuum Einstein equations with cosmological constant $\Lambda = 6$.

Of most importance to us are the following two facts. 

First, $\mathbb{CP}^2$ is a K\" ahler manifold, with an anti-selfdual  K\"{a}hler $2$-form. In the coordinates we use, it is
\begin{eqnarray}\label{2-form in polar} 
K=\frac{r}{(1+r^2)^2}dr\wedge \left(d\psi+\cos\theta d\phi\right)-\frac{1}{2}\frac{r^2}{1+r^2}\sin\theta d\theta\wedge d\phi, \end{eqnarray}and obeys
\begin{equation}\label{2formintegral}
~\int_{\mathbb{CP}^2} K\wedge K=\frac{8 \pi^2}{2} ~ {\rm and} ~ \oint_{\mathbb{S}^2} K = - 2 \pi~.
\end{equation}
The first integral above is a straightforward integration of $K$ over $\mathbb{CP}^2$, while the second  is an integral over the $\mathbb{S}^2$ (or $ \mathbb{CP}^1$) located at $r \rightarrow \infty$ in the coordinates of (\ref{metricpolar}) (take the limit $r\rightarrow \infty$ and integrate  $K$  over the $\S^2$ parametrized by $\theta$ and $\phi$). The importance of the $\S^2$ is that one can thread a 't Hooft flux through it. 

Second, $\mathbb{CP}^2$ is a classic example of a non-spin manifold \cite{Geroch:1968zm,Geroch:1970uv,Hawking:1977ab}. A quick way to see the difficulty of defining spinors is to calculate the index of the Dirac operator in the $\mathbb{CP}^2$ gravitational background via the index theorem and find that it has the non-integer value $-1/8$, clearly implying an inconsistency (see, e.g. \cite{Anber:2020gig} for the relevant formulae). Here, we will  use the procedure of \cite{Anber:2020gig} to turn on 't Hooft fluxes, consistent with the transition functions on $\mathbb{CP}^2$ with gauged 1-form symmetry. As discussed in that reference, to avoid backreaction on the manifold, we  turn on an anti-self dual field strength proportional to the K\" ahler form (its energy momentum tensor is zero owing to the self-duality).

We shall now show that the background generalizing the $\T^4$ background from eqn.~(\ref{tflux2}), for groups with cyclic center (see below for a generalization to $Spin(4N)$), is 
\begin{equation}\label{tfluxcp2}
  F =  C \vecb\mu^* \cdot \vecb{H}   \; K  ~,
  \end{equation}
where $C$ is a constant and $K$ is the K\" ahler form (\ref{2-form in polar}).
One way to argue\footnote{A quick consistency check is to note that the expression (\ref{tfluxcp2})   for $F$, upon integration over the non-contractible $\S^2 \in \mathbb{CP}^2$ yields $e^{i \oint_{\S^2} F} = e^{ - 2 \pi i C \vecb\mu^* \cdot \vecb{H}}$ $\in Z(G)$ for $C \in\Z$, as appropriate for a 't Hooft flux.} for the value of $C$ is to consider the $\S^2$ at $r \rightarrow \infty$ and study the transition functions for the gauge potential. On $\S^2$,  $K = -{1 \over 2} \sin \theta d \theta \wedge d \phi =  d \;( {\cos\theta d \phi\over 2})$. Thus, $F = d A_\pm$, where $A_\pm = C \vecb\mu^* \cdot \vecb{H} \; {1 \over 2} \; (\pm 1+ \cos \theta) d \phi$. The connection $A_+$ should be taken at  $\theta \ne 0$ (the southern hemisphere) and $A_-$ at $\theta \ne \pi$ (the northern hemisphere). The transition function $\Omega(\phi)$ on the equator can be found from $A_+ - A_- = - i \Omega d \Omega^{-1}$ to equal $\Omega(\phi) = e^{ i  C  \vecb\mu^* \cdot \vecb{H} \phi}$; it is not periodic, $\Omega(2 \pi) = e^{i  C 2 \pi  \vecb\mu^* \cdot \vecb{H}} \Omega(0)$. However, it is periodic up to a center element  provided that $ C $ is an integer. Thus, from now on we take $C = n$, $n \in \Z$.

Now we can repeat the computation of  the topological charge (\ref{qtopt4flux}) of the background (\ref{tflux2})   on $\T^4$ for the case of on $\mathbb{CP}^2$ in  the background (\ref{tfluxcp2}), making use of  (\ref{2formintegral}),
\begin{equation} \label{qtopcp2flux}
  Q_\text{top} = {\vecb\alpha_{\text{max}}^2 \over C(R)} \int\limits_{\mathbb{CP}^2} Tr_R \;{ F \wedge F \over 16 \pi^2} = {\vecb\alpha_{\text{max}}^2 \; (\vecb\mu^*)^2}~{n^2 \over 16 \pi^2} \int_{\mathbb{CP}^2} K\wedge K = {\vecb\alpha_{\text{max}}^2 \; (\vecb\mu^*)^2 \over 4}~{n^2} ~.
  \end{equation} 
For the groups with cyclic center, this is $1/2$ the expression (\ref{qtopt4flux}) obtained on $\T^4$, with $\operatorname{Pf}(n) \rightarrow n^2$. To translate this into the actual fractional value of $Q_\text{top}$ on $\mathbb{CP}^2$ shown in the third column of table~\ref{tab:top_charges} requires some care (notably for $E_6$).

For $Spin(4N)$, with even $N$, we take instead  $F =  (n^+ \vecb{w}_{2N}^* + n^- \vecb{w}_{2N-1}^*) \cdot \vecb{H} \; K$, where $n^{\pm}$ are now two integers; and we replace $(\vecb\mu^*)^2 {n^2}$ in (\ref{qtopcp2flux}) by $(n^+ \vecb{w}_{2N}^* + n^- \vecb{w}_{2N-1}^*)^2$. For odd $N$, we instead replace 
$(\vecb\mu^*)^2 {n^2}$ by $(n^- \vecb{w}_{2N}^* + n^+ \vecb{w}_{2N-1}^*)^2$, i.e. interchange $n^+$ and $n^-$. Collecting everything, we now summarize the result for the groups with $\Z_2 \times \Z_2$ center
\begin{eqnarray}
\label{qtopcp2spin4n}  
Q_\text{top}[{Spin(8p)}] &=& {p\over 2} (n_+^2 + n_-^2) - {1 \over 2} n_+ n_- + \Z~, \\
Q_\text{top}[{Spin(8p+4)}] &=&({p\over 2} + {1 \over 4})(n_+^2 + n_-^2)  + \Z~~.
\end{eqnarray}
Our results for $\mathbb{CP}^2$ topological charges  in the backgrounds with 't Hooft fluxes (labeled by $n$ for the groups with cyclic centers and $n^\pm$ for $Spin(4N)$) summarized in table~\ref{tab:top_charges} agree with the results quoted in \cite{Cordova:2019uob}.

   \bibliography{flux1.bib}
 
  \bibliographystyle{JHEP}
\end{document}